\documentclass[11pt,letter]{article}

\usepackage[affil-it]{authblk}
\usepackage{natbib}
\usepackage{amssymb,amsmath,amsfonts,amstext}
\usepackage{graphicx}
\usepackage{subfigure}
\usepackage{mathtools}
\usepackage{stmaryrd}
\usepackage{color}
\usepackage{fullpage}

\usepackage{amsthm}
\usepackage{esint}

\usepackage{float}
\usepackage{subfloat}
\usepackage{rotating}
\newcommand{\X}{\mathbf{X}}
\newcommand{\x}{\mathbf{x}}
\newcommand{\Pb}{\mathbf{P}}
\newcommand{\B}{\mathbf{B}}
\newcommand{\p}{\boldsymbol \varphi}
\newcommand{\F}{\mathbf{F}}
\newcommand{\N}{\mathbf{N}}
\newcommand{\M}{\text{M}}

\title{\textbf{Computational homogenization of heterogeneous media under dynamic loading}}

\author{Chenchen Liu, Celia Reina\footnote{Corrseponding author: creina@seas.upenn.edu}}
\affil{Department of Mechanical Engineering and Applied Mechanics, University of Pennsylvania, Philadelphia, PA 19104, USA}

\begin{document} 

\maketitle
\begin{abstract}
A variational coarse-graining framework for heterogeneous media is developed that allows for a seamless transition from the traditional static scenario to a arbitrary loading conditions, including inertia effects and body forces. The strategy is formulated in the spirit of computational homogenization methods (FE$^2$) and is based on the discrete version of Hill's averaging results recently derived by the authors. In particular, the traditional static multiscale scheme is proved here to be equivalent to a direct homogenization of the principle of minimum potential energy and to hold exactly under a finite element discretization. This perspective provides a unifying variational framework for the FE$^2$ method, in the static setting, with Dirichlet or Neumann boundary conditions on the representative volume element; and it directly manifests the approximate duality of the effective strain energy density obtained with these two types of boundary conditions in the sense of Legendre transformation. Its generalization to arbitrary loading conditions and material constitutive relations is then immediate through the incremental minimum formulation of the dynamic problem {\`a} la Radovitzky and Ortiz (1999, Comput. Methods in Appl. Mech. and Eng. 172(1), 203-240), which, in the discrete setting, is in full analogy to the static problem.  Interestingly, this coarse-graining framework is also applicable to atomistic simulations, directly revealing the analogy between the stress tensor resulting from the homogenization procedure in the presence of micro inertia effects, and the Virial stress tensor commonly used to characterize molecular systems. These theoretical developments are then translated into an efficient multiscale {$\textup{FE}^2$} computational strategy for the homogenization of a microscopic explicit dynamics scheme, with two noteworthy properties. Firstly, each time incremental problem can be solved exactly with a single Newton-Raphson iteration with a constant Hessian, regardless of the specific non-linearities or history-dependence of the micro-constituents' behavior; and the method thus represents a quasi-explicit multiscale solver (QEMS). Secondly, the scheme concurrently solves for the microscopic and macroscopic degrees of freedom, in contrast to standard approaches based on sequential or nested minimizations. The method is illustrated over a simple one-dimensional layered composite and its dynamic response is compared to that obtained via a standard single scale finite element procedure. The results indicate the capability of QEMS to capture the dynamic behavior in the range of frequencies permitted by the premises of separation of length scales. In particular the dispersion properties are accurately predicted and the expected convergence behavior to the exact solution is observed with respect to both, the spatial and temporal discretizations.
\end{abstract}


\section{Introduction}


Direct simulation of materials with fine microstructure is often prohibitively expensive, despite the advent of increasingly large high performance computing platforms. This is even more true for the recently developed metamaterials with engineered nanostructure, where the constituents' length scales may even go below the micron range \citep{schaedler2011ultralight}. Techniques that take advantage of such separation of length scales and provide the homogenized behavior are therefore of high interest. Among the analytical methods, one can highlight, without being comprehensive, variational methods such as $\Gamma-$convergence, asymptotic homogenization schemes, variational bounds, and exact relations for specific microstructures. For details on these techniques we refer the readers to \citet{braides2002gamma,bensoussan1978,sanchez1980non,milton2002theory,tartar2009general,weinan2011principles,fish2013practical} and references there in. These methods have been extremely successful, primarily in the static case, with a recent increased interest in the dynamic setting where micro-inertia plays a crucial role, e.g.~\cite{sabina1988simple,willis1997dynamics,chen2001dispersive,andrianov2008higher,srivastava2012overall,fish2012micro,nassar2015willis}. They often provide closed-form solutions for the effective behavior, which may directly be used in computations. However, in scenarios where the microstructure undergoes complex evolutions, or when the dynamic behavior is influenced by the finite size or non-periodic nature of the sample, these methods may no longer be suitable. Computational homogenization methods, such as FE$^2$ techniques, represent an alternative 
to directly incorporate micromechanical information into a finite element procedure \citep{suquet1987elements,hou1997multiscale,kouznetsova2002multi}. In these schemes, the effective behavior is obtained at each quadrature point of the macroscopic discretization by solving a coupled boundary value problem on a representative volume element (RVE) of the microstructure. The {$\textup{FE}^2$} procedure has been applied to various fields, including composites \citep{feyel2000fe,terada2000simulation}, polycrystalline materials \citep{miehe1999computational,miehe2002homogenization,blanco2014variational}, elastic and plastic porous media \citep{smit1998prediction,reina2013micromechanical}, adhesives  \citep{matouvs2008multiscale}, quasi-brittle separation laws  \citep{nguyen2011homogenization} and wave propagation in metamaterials \citep{pham2013transient}.


The theoretical foundations of the micro-macro coupling in FE$^2$ methods lies on the celebrated averaging theorems of \citet{hill1963elastic,hill1972constitutive}. These original statements were framed in the static setting, and are thus only valid when the wavelength of the waves traveling through the material are much larger than the size of the heterogeneities. However, with the growing interest in the effects of micro-inertia on the macroscopic transient response of both, natural \citep{jacques2012effects} and man made materials \citep{kushwaha1993acoustic,camley1983transverse}, various extensions of Hill's averaging relations to the dynamic setting have been proposed \citep{molinari2001micromechanical,wang2002modeling,reina2011multiscale,jacques2012effects,pham2013transient,de2015rve}. 

\begin{figure}[htbp]
\begin{center}
    {\includegraphics[width=\textwidth]{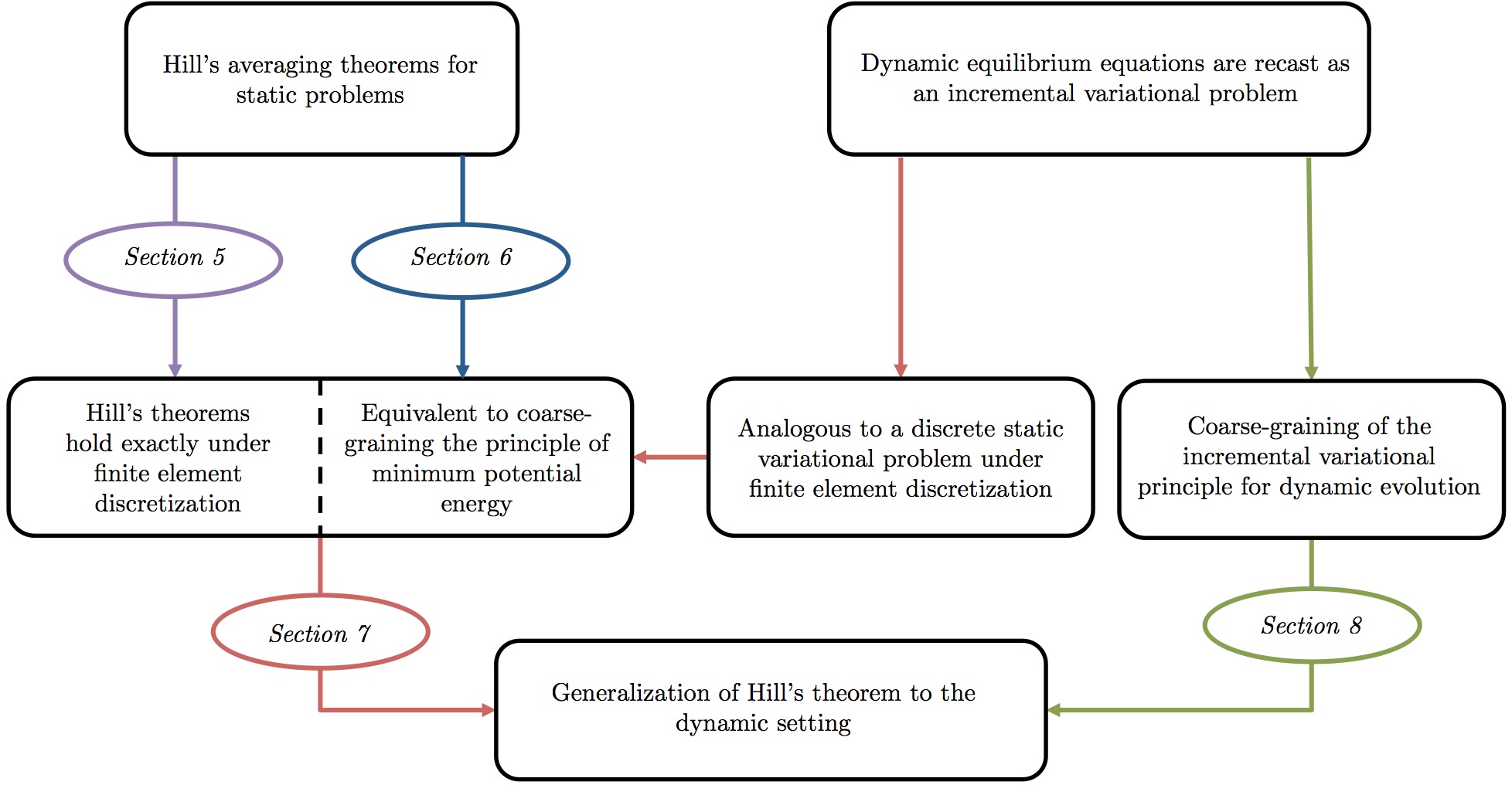}}
    \caption[]{Rational for extending Hill's averaging relations to the dynamic case. } \label{fr}
\end{center}
\end{figure}

In this work we provide a general framework for the micro-to-macro transition that allows to seamlessly consider arbitrary loading conditions (static or dynamic, with or without body forces); material behavior (elastic, viscoelastic or plastic) as well as continuous descriptions or discrete finite element approximations; and it does not require the introduction of any additional hypothesis or kinematic relation with respect to the traditional static scenario. It therefore provides a unified treatment for the various extensions of Hill's theorems that have been proposed in the literature, cf. \cite{ricker2009multiscale,reina2011multiscale,de2015rve} for problems with body forces; and \citep{chenchen2015} for a discrete version of the averaging relations. The rational for the framework is depicted in Figure \ref{fr}, and it is based on a variational coarse-graining of the principle of minimum potential energy (incremental principle for the dynamic case, cf.~\cite{radovitzky1999error}), which is justified both mathematically and physically. Interestingly, it encloses the approximate duality between the static problem with Dirichlet or Neumann boundary conditions for the RVE in the sense of Legendre transformation; and it can also be applied to atomistic models recovering the well-known Virial formula for the stresses. The variational nature of the proposed multiscale framework naturally leads to a FE$^2$ computational homogenization strategy. In particular, the coarse-graining of an explicit dynamic integration scheme for the dynamic problem leads to a numerical approach with two noteworthy properties. Firstly, the method is quasi-explicit, in the sense that it may be solved within a single Newton-Raphson iteration. The Hessian of the incremental variational problem is constant, even for non-linear and history dependent materials, and exclusively depends on the spatio-temporal discretization. It may therefore be pre-computed at the beginning of the simulation delivering a highly-efficient numerical scheme. Secondly, the quasi-explicit multiscale solver (QEMS) concurrently solves for the micro and macro degrees of freedom, in contrast to the traditional sequential or nested solvers. 

The manuscript is organized as follows. After introducing some notation in Section \ref{Sec:Notation}, the classical averaging results for the static problem are reviewed in Section \ref{Sec:Review}. These are recast as a variational problem in Section \ref{Sec:Rational}, providing a rational for its extension to general loading conditions. Such generalization is achieved in 3 steps, cf.~Figure \ref{fr}. Firstly,  in Section \ref{Sec:Discrete}, the static averaging results are shown to hold exactly under a finite element discretization. Secondly, the variational framework is naturally extended, in the discrete case, to the case of body forces in Section \ref{Sec:BodyForces}, and to the dynamic setting in Section \ref{Sec:DiscreteDynamicProblem}. Finally, in Section \ref{Sec:GeneralContinuum} the general micro-macro relations are derived in the continuum framework and are shown to coincide with their discrete counterparts, indicating that the averaging and discretization operations commute. Next, in Section \ref{Sec:VirialStress}, the formula for the Virial stresses associated to atomistic simulations is recovered from this general formulation. The theoretical developments are then used to formulate a concurrent quasi-explicit multiscale scheme (QEMS), which is explained in detail in Section \ref{Sec:MultiscaleSolver}. The numerical strategy is applied in Section \ref{Sec:1DProblem} to a one-dimensional stratified composite and compared to the solutions obtained with a traditional single scale finite element method. The results indicate that the FE$^2$ method is capable of accurately predicting the dispersion properties of the material. Finally, Section \ref{Sec:Conclusions} provides some conclusions for the present investigation.
 
\section{Notation} \label{Sec:Notation}
The analyses in the following sections are cast in the standard Lagrangian description of deformable bodies, where the position at time $t$ of every material point $\X$ in the reference configuration $\Omega_0$ is given by $\x=\p(\X,t)$. The deformation gradient $\F$ is defined as $\F = \nabla \p$. In the dynamic case (the static case being an obvious particularization), the deformation mapping $\p(\X,t)$ is required to satisfy the following initial boundary value problem
\begin{align} \label{Eq::EquilibriumEquationsDynamic}
& \nabla \cdot \Pb + \rho_0 \B=\rho_0 \ddot{\p} , \qquad \text{in }  \Omega_0, \\
& \boldsymbol \varphi = \bar{\boldsymbol \varphi}, \qquad \text{on}\ \partial \Omega_{0,1}, \\ \label{Eq:Traction_bcs}
&  \mathbf{P} \mathbf{N} =  \bar{\mathbf{T}}, \qquad \text{on}\ \partial \Omega_{0,2}, \\
&\p(\X,t=0)=\p_0(\X),\ \dot{\p}(\X,t=0)=\mathbf{V}_0(\X), \qquad \text{in }  \Omega_0,
\end{align}
where $\bar{\boldsymbol \varphi}$ and $\bar{\mathbf{T}}$ are the prescribed deformation mapping and the prescribed traction, respectively, and $\p_0(\X)$ and $\mathbf{V}_0(\X)$ are the initial position and velocity fields. In the equations above, $\Pb$ denotes the first Piola-Kirchhoff stress tensor, $\rho_0$ is the density per unit volume in the reference configuration, $\B$ are the body forces per unit mass, $\mathbf{N}$ is the outward unit normal to the domain $\Omega_0$, and $\dot{\p}$ and $\ddot{\p}$ denote the material velocity and acceleration respectively, i.e.,~$\dot{\p}=\frac{\partial \p}{\partial t}(\X,t)$ and $\ddot{\p}=\frac{\partial^2 \p}{\partial t^2}(\X,t)$. As usual, it is assumed that the boundary of the domain satisfies $\partial \Omega_0 = \partial \Omega_{0,1} \cup \partial \Omega_{0,2}$, and $\partial \Omega_{0,1} \cap \partial \Omega_{0,2} = \emptyset$. 


Two levels of description will be considered in the following: macroscopic and microscopic. In order to differentiate them, the superscript $\M$ will be used to identify the macroscopic fields, whereas no superscript will be used for the microscopic quantities. Similarly, $\nabla^\M$ will be used to denote the gradient with respect to the coordinates $\X^\M$, and $\Omega_0^\M$ will denote the macroscopic domain in the reference configuration. Often, the macroscopic quantities will result as the average of the corresponding microscopic fields. This average operation will be denoted by
\begin{equation}
\langle \cdot \rangle = \frac{1}{|\Omega_0|}\int_{\Omega_0} \cdot\ dV, 
\end{equation}
where $|\Omega_0|$ denotes the volume associated to $\Omega_0$ and $dV$ is the volume differential. When needed, the surface differential will be denoted by $dS$. 
Standard indicial notation will be used in the derivations. Lower case indices will refer to the deformed configuration and upper case indices for the reference configuration. When derivatives are present, no distinction will be made in the indices with regard to the scale (macroscopic or microscopic). It will be clear from the context whether the indices refer to $\nabla$ or $\nabla^\M$.

\section{Review of computational homogenization in the static case with no body forces} \label{Sec:Review}
If the deformation of a heterogeneous material satisfies separation of length scales, i.e.,~characteristic size of the constituents (\emph{micro-length-scale}) small compared to characteristic length of the macroscopic deformation (\emph{macro-length-scale}), the stresses and strains at a macroscopic material point of the continuum may be regarded as uniform. Yet, the different constitutive behavior of the micro constituents, naturally lead to rapidly changing fields at the microstructural level, which are responsible for the macroscopic observables. The relation between the micro and macro fields and the constitutive relations at a given macroscopic material point may be derived with the help of a representative volume element (RVE) \citep{hill1963elastic}. The RVE represents the material neighborhood and needs to be sufficiently large to be statistically representative of the material's heterogeneities \citep{ostoja2006material,Balzani2014}, but small compared with the macro-length scale. This is of course possible under the assumption of separation of length scales previously stated.

Computational homogenization techniques are based on the formulation of appropriate boundary value problems on the RVE associated to macroscopic material points to obtain the effective constitutive relations, and derive, if needed, the evolution of the microstructure. The choice of the boundary conditions and the micro-macro relations are based on the seminal papers of Hill \citep{hill1967essential,hill1972constitutive} and later developments and applications \citep{rice1970structure,hashin1972theory,mandel1972plasticite,havner1982theory,willis1981variational,suquet1987elements,nemat1999averaging}. These were originally devised for the static case with no body forces and are recalled below in the context of finite deformations. See \cite{nemat2013micromechanics}, for instance, for equivalent results in the linearized kinematic setting.

Given an arbitrary stress tensor $\Pb$ in equilibrium ($\nabla \cdot \Pb = 0$), and an arbitrary  compatible deformation gradient $\F$ ($\F = \nabla \p$), not necessarily related to each other, it is readily obtained by successive application of the divergence theorem, that \citep{nemat1999averaging}
\begin{align} \label{Eq:F_ave}
&|\Omega_0| \langle F_{iJ} \rangle = \int_{\partial \Omega_0} \varphi_i N_J\, dS,\\ \label{Eq:P_ave}
&|\Omega_0| \langle P_{iJ} \rangle = \int_{\partial \Omega_0} P_{iQ} X_J N_Q\, dS,\\ \label{Eq:micro_macro_work}
&|\Omega_0|\big(\langle P_{iJ} \dot{F}_{iJ} \rangle - \langle P_{iJ}  \rangle \langle \dot{F}_{iJ} \rangle\big) = \int_{\partial \Omega_0} \left(P_{iJ}N_J - \langle P_{iJ} \rangle N_J \right) \big(\dot{\varphi}_i- \langle \dot{F}_{iQ} \rangle X_Q \big)\, dS.   
\end{align}

From classical results for oscillating functions (see for example \cite{cioranescu1999introduction}, Chapter 2), average quantities of the microscopic fields are the natural choice for the macroscopic counterparts. It is then customary to require $\Pb^\M=\langle \Pb \rangle$ and $\F^\M=\langle \F \rangle$. Equation \eqref{Eq:micro_macro_work} thus indicates that the average rate of stress work rate is equal to macroscopic stress work rate under any of the two following boundary conditions for the RVE
\begin{equation} \label{Eq:D_or_N_bc}
\p = \F^\M \X,  \quad \text{on } \partial \Omega \quad \text{or} \quad \Pb \N = \Pb^\M \N,  \quad \text{on } \partial \Omega.
\end{equation}
Furthermore, these conditions imply, cf.~Eqs.~\eqref{Eq:F_ave} and \eqref{Eq:P_ave}, that $\Pb^\M=\langle \Pb \rangle$ and $\F^\M=\langle \F \rangle$, as desired. As a result, either, the affine displacement or traction boundary conditions stated in Eq.~\eqref{Eq:D_or_N_bc} lead to a well-defined boundary value problem on the RVE from which to obtain the effective constitutive behavior. It is well known that displacement boundary conditions lead to a stiffer result than stress boundary conditions. Best results are typically obtained upon the application of periodic boundary conditions, which also lead to a null value of the right hand side of \eqref{Eq:micro_macro_work}, and thus satisfy equality of micro and macro work rate. We further note that including the macroscopic translation into the boundary conditions, i.e.,
\begin{equation} \label{Eq:BC_periodic}
\p= \p^\M + \F^\M \X + \tilde{\p},  
\end{equation}
with $\tilde{\p}$ periodic, leaves the above results invariant, cf.~Chapter 1 \citet{nemat2013micromechanics}. The term $\p^\M$ informs the RVE of the macroscopic translation, whereas the macroscopic rotation is included in $\F^\M$.

If the microconstituents of the RVE are hyperelastic, the Piola-Kirchhoff stress tensors can be derived from the free energy $W(\F)$ as $\Pb=\frac{\partial W}{\partial \F}$, and Hill's averaging results may be recast in a variational framework \citep{hill1967essential,hashin1972theory,willis1981variational}. In particular, $W^\M=\langle W (\F^*,\X)\rangle$ evaluated at the equilibrium solution for the microscopic problem, denoted by superscript *, acts as a macroscopic energy density from which the macroscopic stresses may be derived,
\begin{equation}
\Pb^\M=\frac{\partial W^\M}{\partial \F^\M}.
\end{equation}

This variational perspective can be generalized to a wide class of non-hyperelastic materials. In particular, materials exhibiting non-Newtonian viscosity, strain rate sensitivity or  arbitrary flow and hardening rules, have a pseudo-elastic strain energy density $W^{n+1} = W(\F^{n+1};\F^n,\mathbf{Q}^n)$ from which the incremental stress-strain relation may be derived, i.e.,~$\Pb^{n+1} = \frac{\partial W^{n+1}}{\partial \F^{n+1}}$, where the indices $n$ and $n+1$ refer to the time $t^n$ and $t^{n+1}=t^n+\Delta t$, respectively, and $\mathbf{Q}^n$ are the value of the internal variables at time $t^n$. These pseudo-elastic potentials can then be used in place of $W$ above, and the equivalence of micro and macro stress work rate holds for the incremental problem \citep{ortiz1999variational,miehe2002homogenization}.


\section{Rational towards a general variational coarse-graining framework} \label{Sec:Rational}
In order to generalize the previous statements to more general situations, we take a different, but equivalent perspective for the above results. In particular, the following statement is considered as the fundamental underlying principle: \emph{the energy of any subdomain is invariant with respect to the level of coarse-graining}. As a result, the macroscopic strain energy shall be defined, for energetic consistency, as $W^\M(\p^\M,\F^\M)=\big \langle W \left( \p^* \left(\p^\M,\F^\M \right),\X\right)\big \rangle$, see for instance Chapter 13 \cite{milton2002theory}. By the separation of length scales discussed on the previous case, such average is independent of the specific choice of the RVE, as long as it is statistically representative of the material.

Once the macroscopic strain energy density is identified, the macroscopic solution can then be obtained by direct application of the principle of minimum potential energy,
\begin{equation} \label{Eq:Pi_macro}
\min_{\p^\M} \Pi^\M =\min_{\p^\M} \left[ \int_{\Omega_0^\M} W^\M(\p^\M,\F^\M ) \, dV^\M - \int_{\partial\Omega_{0,2}^\M} \bar{\mathbf{T}}^\M \cdot \p^\M\, dS^\M \right]
\end{equation}
where the minimization is performed over all fields $\p^\M$ compatible with the macroscopic displacement boundary condition on $\partial \Omega^\M_{0,1}$. In particular, 
\begin{equation}\label{Eq:Macro_eq_2}
\int_{\Omega_0^\M} \left[ \frac{\partial W^\M}{\partial \p^\M} - \nabla^\M \cdot \left(\frac{\partial W^\M}{\partial \F^\M} \right) \right] \delta \p^\M \, dV^\M + \int_{\partial \Omega_{0,2}^\M} \left[ \left(\frac{\partial W^\M}{\partial \F^\M} \right) \N^\M- \bar{\mathbf{T}}^\M  \right] \delta \p^\M \, dS^\M= 0, 
\end{equation}
for all variations $\delta \p^\M$ such that $\delta \p^\M=0$ on $\partial \Omega_{0,1}^\M$. Therefore,
\begin{equation} \label{Eq:Macro_eq}
\frac{\partial W^\M}{\partial \p^\M} - \nabla^\M \cdot \left(\frac{\partial W^\M}{\partial \F^\M} \right) = 0, \quad \text{in } \Omega_0^\M \quad \text{and} \quad \left(\frac{\partial W^\M}{\partial \F^\M} \right) \N^\M= \bar{\mathbf{T}}^\M , \quad \text{on } \partial \Omega_{0,2}^\M,
\end{equation} 
where $\partial W^\M/\partial \p^\M $ and $\partial W^\M/\partial \F^\M $ can be evaluated given the microscopic equilibrium equations. For the static case in the absence of body forces, these are $\partial W^\M/\partial \p^\M = 0$ and $\partial W^\M/\partial \F^\M = \langle \Pb \rangle $. That is,
\begin{equation} \label{Eq:Macro_eq_3}
\nabla^\M \cdot  \langle \Pb \rangle  = 0, \quad \text{in } \Omega_0^\M \quad \text{and} \quad \langle \Pb \rangle  \N^\M= \bar{\mathbf{T}}^\M , \quad \text{on } \partial \Omega_{0,2}^\M,
\end{equation}
The equilibrium with the external tractions, cf.~Eq.~\eqref{Eq:Macro_eq_2}, implies that the macroscopic stress tensor shall be defined as $\Pb^\M =  \partial W^\M/\partial \F^\M$, which in this case is $\langle \Pb \rangle$. The relation $\Pb^\M =  \langle \Pb \rangle  $ thus becomes a consequence of the conservation of energy upon coarse graining.

In a similar spirit to the extension of the classical Hill theorems, in their variational form, to inelastic materials \citep{ortiz1999variational,miehe2002homogenization}, the strategy to generalize the procedure just outlined to situations where body forces and inertia are present will consist on finding an effective microscopic strain energy density from which the microscopic equilibrium equations can be obtained. This is done in three steps. 
\begin{enumerate}
\item[(i)] We first show in Section \ref{Sec:Discrete} that the above results hold true and exactly under a standard finite element discretization, i.e.,\begin{equation}\label{Eq:FE_discretization}
\p^h(\X) = \sum_a \p_a N_a(\X),
\end{equation}
where $\p_a$ are the nodal values of the deformation mapping and $N_a$ are the shape functions, smooth within each element. In this case, the microscopic strain energy density is then a function of the nodal values of the deformation mapping $\p_a$, i.e.~$W(\p_a)$. 
\item[(ii)] Next, in Section \ref{Sec:BodyForces}, we discuss for the static case with body forces $\B$ that the potential energy density constitutes an effective microscopic energy density, i.e.,~$W_{\text{eff}} (\p_a) = W(\p_a) - \rho_0\B\cdot \left(\sum_a \p_a N_a \right) $. Note, that thanks to the finite element discretization, the effective energy density depends exclusively on the nodal values $\p_a$ and is thus analogous to case (i). In the continuum setting, such effective strain energy density would depend both on $\p$ and $\F$, and would therefore not be in full analogy to a general hyperelastic strain energy density $W(\F)$.
\item[(ii)] Finally, in Section \ref{Sec:DiscreteDynamicProblem}, we account for inertia by recasting the dynamic problem as an incremental variational problem, following the work of \cite{radovitzky1999error}. The potential energy density for this dynamic problem, upon discretization, solely depends on the nodal values of the deformation mapping and can thus be considered as an effective strain energy density analogously to (i) and (ii).
\end{enumerate}

Once Hill's theorems have been generalized in a discrete finite element setting, the averaging results are rederived in the continuum framework in Section \ref{Sec:GeneralContinuum}, and the operations of discretization and coarse-graining are shown to commute, cf.~Figure~\ref{fr}.

\section{Finite element discretization and discrete micro-macro relations for the static problem with no body forces}\label{Sec:Discrete}

The derivation of the classical averaging results of Hill makes recurrent use of divergence theorem, both for the stress field and the deformation gradient. In a recent publication by the authors \citep{chenchen2015}, it is shown that relation \eqref{Eq:F_ave}-\eqref{Eq:micro_macro_work} still hold exactly (and not approximately) for a finite element formulation, where the equilibrium equations are only satisfied in a weak sense and the divergence theorem is, in general, no longer applicable. In particular, for a RVE with displacement boundary conditions
\begin{equation}\label{Eq:BC_discrete}
\p^h_a = \p^\M + \F^\M \X_a , \quad \text{or }\quad \p^h_a = \p^\M + \F^\M \X_a + \tilde{\p}_a, \quad \text{for } a \in \partial \Omega
\end{equation}
the microscopic stress tensor $\Pb^h$, which satisfies the weak equilibrium equations ($\int_{\Omega_0}P_{iJ} N_{a,J} \, dV = 0$ for the interior nodes) and the deformation gradient $\F^h = \sum_a \p_a \nabla N_a$, satisfy 
\begin{align} \label{Eq:Fave_discrete}
&\langle \F^h \rangle = \F^\M, \\ \label{Eq:Hill_discrete}
&\langle \Pb^h: \dot{\F}^h \rangle = \langle \Pb^h \rangle: \langle \dot{\F}^h \rangle, \quad \text{with }  \Pb^\M = \langle \Pb^h \rangle.
\end{align}
We note that $\Pb^h$ may fail to be continuous, as is the case, for instance, for piecewise linear shape functions, and thus the proof of Eq.~\eqref{Eq:Hill_discrete} cannot proceed as in the smooth case, by successive application of the divergence theorem. Rather, the results are obtained \citep{chenchen2015} from the weak statement of the equilibrium equations and standard properties of the shape functions \citep{hughes2012finite}: local support (each $N_a$ vanishes over any element not containing the node $a$); Kronecker delta property, $N_a(\X_{a'})=\delta_{aa'}$; partition of unity, $\sum_a N_a(\X) = 1$; and linear field reproduction, $\sum_a N_a(\X) \X_a = \X$. For $C^0$ finite elements, the displacements are continuous and Eq.~\eqref{Eq:Fave_discrete} can be derived with the help of the divergence theorem. 


This discrete version of Hill's averaging results may also be recast in a variational framework for hyperelastic materials. To show this, we separate the node set $\{ a\}$ into interior nodes (\underline{b}ody nodes) $\{b\}$ and boundary nodes (at the \underline{c}ontour of the domain) $\{c\}$. For simplicity, we consider the boundary conditions $\p^h_c = \p^\M + \F^\M \X_c$ and discuss its generalization to the case of periodic boundary conditions later, in Section \ref{Sec:PeriodicBoundaryConditions}.  Then, by the principle of minimum potential energy, the microscopic equilibrium equations are obtained as
\begin{equation} \label{Eq:eq_discrete}
\min_{\p_b} \Pi [\p_a]=\int_{\Omega_0} W\Big(\sum_a \p_a \otimes \nabla N_a\Big)\,dV\quad  \rightarrow \quad\int_{\Omega_0} P_{iJ} N_{b,J}\, dV = 0, \quad \text{for all the interior nodes $b$}.
\end{equation}
The nodal displacements at equilibrium are a function of the boundary data $\p^\M$ and $\F^\M$, and we denote them as $\p_a^*\left(\p^\M,\F^\M \right)$. By the conservation of the energy density upon upscaling procedures, the macroscopic energy density is defined as  $W^\M(\p^\M,\F^\M ) = \langle W(\p_a^*(\p^\M,\F^\M )) \rangle$, ant it allows to construct the macroscopic variational problem as in Eqs.~\eqref{Eq:Pi_macro} and \eqref{Eq:Macro_eq}. Quantities $\partial W^\M/\partial \p^\M$ and $\partial W^\M / \partial \F^\M$ can then be obtained as
\begin{align} \nonumber
|\Omega_0|\delta \big\langle W(\p_a^*(\p^\M,\F^\M )) \big\rangle &= \int_{\Omega_0} P_{iJ}\sum_a \delta \varphi_{ai} N_{a,J} \, dV = \sum_b  \int_{\Omega_0} P_{iJ} \delta \varphi_{bi} N_{b,J}\, dV + \sum_c  \int_{\Omega_0} P_{iJ} \delta \varphi_{ci} N_{c,J}\, dV \\ \nonumber
&=\sum_c  \int_{\Omega_0} P_{iJ} \delta \varphi_{ci} N_{c,J}\, dV \\  \label{Eq:VarPi}
&= \bigg[\sum_c \int_{\Omega_0} P_{iJ}  N_{c,J}\, dV\bigg]  \delta \varphi^\M_{i}  +  \bigg[ \sum_c\int_{\Omega_0} P_{iJ} X_{cQ} N_{c,J}\, dV \bigg]\delta F^\M_{iQ},
\end{align}
where we have separated the set of nodes $\{a\}$ into the sets $\{b\}$ and $\{c\}$, and we have made use of the microscopic equilibrium equations, cf.~Eqs.~\eqref{Eq:eq_discrete}, and the boundary conditions for nodes $\{c\}$. Then, from the properties of the shape functions, it is obtained that
\begin{align} \label{Eq:Properties_Shape_Functions}
&\sum_a N_{a} = 1 \quad \rightarrow \quad \sum_a N_{a,J} = 0 \quad \rightarrow \quad \sum_c N_{c,J} = -\sum_b N_{b,J}, \\ \label{Eq:ShapeFunctions_properties}
&\sum_a X_{aQ} N_a = X_Q  \quad \rightarrow \quad \sum_a X_{aQ} N_{a,J} = \delta_{QJ},
\end{align}
which allows to simply Eq.~\eqref{Eq:VarPi} as
\begin{align} \nonumber
|\Omega_0|\delta \big\langle W(\p_a^*(\p^\M,\F^\M )) \big\rangle &=   -\bigg[\sum_b \int_{\Omega_0} P_{iJ}  N_{b,J} \, dV\bigg] \delta \varphi^\M_i + \bigg[ \sum_c X_{cQ} \int_{\Omega_0} P_{iJ} N_{c,J}\, dV \bigg] \delta F^\M_{iQ} \\ \label{Eq:VarPi_2}
&= \bigg[  \int_{\Omega_0} P_{iJ} \sum_a X_{aQ} N_{a,J} \, dV \bigg] \delta F^\M_{iQ} = \bigg[  \int_{\Omega_0} P_{iQ} \, dV \bigg] \delta F^\M_{iQ} = |\Omega_0| \langle P_{iJ}\rangle \delta F^\M_{iJ},
\end{align}
where the equilibrium equations, cf.~Eq.~\eqref{Eq:eq_discrete}, have again been used.  Therefore $\partial W^\M/\partial \p^\M=0$ and $\Pb^\M=\partial W^\M/\partial \F^\M= \langle \Pb \rangle$ and the macroscopic equations read, as expected, $\nabla \cdot \Pb^\M = 0$ in $\Omega_0^\M$ and $\Pb^\M \N^\M=\bar{\mathbf{T}}^\M$ on $\partial \Omega_{0,2}^\M$.

\section{Body forces and the potential energy density} \label{Sec:BodyForces}

In the presence of body forces, $\mathbf{B}$, the solution of the RVE, when discretized as in Eq.~\eqref{Eq:FE_discretization} and subjected to affine displacement boundary conditions, can be obtained via the principle of minimum potential energy
\begin{equation}\label{Eq:Pi_body_discrete}
\min_{\p_b} \Pi[\p_a] = \int_{\Omega_0} \bigg[ W\Big(\sum_a \p_a \otimes  \nabla N_a\Big)-\rho_0 \mathbf{B}\cdot \Big(\sum_a \p_a N_a \Big)\bigg]\, dV.
\end{equation}
The minimization is carried out for the interior nodes, whereas the nodes at the boundary are required to satisfy $\p_c = \p^\M + \F^\M \X_c$. Again, the case of periodic boundary conditions will be discussed later, in Section \ref{Sec:PeriodicBoundaryConditions}.

Introducing the effective strain energy density $W_{\text{eff}}(\p_a)=W(\sum_a \p_a \otimes \nabla N_a)-\rho_0 \mathbf{B}\cdot \left(\sum_a \p_a N_a \right)$, the variational problem described in Eq.~\eqref{Eq:Pi_body_discrete} is fully analogous to Eq.~\eqref{Eq:eq_discrete}. One can then apply an identical coarse-graining procedure to that of Section \ref{Sec:Discrete} to obtain the macroscopic equilibrium equations and the appropriate definition of the macroscopic stress tensor. We begin by deriving the weak form of the equilibrium equations. These read
\begin{equation} \label{Eq:equ_b_dis}
\min_{\p_b} \Pi[\p_a] = \int_{\Omega_0} W_{\text{eff}}(\p_a) \, dV \quad  \rightarrow \quad\int_{\Omega_0} \left( P_{iJ} N_{b,J} -\rho_0B_iN_b\right)\, dV = 0, \quad \text{for all the interior nodes $b$},
\end{equation}
and the equilibrium solution can be written as $\p_a^*=\p_a^*\left(\p^\M,\F^\M\right)$.

On their side, the macroscopic equilibrium equations are given by Eqs.~\eqref{Eq:Macro_eq}, with $W^\M(\p^\M,\F^\M)=\langle W_\text{eff} (\p_a^*(\p^\M,\F^\M))\rangle$. Proceeding similarly to Eqs.~\eqref{Eq:VarPi} and \eqref{Eq:VarPi_2},  $\partial W^\M/\partial \p^\M$ and $\partial W^\M / \partial \F^\M$ can be obtained by taking variations over the macroscopic energy density $W^\M$. Separating the nodes $\{a\}$ into the sets $\{b\}$ and $\{c\}$, and using Eqs.~\eqref{Eq:equ_b_dis} and the boundary conditions, one obtains
\begin{equation}
\begin{split}
|\Omega_0|\delta \big\langle &W_{\text{eff}}(\p_a^*(\p^\M,\F^\M )) \big\rangle= \int_{\Omega_0} P_{iJ} \sum_a \delta \varphi_{ia} N_{a,J} \, dV -\int_{\Omega_0} \rho_0 B_i \sum_a \delta \varphi_{ia} N_a \, dV \\
&=\sum_b \delta \varphi_{bi} \int_{\Omega_0} \left(P_{iJ} N_{b,J}-\rho_0 B_i N_b \right)\, dV  + \sum_c \delta \varphi_{ci} \int_{\Omega_0} \left(P_{iJ} N_{c,J}-\rho_0 B_i N_c \right)\, dV \\
&=\sum_c \delta \varphi_{ci} \int_{\Omega_0} \left(P_{iJ} N_{c,J}-\rho_0 B_i N_c \right)\, dV \\
& =  \sum_c \delta \varphi^\M_{i} \left[\int_{\Omega_0} \left(P_{iJ} N_{c,J}-\rho_0 B_i N_c \right) \, dV \right]+ \sum_c \delta F^\M_{iQ} X_{cQ} \left[\int_{\Omega_0} \left(P_{iJ} N_{c,J}-\rho_0 B_i N_c \right) \, dV \right]  \\ 
& =  \sum_a \delta \varphi^\M_{i} \left[\int_{\Omega_0} \left(P_{iJ} N_{a,J}-\rho_0 B_i N_a \right) \, dV \right]  + \sum_a \delta F^\M_{iQ} X_{aQ} \left[\int_{\Omega_0} \left(P_{iJ} N_{a,J}-\rho_0 B_i N_a \right) \, dV \right].  
\end{split}
\end{equation}
Next, by the partition of unity and linear field reproduction properties of the shape functions, as well as the properties obtained in Eqs.~\eqref{Eq:Properties_Shape_Functions} and \eqref{Eq:ShapeFunctions_properties},
\begin{equation}
\begin{split}
|\Omega_0|&\delta \big\langle W_{\text{eff}}(\p_a^*(\p^\M,\F^\M )) \big\rangle \\
&=  \delta \varphi^\M_{i} \left[\int_{\Omega_0} \Big(P_{iJ} \sum_a N_{a,J}-\rho_0 B_i \sum_a N_a \Big) \, dV \right] +  \delta F^\M_{iQ}  \left[\int_{\Omega_0} \Big(P_{iJ} \sum_a N_{a,J} X_{aQ}-\rho_0 B_i \sum_a N_a X_{aQ} \Big) \, dV \right]  \\
& =  \delta \varphi^\M_{i} \left[\int_{\Omega_0} -\rho_0 B_i  \, dV \right]  +  \delta F^\M_{iJ}  \left[\int_{\Omega_0} \left(P_{iJ} -\rho_0 B_i X_J \right) \, dV \right].  
\end{split}
\end{equation}
One then obtains, in accordance with previous results in the literature \citep{ricker2009multiscale,reina2011multiscale,de2015rve}
\begin{equation} \label{Eq:Variations_bodyforces}
\frac{\partial W^\M}{\partial \boldsymbol \varphi^\M} =\langle - \rho_0 \mathbf{B}\rangle \quad \text{and} \quad  \Pb^\M=\frac{\partial W^\M}{\partial \mathbf{F}^\M} = \langle \mathbf{P} - \rho_0 \mathbf{B} \otimes \mathbf{X} \rangle
\end{equation}
and the macroscopic equilibrium equations, cf.~Eqs.~\eqref{Eq:Macro_eq}, read
\begin{equation} \label{Eq:eq_bodyforces}
\nabla^\M \cdot \Pb^\M =  \langle \rho_0 \mathbf{B}\rangle \text{ in } \Omega_0^\M \quad \text{and}\quad \Pb^\M \mathbf{N}^\M = \bar{\mathbf{T}}^\M \text{ on } \partial \Omega^\M_{0,2}.
\end{equation}
where the macroscopic body forces are, as expected, $\rho_0^\M \mathbf{B}^\M= \langle \rho_0 \mathbf{B}\rangle$. When the body forces per unit mass are constant throughout the body and the origin of the reference for the RVE is located at its center of mass, the macroscopic stresses reduce to $\Pb^\M=\langle \Pb \rangle$, as expected. On the contrary, if the body forces are not uniform, a total torque may be induced, that would affect the value of the macroscopic stress tensor in accord with our intuition.

\subsection{Remark on the potential energy density}

While averaging the principle of minimum potential energy for displacement boundary conditions in the absence of body forces can be seen as natural, its generalization to other situations, such as the above, may appear to be a mathematical construct with no intuitive physical support. We provide some insight into this issue by reasoning on the potential energy of an RVE with Neumann boundary conditions, i.e.,
\begin{equation}
\Pi[\p] = \int_{\Omega_0} W(\nabla \p) \, dV -\int_{\partial \Omega_{0}} \bar{\mathbf{T}}\cdot \p \, dS.
\end{equation}
Its extension to the case where body forces are included can then be done by analogy between the term pertaining to external boundary loads, $\int_{\partial \Omega_{0}} \bar{\mathbf{T}}\cdot \p \, dS$, and that of external body forces, $\int_{\Omega_0} \rho_0\mathbf{B} \cdot \p \, dV$.

\begin{figure}[htbp]
\begin{center}
    {\includegraphics[width=0.9\textwidth]{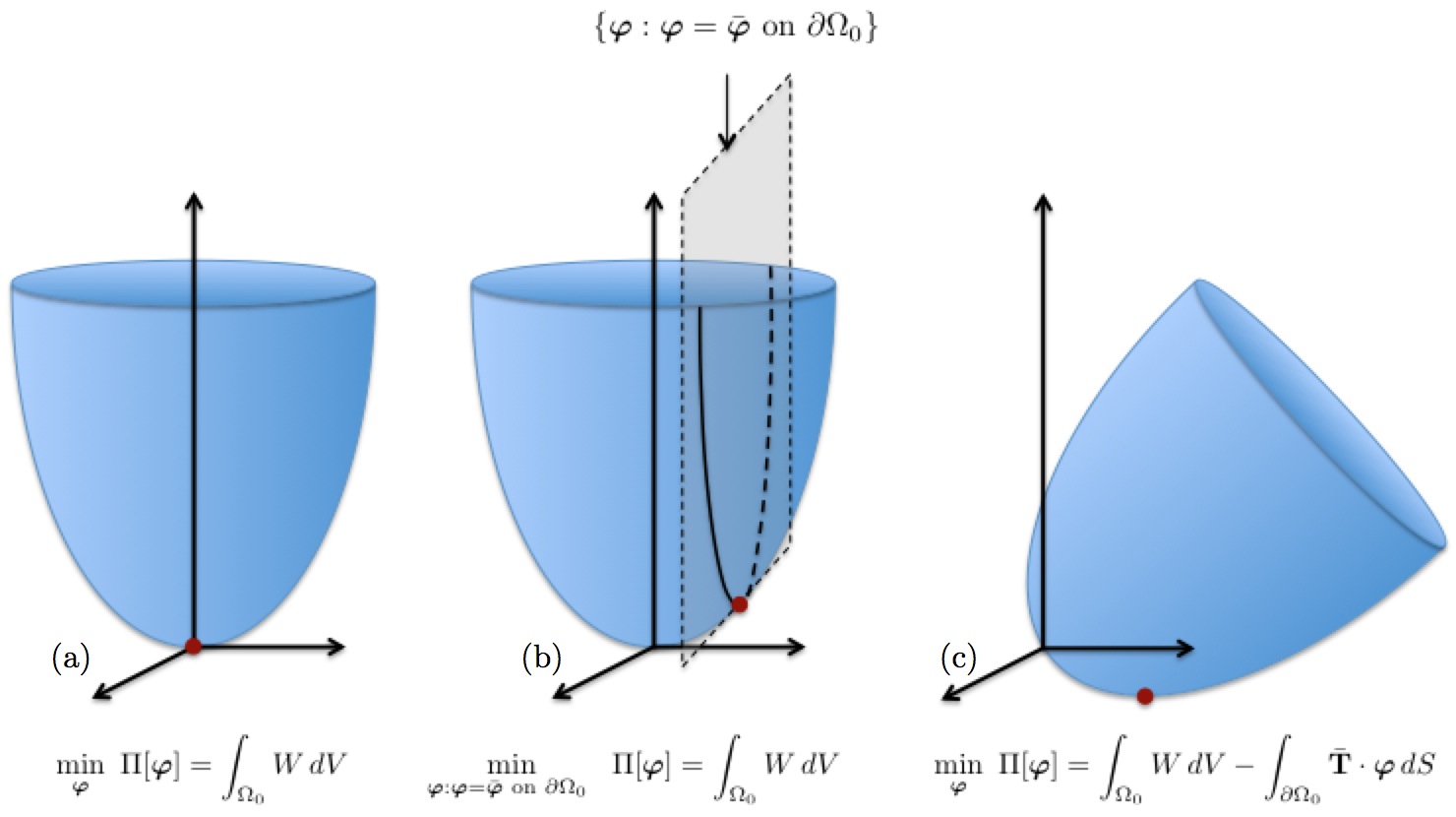}}
    \caption[]{Schematic representation of the role of displacement and traction boundary conditions in the principle of minimum potential energy.} \label{Fig:MinPotEn}
\end{center}
\end{figure}

The principle of minimum potential energy has an unequal treatment of displacement and traction boundary conditions. In particular, the displacement boundary conditions restrict the variations of the deformation mapping, while the external loads appear explicitly in the expression of the potential energy. Yet, it is well known that displacement boundary conditions could be physically replaced by the forces needed to achieve such displacement at the boundary, leading to an identical solution.  Such strategy is common in the method of Lagrange multipliers, as well as the standard method for solving hyperstatic (statically indeterminate) structures in structural analysis \citep{carpinteri2002structural}. This apparent asymmetry between the two types of boundary conditions is explained with the help of the sketch in Figure~\ref{Fig:MinPotEn}. In the absence of any boundary conditions, the minimum of the potential energy corresponds to the infimum of the energy density, where the function takes a zero value\footnote{In general $W(\mathbf{I})=\text{constant}$. Without loss of generality we take that constant to be $0$.}, cf.~Figure \ref{Fig:MinPotEn}(a) (in reality, the zero energy corresponds to the space of rotations but has been represented, for simplicity, as a point in the figure). However, if displacement boundary conditions are imposed, the minimization needs to be performed in a the subspace of deformation mappings that satisfy such constraint. This is sketched in Figure \ref{Fig:MinPotEn}(b), where the subspace is represented, for simplicity, as a plane. The zero energy configuration(s) is now unattainable and the new minimum, represented in red in the figure, is non-trivial (not a rotation). For the case of traction boundary conditions, the minimization is again unconstrained, but a new term contributes to the potential energy, namely $ \int_{\partial \Omega_{0}} \bar{\mathbf{T}}\cdot \p \, dS$, whose effect can be considered equivalent to `tilting' the strain energy density, cf.~Figure \ref{Fig:MinPotEn}(c). As it can be seen schematically, it is possible to `tilt' the strain energy density so that its minimum configuration coincides with the case of Dirichlet boundary conditions.

To further support the averaging of the potential energy density, we note that if one chooses traction boundary conditions for the RVE corresponding to a given macroscopic material point, i.e.~$\mathbf{T}=\Pb\N = \bar{\mathbf{T}}=\Pb^\M \N $, then the microscopic problem solves, in the continuum setting,
\begin{equation}
\min_{\p}\ \Pi[\p] = \int_{\Omega_0} W(\nabla \p)\, dV - \int_{\partial \Omega_0} \bar{\mathbf{T}}\cdot \p\, dS.
\end{equation}
From the aforementioned reasonings, the macroscopic strain energy density is then the average of the potential energy density, which, upon replacement of the boundary conditions and application of the divergence theorem, reads
\begin{equation}
W^\M(\Pb^\M) = \langle W(\Pb^\M) \rangle - \Pb^\M :\langle \F \rangle= \langle W(\Pb^\M) \rangle - \Pb^\M: \F^\M.
\end{equation}
But this is nothing else than the Legendre transform of $\langle W(\F^\M)\rangle$ \citep{rockafellar1970convex}, i.e.,~$W^\M(\Pb^\M)\simeq W^{\M,*}(\F^\M)$. We note that the equality is only approximate, since the fields with displacement and traction boundary conditions are generally different, and so are their averaged quantities; see for instance Chapter 2 of \citet{nemat2013micromechanics} for a discussion on the macropotentials in the linearized kinematic setting. We further note that full duality between the strain and stress formulation only occurs in the linearized kinematic setting, where the energy is convex and $W^{\M,**}=W^\M$ (superscript $**$ indicates twice the application of the Legendre transform) \citep{dacorogna2007direct}. In general $W^{\M,**}$ leads to the convex envelope of $W^\M$.

\section{Coarse-graining of the discrete dynamic problem} \label{Sec:DiscreteDynamicProblem}

\citet{radovitzky1999error} showed that the time-discretization of the equations of elastodynamics, cf.~Eq.~\eqref{Eq::EquilibriumEquationsDynamic}, with a Newmark's algorithm, follow an incremental minimum principle. In particular, for the specific case of explicit dynamics, the solution of the RVE with displacement boundary conditions at time $t^{n+1}$ is given by
\begin{equation} \label{Eq:Pi_dynamic}
\begin{split}
\Pi[\boldsymbol \varphi^{n+1}] &= \int_{\Omega_0} \Big[ \rho_0 \frac{|\boldsymbol \varphi^{n+1}-\boldsymbol \varphi^{n+1,\text{pre}}|^2}{\Delta t^2}+\mathbf{P}^n:\nabla \boldsymbol \varphi^{n+1}- \rho_0 \mathbf{B}^n \cdot \boldsymbol \varphi^{n+1} \Big]\, dV,\\
&\rightarrow \quad 2 \rho_0 \frac{\boldsymbol \varphi^{n+1}-\boldsymbol \varphi^{n+1,pre}}{\Delta t^2} = \nabla \cdot \mathbf{P}^n + \rho_0 \mathbf{B}^n, \quad \text{for } \mathbf{X} \in \Omega_0,
\end{split}
\end{equation}
where $\boldsymbol \varphi^{n+1,\text{pre}} := \boldsymbol \varphi^n + \Delta t\ \dot{\boldsymbol \varphi}^n$ and $\Pb^n$ and $\B^n$ are the stress tensor and body forces, respectively, at time $t^n = t^{n+1}-\Delta t$. Using a finite element formulation, Eqs.~\eqref{Eq:Pi_dynamic} read as a discrete static problem, cf.~Eq.~\eqref{Eq:eq_discrete}, with an effective (incremental) strain energy density 
\begin{equation} \label{Eq:Weff_dynamic}
W_{\text{eff}}^{n+1}(\varphi_{ai}) = \rho_0 \frac{ |\sum_a (\boldsymbol\varphi^{n+1}_{a}-\boldsymbol\varphi^{n+1,\text{pre}}_{a})N_a|^2}{\Delta t^2} + P^n_{iJ} \bigg(\sum_a \varphi^{n+1}_{ai} N_{a,J} \bigg) - \rho_0 B_i^{n} \bigg(\sum_a \varphi^{n+1}_{ai} N_a \bigg).
\end{equation}
The corresponding weak version of the equilibrium equations for the interior nodes are
\begin{equation} \label{Eq:Pi_dynamic_discrete}
 \int_{\Omega_0} \bigg[ 2\rho_0 \frac{ \sum_a (\varphi^{n+1}_{ai}-\varphi^{n+1,pre}_{ai})N_a}{\Delta t^2}  N_b + P^n_{iJ} N_{b,J} - \rho_0 B_i^{n} N_b \bigg] \, dV = 0.
 \end{equation}

By the conservation of the energy density upon upscaling procedures, the incremental macroscopic energy density is defined as  $W^{\M,n+1}\left(\p^\M,\F^\M \right) = \langle W_{\text{eff}}^{n+1}\left(\p_a^*\left(\p^\M,\F^\M \right)\right) \rangle$, and it allows the construction of 
the macroscopic incremental variational problem as in Eqs.~\eqref{Eq:Pi_macro} and \eqref{Eq:Macro_eq}. Quantities $\partial W^\M/\partial \p^\M$ and $\partial W^\M / \partial \F^\M$ can be obtained from the microscopic equilibrium equations, cf.~Eq.~\eqref{Eq:Pi_dynamic_discrete}, the boundary conditions, considered here to be of Dirichlet type, $\delta\varphi^{n+1}_{ci} = \delta\varphi^{\M,n+1}_{i} +\delta F^{\M,n+1}_{iQ} X_{cQ}$, and the properties of the shape functions. More specifically,
\begin{equation}
\begin{split}
|\Omega_0|\delta  \langle &W_{\text{eff}}^{n+1}(\p_a^*(\p^\M,\F^\M )) \rangle = \int_{\Omega_0} \sum_b \delta \varphi^{n+1}_{bi} \left[ 2\rho_0 \frac{ \sum_a (\varphi^{n+1}_{ai}-\varphi^{n+1,\text{pre}}_{ai})N_a}{\Delta t^2}  N_b + P^n_{iJ} N_{b,J} - \rho_0 B_i^{n} N_b \right] \, dV \\
&+ \int_{\Omega_0} \sum_c \delta \varphi^{n+1}_{ci} \left[ 2\rho_0 \frac{ \sum_a (\varphi^{n+1}_{ai}-\varphi^{n+1,\text{pre}}_{ai})N_a}{\Delta t^2}  N_c + P^n_{iJ} N_{c,J} - \rho_0 B_i^{n} N_c \right] \, dV \\
=& \int_{\Omega_0} \sum_c \delta \varphi^{n+1}_{ci} \left[ 2\rho_0 \frac{ \sum_a (\varphi^{n+1}_{ai}-\varphi^{n+1,\text{pre}}_{ai})N_a}{\Delta t^2}  N_c + P^n_{iJ} N_{c,J} - \rho_0 B_i^{n} N_c \right] \, dV \\
= & \int_{\Omega_0} \sum_c \delta \varphi^{\M,n+1}_{i} \left[ 2\rho_0 \frac{ \sum_a (\varphi^{n+1}_{ai}-\varphi^{n+1,\text{pre}}_{ai})N_a}{\Delta t^2}  N_c + P^n_{iJ} N_{c,J} - \rho_0 B_i^{n} N_c \right] \, dV \\
&+ \int_{\Omega_0} \sum_c\delta F^{\M,n+1}_{iQ} X_{cQ} \left[ 2\rho_0 \frac{ \sum_a (\varphi^{n+1}_{ai}-\varphi^{n+1,\text{pre}}_{ai})N_a}{\Delta t^2}  N_c + P^n_{iJ} N_{c,J} - \rho_0 B_i^{n} N_c \right] \, dV \\
= & \int_{\Omega_0} \sum_a \delta \varphi^{\M,n+1}_{i} \left[ 2\rho_0 \frac{ \sum_{a'} (\varphi^{n+1}_{a'i}-\varphi^{n+1,\text{pre}}_{ai})N_{a'}}{\Delta t^2}  N_a + P^n_{iJ} N_{a,J} - \rho_0 B_i^{n} N_a \right] \, dV \\
&+ \int_{\Omega_0} \sum_a\delta F^{\M,n+1}_{iQ} X_{aQ} \left[ 2\rho_0 \frac{ \sum_{a'} (\varphi^{n+1}_{a'i}-\varphi^{n+1,\text{pre}}_{a'i})N_{a'}}{\Delta t^2}  N_a + P^n_{iJ} N_{a,J} - \rho_0 B_i^{n} N_a \right] \, dV \\
=&\delta \varphi^{\M,n+1}_{i} \int_{\Omega_0} \ \left[ 2\rho_0 \frac{ \sum_{a'} (\varphi^{n+1}_{a'i}-\varphi^{n+1,\text{pre}}_{a'i})N_{a'}}{\Delta t^2}   \sum_a N_a + P^n_{iJ}   \sum_a N_{a,J} - \rho_0 B_i^{n}   \sum_a N_a \right] \, dV \\
&+ \delta F^{\M,n+1}_{iQ} \int_{\Omega_0}  \left[ 2\rho_0 \frac{ \sum_{a'} (\varphi^{n+1}_{a'i}-\varphi^{n+1,\text{pre}}_{a'i})N_{a'}}{\Delta t^2}  \sum_a X_{aQ} N_a + P^n_{iJ} \sum_a X_{aQ} N_{a,J} - \rho_0 B_i^{n}  \sum_a X_{aQ} N_a \right] \, dV \\
 =&\delta \varphi^{\M,n+1}_{i} \int_{\Omega_0} \ \left[ 2\rho_0 \frac{ \sum_{a'} (\varphi^{n+1}_{a'i}-\varphi^{n+1,\text{pre}}_{a'i})N_{a'}}{\Delta t^2}  - \rho_0 B_i^{n}  \right] \, dV \\
&+ \delta F^{\M,n+1}_{iQ} \int_{\Omega_0}  \left[ 2\rho_0 \frac{ \sum_{a'} (\varphi^{n+1}_{a'i}-\varphi^{n+1,\text{pre}}_{a'i})N_{a'}}{\Delta t^2}  X_{Q}  + P^n_{iQ}  - \rho_0 B_i^{n} X_{Q} \right] \, dV\, .
\end{split}
\end{equation}
Therefore,
\begin{align} \label{Eq:Variations:dynamic_1}
&\frac{\partial W^{\M,n+1}}{\partial \p^{\M,n+1}_i} =\bigg\langle 2\rho_0 \frac{\p^{n+1}-\p^{n+1,\text{pre}}}{\Delta t^2}  - \rho_0 \B^{n} \bigg \rangle = \big\langle \rho_0 \ddot{\p}^{n}  - \rho_0 \B^{n} \big \rangle, \\ \label{Eq:Variations:dynamic_2}
&\Pb^{\M,n} = \frac{\partial W^{\M,n+1}}{\partial \F^{\M,n+1}} =  \bigg \langle  \Pb^n + \left(2\rho_0 \frac{ \p^{n+1}-\p^{n+1,\text{pre}}}{\Delta t^2} - \rho_0 B_i^{n} \right) \otimes \X \bigg \rangle = \big \langle  \Pb^n + \left(\rho_0 \ddot{\p}^{n} - \rho_0 \B^{n} \right) \otimes \X \big \rangle.
\end{align}
The superindex $n+1$ in the effective macroscopic energy density $W^\M$ and deformation tensor $\F^\M$ originates from the fact that Eqs.~\eqref{Eq:Pi_dynamic_discrete} allow solving for $\p^{n+1}$; and such solution has to be compatible, for consistency, with the boundary conditions given by $\p^\M$ and $\F^\M$ at time $t^{n+1}$. However, due to the explicit nature of the time integration scheme used at the microscopic level, $2\rho_0 \frac{\p^{n+1}-\p^{n+1,\text{pre}}}{\Delta t^2}=\rho_0 \ddot{\p}^n$, or equivalently $\p^{n+1} = \p^n+\dot{\p}^n \Delta t + \frac{1}{2} \ddot{\p}^n \Delta t^2$. Therefore, $\frac{\partial W^{\M,n+1}}{\partial \F^{\M,n+1}} $ physically represents the macroscopic stress at time $n$, and has been denoted as such in the equation above.

Expression \eqref{Eq:Variations:dynamic_2} obtained for the macroscopic stress tensor consists of the average of the microscopic stress tensor, as in the static case, and an often-called \emph{dynamic} part \citep{molinari2001micromechanical,jacques2012effects}, as it includes the effects of micro-inertia. The importance of each of these two terms in their contributions to $\Pb^\M$ will be analyzed in Section \ref{Sec:StaticDynamicStress}, where the simple case of wave propagation in a layer media is studied in detail with standard finite elements and via computational homogenization. The relation of Eq.~\eqref{Eq:Variations:dynamic_2} to the continuum stress measure from atomistic simulations is discussed in Section \ref{Sec:VirialStress}.

Equations~\eqref{Eq:Variations:dynamic_1} and \eqref{Eq:Variations:dynamic_2} deliver the macroscopic equilibrium equations 
\begin{equation} \label{Eq:macro_dynamic_eq}
\nabla \cdot \Pb^{\M,n}+ \langle \rho_0 \B^{n} \rangle = \bigg\langle 2\rho_0 \frac{ \p^{n+1}-\p^{n+1,\text{pre}}}{\Delta t^2} \bigg  \rangle = \langle \rho_0 \ddot{\p}^{n} \rangle.
\end{equation}
The right hand side of the expression is the time derivative of the average momentum, i.e.~$ \langle \rho_0 \ddot{\p}^{n}\rangle = \frac{d}{dt}\langle \rho_0 \dot{\p}\rangle$. 
We note that Eq.~\eqref{Eq:macro_dynamic_eq} and the definition of the macroscopic momentum result exclusively from the conservation of energy with the coarse-groaning procedure, cf.~Section \ref{Sec:Rational}, and did not require any further hypothesis on the micro-macro kinematics, which are in general non-trivial due to the lack of averaging theorems for the deformation mapping (even in the static case). The precise way in which the coupled micro-macro problem can be consistently solved is explained in Section \ref{Sec:MultiscaleSolver}, and it exemplified for a one-dimensional stratified composite in Section \ref{Sec:1DProblem}.

\section{General micro-macro relations for computational homogenization} \label{Sec:GeneralContinuum}

The rational described in Section \ref{Sec:Rational} has provided a general variational framework for coarse-graining the microscopic equilibrium equations under arbitrary loading conditions and a wide range of constitutive relations, namely, those that can be recast via variational constitutive updates \citep{ortiz1999variational}. Its only hypothesis is the separation of length scales and it is exclusively based on the conservation of energy upon coarse-graining. The method, which so far considers a finite element discretization for the RVE, has been applied to both static and dynamic conditions in Sections \ref{Sec:BodyForces} and \ref{Sec:DiscreteDynamicProblem}, respectively, for affine boundary conditions and the results are summarized in Table \ref{Table:discrete}.


\begin{sidewaystable} \footnotesize
\caption{Variational coarse-gaining framework under a finite element discretization with affine displacement boundary conditions for the RVE.
\newline
}
\begin{tabular}{ |c|c|c| } 
  \hline
  {\bf Static no body forces} & {\bf Static with body forces} & {\bf Dynamic with body forces} \\ 
  \hline
      \multicolumn{3}{|c|}{} \\
  \hline
      \multicolumn{3}{|c|}{} \\
  \multicolumn{3}{|c|}{ Microscopic problem: $\quad \quad \min_{\p_b} \Pi = \int_{\Omega_0} W_{\text{eff}}\left(\p_a \right)\, dV, \quad \quad \{a\} =\{b\}\cup \{c\}, \quad \p_c=\p^\M+\F^\M \X_c $}\\ [0.5cm]
  \multicolumn{3}{|c|}{ $ \frac{\partial W_{\text{eff}}}{\partial \p_b}=0 \rightarrow \p_a=\p_a^*\left(\p^\M,\F^\M \right) $} \\ [0.5cm]
  \hline
  & & \\
 $W_{\text{eff}}= W\left(\sum_a \p_a \nabla N_a \right) $&  $W_{\text{eff}}= W\left(\sum_a \p_a \nabla N_a \right) -\rho_0 \B \cdot \left( \sum_a \p_a N_a\right)$ & $W_{\text{eff}}^{n+1} = \rho_0 \frac{ |\sum_a (\p^{n+1}_{a}-\p^{n+1,\text{pre}}_{a})N_a|^2}{\Delta t^2} + \Pb^n : \left(\sum_a \p^{n+1}_{a}\otimes \nabla N_{a} \right) $\\ [0.5cm]
  && $ - \rho_0 \B^{n}\cdot \left(\sum_a \p^{n+1}_{a} N_a \right)$ \\
    \hline
& & \\  
  $\int_{\Omega_0} \Pb \nabla N_{b}\, dV = 0$ & $\int_{\Omega_0} \left( \Pb \nabla N_{b} + \B N_b\right)\, dV = 0$ & $\int_{\Omega_0} \left[ 2\rho_0 \frac{ \sum_a (\p^{n+1}_{a}-\p^{n+1,\text{pre}}_{a})N_a}{\Delta t^2}  N_b + \Pb^n \nabla N_{b} - \rho_0 \B^{n} N_b \right] \, dV = 0$\\ [0.5cm]
  \hline
      \multicolumn{3}{|c|}{} \\
  \hline 
    \multicolumn{3}{|c|}{} \\
   \multicolumn{3}{|c|}{ Macroscopic problem: $\quad \quad \min_{\p^\M} \Pi^\M = \int_{\Omega_0^\M} W^\M\left(\p^\M,\F^\M \right)\, dV^\M - \int_{\partial \Omega_{0,2}^\M} \bar{\mathbf{T}}^\M\cdot \p^\M \, dS^\M$, \quad \quad $W^\M\left(\p^\M,\F^\M \right)= \big\langle W_{\text{eff}} \left(\p_a^* \left(\p^\M,\F^\M \right) \right)\big\rangle$ } \\
    \multicolumn{3}{|c|}{} \\
 \multicolumn{3}{|c|}{ \hspace{4cm}  $\frac{\partial W^\M}{\partial \p^\M} - \nabla^\M \cdot \left(\frac{\partial W^\M}{\partial \F^\M} \right) = 0, \quad \text{in } \Omega_0^\M \quad \text{and} \quad \left(\frac{\partial W^\M}{\partial \F^\M} \right) \N^\M= \bar{\mathbf{T}}^\M , \quad \text{on } \partial \Omega_{0,2}^\M, \quad \quad \Pb^\M = \frac{\partial W^\M}{\partial \F^\M}$ } \\ [0.5cm] 
   \hline
      & & \\ 
   $\frac{\partial W^\M}{\partial \p^\M}=0$ & $\frac{\partial W^\M}{\partial \p^\M}=\langle\rho_0 \B \rangle$ & $\frac{\partial W^{\M,n+1}}{\partial \p^{\M,n+1}}=\Big \langle\rho_0 \B^n - 2\rho_0 \frac{ \p^{n+1}-\p^{n+1,\text{pre}}}{\Delta t^2}\Big \rangle= \langle\rho_0 \B -  \rho_0\ddot{\p}^{n}\rangle$\\    [0.5cm]
   \hline
   & & \\
 $\Pb^\M =\langle \Pb \rangle$ &$\Pb^\M =\langle \Pb -\rho_0 \B \otimes \X \rangle$ &  $\Pb^{\M,n} =\bigg \langle \Pb^n+ \left( 2\rho_0 \frac{ \p^{n+1}-\p^{n+1,\text{pre}}}{\Delta t^2}-\rho_0 \B^n \right) \otimes \X \bigg \rangle= \langle \Pb^n+ \rho_o \left( \ddot{\p}^n- \B^n \right) \otimes \X \rangle$ \\   [0.5cm] 
 \hline
    & & \\
 $\nabla \cdot \Pb^\M = 0$ &$\nabla \cdot \Pb^\M +\langle\rho_0 \B \rangle= 0$ & $\nabla \cdot \Pb^{\M,n} +\langle\rho_0 \B^n \rangle = \Big\langle 2\rho_0 \frac{ \p^{n+1}-\p^{n+1,\text{pre}}}{\Delta t^2} \Big  \rangle = \langle\rho_0 \ddot{\p}^{n} \rangle$ \\  [0.5cm]
 \hline
\end{tabular} \label{Table:discrete}
\end{sidewaystable} 

In this section we postulate that such coarse-graining procedure holds true in the continuum case (with no finite element discretization for the RVE). Although, in such case, the microscopic effective potential energy depends both on $\p$ and $\F=\nabla \p$, the macroscopic stress tensor and the macroscopic equilibrium equations resulting from averaging the (incremental) principle of minimum potential energy are identical, after discretization, to those depicted in the last row of Table \ref{Table:discrete}. We show such equality in the sections below for both the static case with body forces and for the dynamic case, both with affine Dirichlet boundary conditions for consistency. The case of periodic boundary conditions is discussed later in Section \ref{Sec:PeriodicBoundaryConditions}.

\subsection{Static case with body forces}
Following the same procedure as in Section \ref{Sec:BodyForces}, we begin by writing the principle of minimum potential energy for the RVE problem and the associated equilibrium equations, i.e., 
\begin{equation}\label{Eq: PiBodyforceCon}
\min_{\p}\Pi[\p] = \min_{\p}\Big\{ \int_{\Omega_0} W(\nabla \boldsymbol \varphi,\mathbf{X})\, dV- \int_{\Omega_0} \rho_0 \mathbf{B} \cdot \boldsymbol \varphi\, dV \Big\} \quad \rightarrow \quad \nabla \cdot \mathbf{P} + \rho_0 \mathbf{B} = 0, \quad \text{for } \mathbf{ X} \in \Omega_0.
\end{equation}
Next, we compute the variations of the average potential energy density of the RVE with respect to the boundary data. Denoting by $W_{\text{eff}}=W(\nabla \boldsymbol \varphi,\mathbf{X})- \rho_0 \mathbf{B} \cdot \boldsymbol \varphi$, one obtains by recursive use of the divergence theorem
\begin{equation}
|\Omega_0|\delta \langle W_{\text{eff}}\rangle= \int_{\Omega_0} P_{iJ} \delta \varphi_{i,J} \, dV -\int_{\Omega_0} \rho_0 B_i \delta \varphi_i\, dV =\int_{\Omega_0} (P_{iJ} \delta \varphi_i)_{,J} \, dV - \int_{\Omega_0} P_{iJ,J} \delta \varphi_i \, dV -\int_{\Omega_0} \rho B_i \delta \varphi_i \, dV.
\end{equation}
Using the equilibrium equations and the boundary conditions, it reduces to
\begin{equation}
\begin{split}
|\Omega_0|\delta \langle W_{\text{eff}}\rangle&= \int_{\partial \Omega_0} P_{iJ} N_J \delta \varphi_i \, dS = \left[ \int_{\partial \Omega_0} P_{iJ} N_J \, dS \right] \delta \varphi^\M_i+ \left[ \int_{\partial \Omega_0} P_{iJ} N_J X_P \, dS \right] \delta \varphi^\M_{i,P}  \\
&= \left[ \int_{\Omega_0} P_{iJ,J} \, dV \right] \delta \varphi_i^\M +\left[\int_{\Omega_0} \left(P_{iJ} +P_{iP,P} X_J\right) \, dV\right] \delta \varphi^\M_{i,J}\\
&= \left[ \int_{\Omega_0} -\rho_0 B_i \, dV \right] \delta \varphi_i^\M +\left[\int_{\Omega_0} \left(P_{iJ} -\rho_0 B_i X_J\right) \, dV\right] \delta \varphi^\M_{i,J}.
\end{split}
\end{equation}
Equations (\ref{Eq:Variations_bodyforces}) are then recovered, and the resulting macroscopic equilibrium equations are identical to those obtained using the finite element discretization, cf.~Eqs.~(\ref{Eq:eq_bodyforces}). 

\subsection{Dynamic case with body forces}\label{SubSec:DynamicContinuum}

Similarly, for the dynamic case, we consider the incremental minimum principle of the potential energy defined in Eq.~\eqref{Eq:Pi_dynamic}, and the associated equilibrium equations
\begin{equation}
2 \rho_0 \frac{\boldsymbol \varphi^{n+1}-\boldsymbol \varphi^{n+1,pre}}{\Delta t^2} = \nabla \cdot \mathbf{P}^n + \rho_0 \mathbf{B}^n, \quad \text{for } \mathbf{X} \in \Omega_0.
\end{equation}

Defining $W^{n+1}_{\text{eff}}= \rho_0 \frac{|\boldsymbol \varphi^{n+1}-\boldsymbol \varphi^{n+1,\text{pre}}|^2}{\Delta t^2}+\mathbf{P}^n:\nabla \boldsymbol \varphi^{n+1}- \rho_0 \mathbf{B}^n \cdot \boldsymbol \varphi^{n+1}$, the variation of its average over the RVE with respect to the macroscopic quantities $\p^\M$ and $\F^\M$ can then be readily computed as

\begin{equation}
\begin{split}
 |\Omega_0|\delta \langle W^{n+1}_{\text{eff}}\rangle&=  \int_{\Omega_0}\bigg[2\rho_0\frac{ \varphi^{n+1}_i-\boldsymbol \varphi^{n+1,pre}_i}{\Delta t^2}\delta \varphi_i^{n+1}+ P^n_{iJ} \delta \varphi_{i,J}^{n+1} - \rho_0 B_i^n \delta \varphi_i^{n+1} \bigg]\, dV \\
 & = \int_{\Omega_0} \bigg[2\rho_0\frac{ \varphi_i^{n+1}-\boldsymbol \varphi^{n+1,pre}_i}{\Delta t^2} - P_{iJ,J}^{n} -  \rho_0 B_i^n \bigg] \delta \varphi_i^{n+1} \, dV + \int_{\partial \Omega_0}P^n_{iJ} N_J \delta \varphi_i^{n+1} \, dS \\
 & = \int_{\partial \Omega_0}P^n_{iJ} N_J  \left(\delta \varphi_i^{\M,n+1} + \delta F^{\M,n+1}_{iQ} X_Q  \right) \, dS \\
 & = \delta \varphi_i^{\M,n+1} \bigg[\int_{\Omega_0}P^n_{iJ,J} \, dV \bigg] + \delta F^{\M,n+1}_{iQ}  \bigg[\int_{\Omega_0} \left(P^n_{iJ}X_Q \right)_{,J} \, dV \bigg].
 \end{split}
\end{equation}
Equations \eqref{Eq:Variations:dynamic_1} and \eqref{Eq:Variations:dynamic_2} are recovered and the results are then identical to those previously obtained by discretizing the microscopic fields via a finite element approximation.
Therefore, for both, the static and the dynamic scenario, the operations of discretization and averaging commute.

\subsection{Periodic boundary conditions} \label{Sec:PeriodicBoundaryConditions}

Both the discrete and continuum results hold for periodic boundary conditions. The proofs are similar to those already outlined, and are thus not redone in the main text to avoid repetition. These may be found, for completeness, in the Appendix A for the discrete setting, and Appendix B for the continuum analogue.

\section{ Virial stress} \label{Sec:VirialStress}


The definition of the macroscopic stresses obtained in Section \ref{Sec:DiscreteDynamicProblem}, cf.~Eq.~\eqref{Eq:Variations:dynamic_2}, is reminiscent of the definition of stress tensor for atomistic systems, see \cite{admal2010unified} for a review on the topic. The connection, which has been previously suggested, is here made precise thanks to the discrete averaging results outlined in Section \ref{Sec:DiscreteDynamicProblem} for elastodynamics. In particular, the atomistic equations of motion in a NVE ensemble follow Newton's law, which discretized in time with a forward Euler scheme read 

\begin{equation} \label{eq:Atoms_eq}
\varphi_i^{n+1}=\varphi_i^{n}+\dot{\varphi}_i^{n}\Delta t+\frac{1}{2}\ddot{\varphi}_i^n\Delta t^2  \quad \quad \text{with}\quad \quad m\ddot{\varphi}_i^n=f_i^n=-\frac{\partial \mathcal{V}}{\partial \varphi_i}\Big\vert^n,
\end{equation}
where $\mathcal{V}$ is the interatomic potential. Forces $f_i^n$ will also include, when present, external forces applied to the atoms. Equations \eqref{eq:Atoms_eq} can be recast in a variational form, as the following incremental minimum principle
\begin{equation}
\Pi[\varphi_{ai}^{n+1}]=\sum_{a}\bigg[m_a\frac{|\boldsymbol\varphi_{a}^{n+1}-\boldsymbol\varphi_{a}^{n+1,\text{pre}}|^2}{\Delta t^2}-f_{ai}^n\varphi_{ai}^{n+1}\bigg],
\end{equation}
where the subindex $a$ refers now to the individual atoms, and $\varphi_{ai}^{n+1,\text{pre}}=\varphi_{ai}^{n}+\dot{\varphi}_{ai}^{n}\Delta t$. For a big system of atoms with periodic boundary conditions, one may again differentiate between atoms at the boundary of the simulation box $\{c\}$, which satisfy
\begin{equation} \label{Eq:eq_atoms_c}
\begin{split}
&\varphi_{ci}^{n+1}=\varphi_i^{\M,n+1}+F_{iJ}^{\M,n+1}X_{cJ}+\tilde\varphi^{n+1}_{ci}, \\
&\frac{\partial \Pi}{\partial\varphi_{ci}^{n+1}}=0 \quad \rightarrow \quad \sum_c\bigg[2m_c\frac{(\varphi_{ci}^{n+1}-\varphi_{ci}^{n+1,\text{pre}})}{\Delta t^2}-f_{ci}^{n}\bigg]\delta\tilde\varphi^{n+1}_{ci}=0
\end{split}
\end{equation}
and interior atoms $\{b\}$, that obey
\begin{equation} \label{Eq:eq_atoms_b}
\frac{\partial \Pi}{\partial\varphi_{bi}^{n+1}}=0 \quad \rightarrow \quad 2 m_b \frac{\varphi_{bi}^{n+1}-\varphi_{bi}^{n+1,\text{pre}}}{\Delta t^2}-f_{bi}^n=0.
\end{equation}

We define, as customary now, the macroscopic (incremental) effective strain energy density as 
\begin{equation}
|\Omega_0|W^{n+1}_{\text{eff}} = \sum_{a}\left[m_a\frac{|\boldsymbol\varphi_{a}^{n+1}-\boldsymbol\varphi_{a}^{n+1,\text{pre}}|^2}{\Delta t^2}-f_{ai}^n\varphi_{ai}^{n+1}\right]
\end{equation}
and compute its variation with respect to the macroscopic fields $\p^\M$ and $\F^\M$ to determine the macroscopic stress tensor
\begin{equation}
\begin{split}
|\Omega_0|\delta W^{n+1}_{\text{eff}}&=\sum_a\left[2m_a\frac{(\varphi_{ai}^{n+1}-\varphi_{ai}^{n+1,\text{pre}})}{\Delta t^2}-f_{ai}^{n}\right]\delta\varphi_{ai}^{n+1}\\
&=\sum_c\left[2m_c\frac{(\varphi_{ci}^{n+1}-\varphi_{ci}^{n+1,\text{pre}})}{\Delta t^2}-f_{ci}^{n}\right]\left (\delta \varphi_i^{\M,n+1}+\delta F_{iJ}^{\M,n+1}X_{cJ}+\delta\tilde\varphi^{n+1}_{ci}\right)\\
&=\sum_c\left[2m_c\frac{(\varphi_{ci}^{n+1}-\varphi_{ci}^{n+1,\text{pre}})}{\Delta t^2}-f_{ci}^{n}\right]\left (\delta \varphi_i^{\M,n+1}+\delta F_{iJ}^{\M,n+1}X_{cJ}\right),
\end{split}
\end{equation}
where we have separated the set of all atoms into interior and boundary atoms, we have made use of the equilibrium equations for atoms $\{b\}$, cf.~Eqs.~\eqref{Eq:eq_atoms_b}, and the boundary conditions given in Eqs.~\eqref{Eq:eq_atoms_c}.
Then, using again the equilibrium equations, one obtains
\begin{align} \nonumber
\frac{\delta W^{n+1}_{\text{eff}}}{\delta\p^{\M,n+1}}&=\frac{1}{|\Omega_0|}\sum_c\left[2m_c\frac{(\p_{c}^{n+1}-\p_{c}^{n+1,\text{pre}})}{\Delta t^2}-\mathbf{f}_{c}^{n}\right]=\frac{1}{|\Omega_0|}\sum_c\left(m_c\ddot{\p}_c^n-\mathbf{f}_{c}^{n}\right)=\frac{1}{|\Omega_0|}\sum_a\left(m_a\ddot{\p}_a^n-\mathbf{f}_{a}^{n}\right),\\ \label{Eq:Piola_atom}
\Pb^{\M,n}&=\frac{\delta W^{n+1}_{\text{eff}}}{\delta \F^{\M,n+1}}=\frac{1}{|\Omega_0|}\sum_a\left[2m_a\frac{(\p_{a}^{n+1}-\p_{a}^{n+1,\text{pre}})}{\Delta t^2}-\mathbf{f}_{a}^{n}\right]\otimes \X_{a}=\frac{1}{|\Omega_0|}\sum_a\left(m_a\ddot{\p}_a^n-\mathbf{f}_{a}^{n}\right)\otimes \X_{a},
\end{align}
where $\frac{\delta W^{n+1}_{\text{eff}}}{\delta\p^{\M,n+1}}=0$ for self-equilibrated external forces, i.e., $\sum_a \mathbf{f}_a=0$, and fixed center of mass, i.e., $\sum_a m_a \ddot{\p}_a=0$. Assuming that is the case, one can then recover the Cauchy stress tensor from Eq.~\eqref{Eq:Piola_atom}. Eliminating, for easiness in the notation, the super index $n$ attendant to the time step, and making use of the equilibrium equations, one obtains
\begin{equation}
\begin{split}
\sigma_{ij}^{\M}&=J^{\M;-1}P_{iJ}^{\M}F_{Jj}^{\M;T}=\frac{1}{|\Omega|}\sum_c\left(m_c\ddot{\varphi}_{ci}-f_{ci}\right)X_{cJ}F_{Jj}^{\M;T} =\frac{1}{|\Omega|}\sum_c\left(m_c\ddot{\varphi}_{ci}-f_{ci}\right)\left(\varphi_{cj}-\varphi_j^{\M}-\tilde{\varphi}_{cj}\right)\\
&=\frac{1}{|\Omega|}\sum_c\left(m_c\ddot{\varphi}_{ci}-f_{ci}\right)\left(\varphi_{cj}-\varphi_j^{\M}\right)=\frac{1}{|\Omega|}\sum_a\left(m_a\ddot{\varphi}_{ai}-f_{ai}\right)\left(\varphi_{aj}-\varphi_j^{\M}\right)\\
&=\frac{1}{|\Omega|}\sum_a m_a\ddot{\varphi}_{ai} \left(\varphi_{aj}-\varphi_j^{\M}\right) - \frac{1}{|\Omega|}\sum_a f_{ai}\left(\varphi_{aj}-\varphi_j^{\M}\right),
\end{split}
\end{equation}
where $|\Omega|$ is the (deformed) volume  and  $J^\M=\det \F^\M=\frac{|\Omega|}{|\Omega_0|}$. By the equilibrium of the external forces,  $\sum_{a} \mathbf{f}_a=0$, and therefore 
\begin{equation}
\boldsymbol \sigma=\frac{1}{|\Omega|}\sum_a m_a \ddot{\p}_a \otimes  (\p_a-\p^\M)-\frac{1}{|\Omega|}\sum_a \mathbf{f}_a \otimes  \p_a.
\end{equation}
Furthermore, if the system in equilibrium is large enough, $\boldsymbol \sigma$ will be constant in time \citep{touchette2009large} and equal to its time average over an interval $\Delta T$. Then, defining $\p^\M \sum_a m_a =\sum_a m_a \p_a$, averaging over time and integrating by parts, one obtains, using a similar strategy to that of \cite{admal2010unified} in Appendix A,
\begin{equation}
\begin{split}
\boldsymbol \sigma&=\frac{1}{|\Omega|}\sum_a \frac{1}{\Delta T}\int_0^{\Delta T} m_a \left(\ddot{\p}_a -\ddot{\p}^\M\right)\otimes  (\p_a-\p^\M)\, dt -\frac{1}{|\Omega|}\sum_a \mathbf{f}_a \otimes  \p_a\\
&\simeq- \frac{1}{|\Omega|}\sum_a \frac{1}{\Delta T}\int_0^{\Delta T} m_a \left( \dot{\p}_a -\dot{\p}^\M\right)\otimes  (\dot{\p}_a-\dot{\p}^\M)\, dt -\frac{1}{|\Omega|}\sum_a \mathbf{f}_a \otimes  \p_a \\
&= -\frac{1}{|\Omega|}\sum_a \overline{\left( m_a \dot{\p}_a^{\text{rel}} \otimes  \dot{\p}_a^{\text{rel}} +\mathbf{f}_a \otimes  \p_a \right)},
\end{split}
\end{equation}
where the boundary terms become negligible as $\Delta T$ increases (assuming $m_a \dot{\p}_a^{\text{rel}} \otimes \p_a^{\text{rel}}$ is bounded), and the bar in the last expression indicates time average. The obtained result is precisely the Virial stress.

\section{Quasi-explicit multiscale solver (QEMS) for the dynamic evolution of heterogeneous media} \label{Sec:MultiscaleSolver}



The discrete dynamic coarse-graining framework described in Section \ref{Sec:DiscreteDynamicProblem} naturally leads to a computational FE$^2$ procedure that is based on the homogenization of a microscopic explicit dynamic evolution. The coupling between the two scales, leads, as expected, to a non-explicit procedure for the coupled micro-macro problem. However, we will show that the apparent implicit problem can be exactly solved within a single global iteration of a Newton-Raphson procedure, where both, the micro and the macro degrees of freedom are concurrently minimized. The Hessian involved is constant and independent of the material constitutive behavior and it can thus be precomputed at the beginning of the simulation. This leads to what we call a \emph{quasi-explicit multiscale solver} (QEMS). 

\subsection{Macroscopic finite element discretization and incremental problem}
We consider a finite element discretization for the macroscopic problem of the form
\begin{equation}
\p^\M(\X^\M) = \sum_A \p^\M_A N^\M_A(\X^\M)
\end{equation}
where $A$ are the nodes of the macroscopic mesh, and $N^\M_A$ are the associated macroscopic shape functions. These may of course be of different nature than those chosen to solve the microscopic problem. We separate for convenience the set of nodes $\{A\}$ into those at the boundary of the macroscopic domain $\{C\}$ and the interior nodes $\{B\}$.

\begin{figure}[t]
\begin{center}
    {\includegraphics[width=0.7\textwidth]{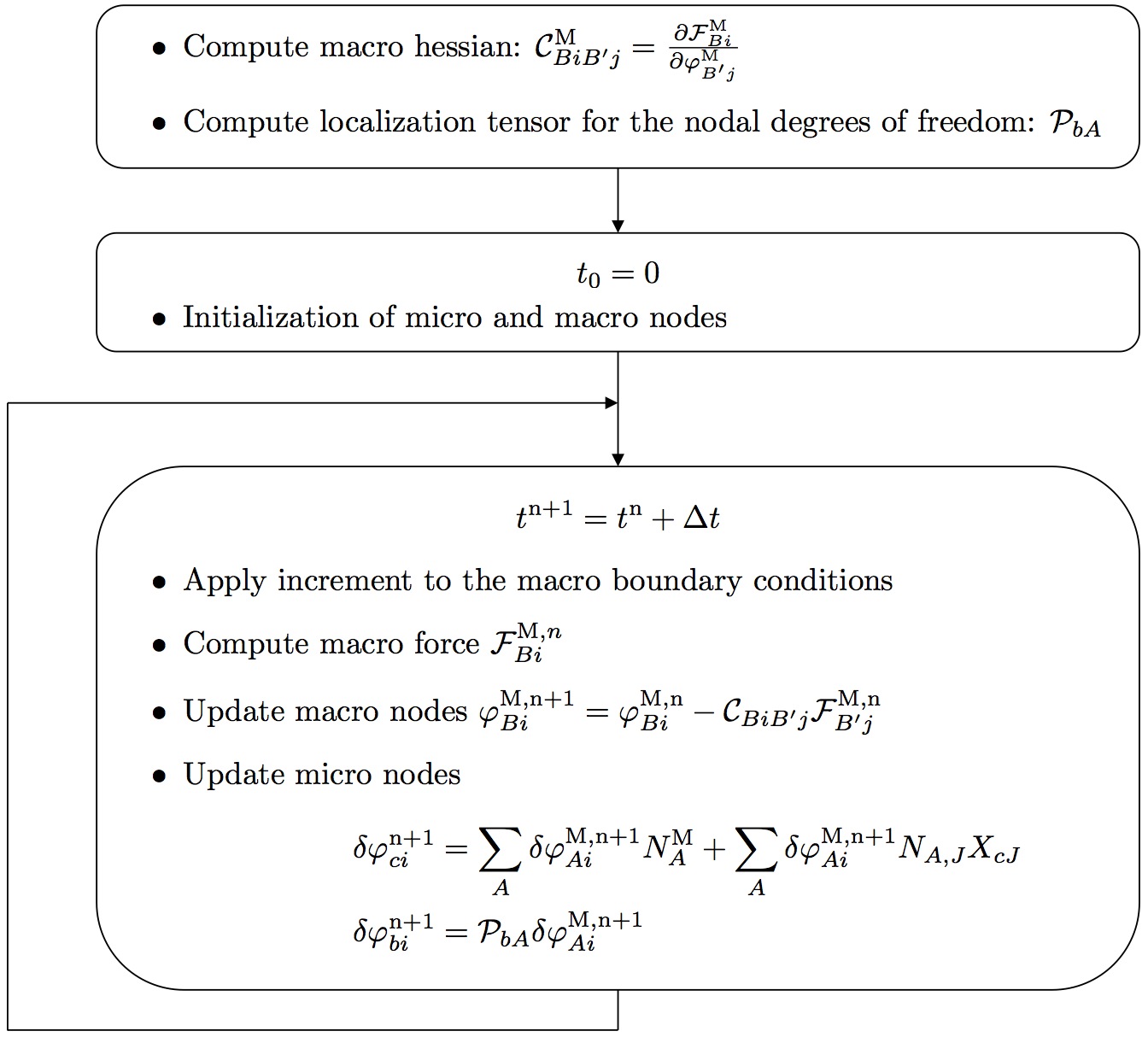}}
    \caption[]{Flow chart of the multiscale scheme. } \label{flow}
\end{center}
\end{figure}

The weak form of the macroscopic equilibrium equations can be obtained by the incremental variational principle given in Eq.~\eqref{Eq:Pi_macro}, where the minimization is done over the nodal values of the macroscopic displacement field. Considering, for simplicity, displacement boundary conditions for the macroscopic problem, the principle with the finite element discretization just described reads as
\begin{equation}
\min_{\p_B^{\M,n+1}} \Pi^{\M,n+1}[\p^{\M,n+1}_A] = \min_{\p_B^{\M,n+1}}   \int_{\Omega^\M_0} W^{\M,n+1} \bigg(\sum_A \p^\M_A N^\M_A, \sum_A \p^\M_A \otimes \nabla^\M N^\M_A\bigg)\, dV^\M\, 
\end{equation}
where $W^{\M,n+1} = \langle W_{\text{eff}}^{n+1}(\p_a^*\left(\p^\M,\F^\M \right)) \rangle$, and $W^{n+1}_{\text{eff}}$ is as in Eq.~\eqref{Eq:Weff_dynamic}. The macroscopic equilibrium equations are then
\begin{equation}
\delta \Pi^{\M,n+1} =\int_{\Omega^\M_0}  \frac{\partial W^{\M,n+1}}{\partial \varphi^{\M,n+1}_i} \sum_B \delta \varphi^{\M,n+1}_{Bi} N^\M_B \, dV^\M + \int_{\Omega^\M_0}  \frac{\partial W^{\M,n+1}}{\partial F^{\M,n+1}_{iJ}} \sum_B \delta \varphi^{\M,n+1}_{Bi} N^\M_{B,J} \, dV^\M = 0 
\end{equation} 
for all $\delta \varphi^{\M,n+1}_{Bi}$, or equivalently, using Eqs.~\eqref{Eq:Variations:dynamic_1} and \eqref{Eq:Variations:dynamic_2}
\begin{equation}\label{Eq::Dynamic}
\mathcal{F}^{\M,n+1}_{Bi}=\frac{\partial \Pi^{\M,n+1}}{\partial \varphi_{Bi}^{\M,n+1}} = \int_{\Omega^\M_0}   N^\M_B  \Big\langle 2\rho_0 \frac{\varphi^{n+1}_i-\varphi_i^{n+1,\text{pre}}}{\Delta t^2}  - \rho_0 B^{n}_i \Big \rangle \, dV^\M +\int_{\Omega^\M_0} N^\M_{B,J} P_{iJ}^{\M,n}  \, dV^\M =0, 
\end{equation} 
with $P^{\M,n}$ defined in Eq.~\eqref{Eq:Variations:dynamic_2}. Equation \eqref{Eq::Dynamic} can be solved by recourse of a Newton-Raphson iterative procedure, namely,
\begin{equation}
\varphi_{Bi}^{\M,n+1,k+1} = \varphi_{Bi}^{\M,n+1,k} -\left[ \frac{\partial \mathcal{F}^{\M,n+1,k}_{Bi}}{\partial \varphi_{B'j}^{\M,n+1}}\right]^{-1} \mathcal{F}^{\M,n+1,k}_{B'j},
\end{equation}
where $k$ is the index corresponding to each iteration. The tensor $ \frac{\partial \mathcal{F}^{\M,n+1,k}_{Bi}}{\partial \varphi_{B'j}^{\M,n+1}}= \frac{\partial^2 \Pi^{\M,n+1,k}}{\partial \varphi_{Bi}^{\M,n+1}\partial \varphi_{B'j}^{\M,n+1}}$ is computed next, and it is shown to be exclusively dependent on the spatio-temporal discretization of the micro and macro problem, and therefore, independent of $k$. This leads to a quasi-explicit computational procedure that simultaneously solves both the micro and macro scale, cf.~Figure~\ref{flow}. 

\subsection{Hessian of the macroscopic problem}
We begin by taking variations of $\mathcal{F}^{\M,n+1}_{Bi}$, given by Eq.~\eqref{Eq::Dynamic},
\begin{equation} \label{Eq:MacroForce}
\delta \mathcal{F}^{\M,n+1}_{Bi} = \int_{\Omega^\M_0}   N^\M_B  \Big \langle \frac{2\rho_0}{\Delta t^2}  \sum_{a} \delta \varphi^{n+1}_{ai} N_{a}\Big\rangle \, dV^\M  +\int_{\Omega^\M_0} N^\M_{B,J}  \Big \langle  \frac{2\rho_0}{\Delta t^2}  \sum_{a}\delta \varphi^{n+1}_{ai}  N_{a}X_{J}  \Big \rangle  \, dV^\M
\end{equation}
and proceed to relate the variation of the microscopic degrees of freedom, $\delta \p_a$ with the macroscopic ones, $\delta \p^\M_A$. The microscopic boundary nodes $\{c\}$ are directly related via the boundary conditions, assumed here to be affine\footnote{Consideration of periodic boundary conditions for the RVE is of course possible and the proofs would proceed in a very similar manner.}, i.e., 
\begin{equation} \label{Eq:QEMS_bc_c}
\delta \varphi^{n+1}_{ci} = \sum_A \delta \varphi^{\M,n+1}_{Ai} N_A^\M + \sum_A  \varphi^{\M,n+1}_{Ai} N^\M_{A,Q} X_{cQ},
\end{equation}
whereas for the interior nodes $\{b\}$, such relation may be derived from the microscopic equilibrium equations
\begin{equation}
0= \int_{\Omega_0} \bigg[ 2\rho_0 \frac{ \sum_a (\varphi^{n+1}_{ai}-\varphi^{n+1,\text{pre}}_{ai})N_a}{\Delta t^2}  N_b + P^n_{iJ} N_{b,J} - \rho_0 B_i^{n} N_b \bigg] \, dV.
\end{equation}
Indeed, variation of this equation gives
\begin{equation}
0 = \int_{\Omega_0} \frac{2\rho_0}{\Delta t^2} \sum_a \delta \varphi^{n+1}_{ai} N_a N_{b'}  \, dV.
\end{equation}
Separating again between interior and boundary nodes, and applying the boundary conditions, cf.~Eqs.~\ref{Eq:QEMS_bc_c}, one obtains
\begin{equation}
\begin{split}
0 =& \sum_b \delta \varphi^{n+1}_{bi} \left[ \int_{\Omega_0} \frac{2\rho_0}{\Delta t^2}  N_b N_{b'}  \, dV \right] + \sum_c \delta \varphi^{n+1}_{ci} \left[ \int_{\Omega_0} \frac{2\rho_0}{\Delta t^2}  N_c N_{b'}  \, dV  \right] \\
=&\sum_b \delta \varphi^{n+1}_{bi} \left[ \int_{\Omega_0} \frac{2\rho_0}{\Delta t^2}  N_b N_{b'}  \, dV \right] +  \sum_A\delta \varphi^{\M,n+1}_{Ai}N^\M_{A} \bigg[ \int_{\Omega_0} \frac{2\rho_0}{\Delta t^2} \sum_c N_c N_{b'}  \, dV  \bigg] \\
&+ \sum_A \delta \varphi^{\M,n+1}_{Ai} N^\M_{A,Q} \sum_c  X_{cQ} \left[ \int_{\Omega_0} \frac{2\rho_0}{\Delta t^2}  N_c N_{b'}  \, dV  \right]\\
=&\sum_b \mathcal{M}_{b'b}\ \delta \varphi^{n+1}_{bi} + \sum_A\mathcal{M}^\M_{b'A}\  \delta \varphi^{\M,n+1}_{Ai},
\end{split}
\end{equation}
%
with
\begin{align}
&\mathcal{M}_{b'b} = \int_{\Omega_0} \frac{2\rho_0}{\Delta t^2}  N_b N_{b'}  \, dV \\
& \mathcal{M}^\M_{b'A} = N^\M_{A} \bigg[ \int_{\Omega_0} \frac{2\rho_0}{\Delta t^2} \sum_c N_c N_{b'}  \, dV  \bigg] + N^\M_{A,Q} \sum_c  X_{cQ} \left[ \int_{\Omega_0} \frac{2\rho_0}{\Delta t^2}  N_c N_{b'}  \, dV  \right],
\end{align}
and therefore
\begin{equation}\label{Eq:QEMS_bc_b}
\delta \varphi^{n+1}_{bi} = -\sum_A \sum_{b'} \mathcal{M}_{bb'}^{-1}\ \mathcal{M}^\M_{b'A}\  \delta \varphi^{\M,n+1}_{Ai}=\sum_A \mathcal{P}_{bA} \delta \varphi^{\M,n+1}_{Ai},
\end{equation}
with $\mathcal{P}_{bA}=- \sum_{b'} \mathcal{M}_{bb'}^{-1}\ \mathcal{M}^\M_{b'A}$.

Then, one can separate the microscopic nodes $\{a\}$ in Eq.~\eqref{Eq:MacroForce}, into nodes $\{b\}$ and $\{c\}$, i.e.
\begin{equation}
\begin{split}
\delta \mathcal{F}^{\M,n+1}_{Bi} =& \int_{\Omega^\M_0}   N^\M_B \Big \langle \frac{2\rho_0}{\Delta t^2}  \sum_{b} \delta \varphi^{n+1}_{bi} N_{b} \Big \rangle \, dV^\M  +\int_{\Omega^\M_0} N^\M_{B,J}   \Big \langle \frac{2\rho_0}{\Delta t^2}  \sum_{b}\delta \varphi^{n+1}_{bi}  N_{b}X_{J}  \Big \rangle \, dV^\M \\
&+ \int_{\Omega^\M_0}   N^\M_B   \Big \langle \frac{2\rho_0}{\Delta t^2}  \sum_{c} \delta \varphi^{n+1}_{ci} N_{c}  \Big \rangle \, dV^\M  +\int_{\Omega^\M_0} N^\M_{B,J}   \Big \langle \frac{2\rho_0}{\Delta t^2}  \sum_{c}\delta \varphi^{n+1}_{ci}  N_{c}X_{J}   \Big \rangle \, dV^\M,
\end{split}
\end{equation}
and apply Eqs.~\eqref{Eq:QEMS_bc_c} and \eqref{Eq:QEMS_bc_b} to obtain
\begin{equation}
\begin{split}
&\delta \mathcal{F}^{\M,n+1}_{Bi} = \int_{\Omega^\M_0}  \sum_A N^\M_B \delta \varphi^{\M,n+1}_{Ai} \bigg[ \sum_{b}  \mathcal{P}_{bA}\   \Big \langle \frac{2\rho_0}{\Delta t^2}  N_{b} \Big \rangle\bigg] \, dV^\M  +\int_{\Omega^\M_0} \sum_A N^\M_{B,J}  \delta \varphi^{\M,n+1}_{Ai} \bigg[\sum_{b}\mathcal{P}_{bA}\   \Big \langle \frac{2\rho_0}{\Delta t^2}   N_{b}X_{J} \Big \rangle \bigg]  \, dV^\M \\
&\phantom{\delta \mathcal{F}}+ \int_{\Omega^\M_0}   N^\M_B \sum_A \delta \varphi^{\M,n+1}_{Ai} N^\M_A  \Big \langle \frac{2\rho_0}{\Delta t^2}  \sum_{c} N_{c} \Big \rangle \, dV^\M  +\int_{\Omega^\M_0} N^\M_{B,J} \sum_A \delta \varphi^{\M,n+1}_{Ai} N^\M_A   \Big \langle \frac{2\rho_0}{\Delta t^2}  \sum_{c}  N_{c}X_{J}   \Big \rangle  \, dV^\M  \\
&\phantom{\delta \mathcal{F}}+ \int_{\Omega^\M_0}   N^\M_B   \sum_A \delta \varphi^{\M,n+1}_{Ai} N^\M_{A,Q}  \Big \langle \frac{2\rho_0}{\Delta t^2}  \sum_{c} X_{cQ} N_{c} \Big \rangle\, dV^\M  +\int_{\Omega^\M_0} N^\M_{B,J}   \sum_A \delta \varphi^{\M,n+1}_{Ai} N^\M_{A,Q}  \Big \langle \frac{2\rho_0}{\Delta t^2}  \sum_{c}X_{cQ}  N_{c}X_{J}  \Big \rangle \, dV^\M.  \\
\end{split}
\end{equation}
The sough-after Hessian then reads
\begin{equation}
\begin{split}
\frac{\partial \mathcal{F}^{\M,n+1}_{Bi}}{\partial \varphi^{\M,n+1}_{B'p}}  =&  \delta_{ip} \frac{1}{\Delta t^2} \bigg\{\int_{\Omega^\M_0} N^\M_B \bigg[ \sum_{b}  \mathcal{P}_{bB'}  \Big \langle 2\rho_0 N_{b} \Big \rangle \bigg]\, dV^\M  +\int_{\Omega^\M_0} N^\M_{B,J}   \bigg[\sum_{b} \mathcal{P}_{bB'}\  \Big \langle 2\rho_0  N_{b}X_{J}  \Big \rangle \bigg] \, dV^\M \\
&+ \int_{\Omega^\M_0} N^\M_B  N^\M_{B'} \Big \langle 2\rho_0  \sum_{c} N_{c} \Big \rangle \, dV^\M  +\int_{\Omega^\M_0} N^\M_{B,J}  N^\M_{B'}  \Big \langle 2\rho_0  \sum_{c}  N_{c}X_{J}   \Big \rangle  \, dV^\M  \\
&+ \int_{\Omega^\M_0} N^\M_B   N^\M_{B',Q}  \Big \langle 2\rho_0  \sum_{c} X_{cQ} N_{c} \Big \rangle \, dV^\M  +\int_{\Omega^\M_0} N^\M_{B,J}  N^\M_{B',Q}  \Big \langle 2\rho_0  \sum_{c}X_{cQ}  N_{c}X_{J}  \Big \rangle  \, dV^\M \bigg\} .\\
\end{split}
\end{equation}
which is constant for a fixed spatial discretization, and it may be easily adapted for changes in the time step $\Delta t$. We emphasize that the quasi-explicit nature of the method still holds true for history-dependent non-linear materials, such as viscoelastoplastic materials. Furthermore, the scheme concurrently minimizes the micro and macro problem, rather than solving for the two scales in a sequential iterative manner. The concurrent versus two-stage minimization has recently been explored in the context of quasi-continuum methods by \citet{sorkin2014local}, leading not only to a faster algorithm, but also to a more accurate characterization of the solution path. In particular, their results indicate that a staggered two-stage minimization can artificially stabilize points that do not correspond to local equilibrium positions. 

\section{Numerical study of wave propagation in a one-dimensional stratified composite} \label{Sec:1DProblem}

In this section we exemplify the dynamic computational homogenization strategy previously outlined, QEMS, over a one-dimensional layered composite with periodic excitation. This problem has been extensively studied in the literature and is thus ideal to assess the reliability of the coarse-grained procedure to capture dispersion effects and their relation to the microstructure. In all cases, the results will be compared with those obtained via standard single scale finite elements to assess their accuracy, and the convergence in space and time will be quantified. Additionally, the effect of the static and dynamic part of the stresses towards the dispersive behavior will be carefully examined.



\subsection{ Description of the case study and numerical discretization} \label{Sec:Description}
\begin{figure}[htbp]
\begin{center}
    {\includegraphics[width=0.7\textwidth]{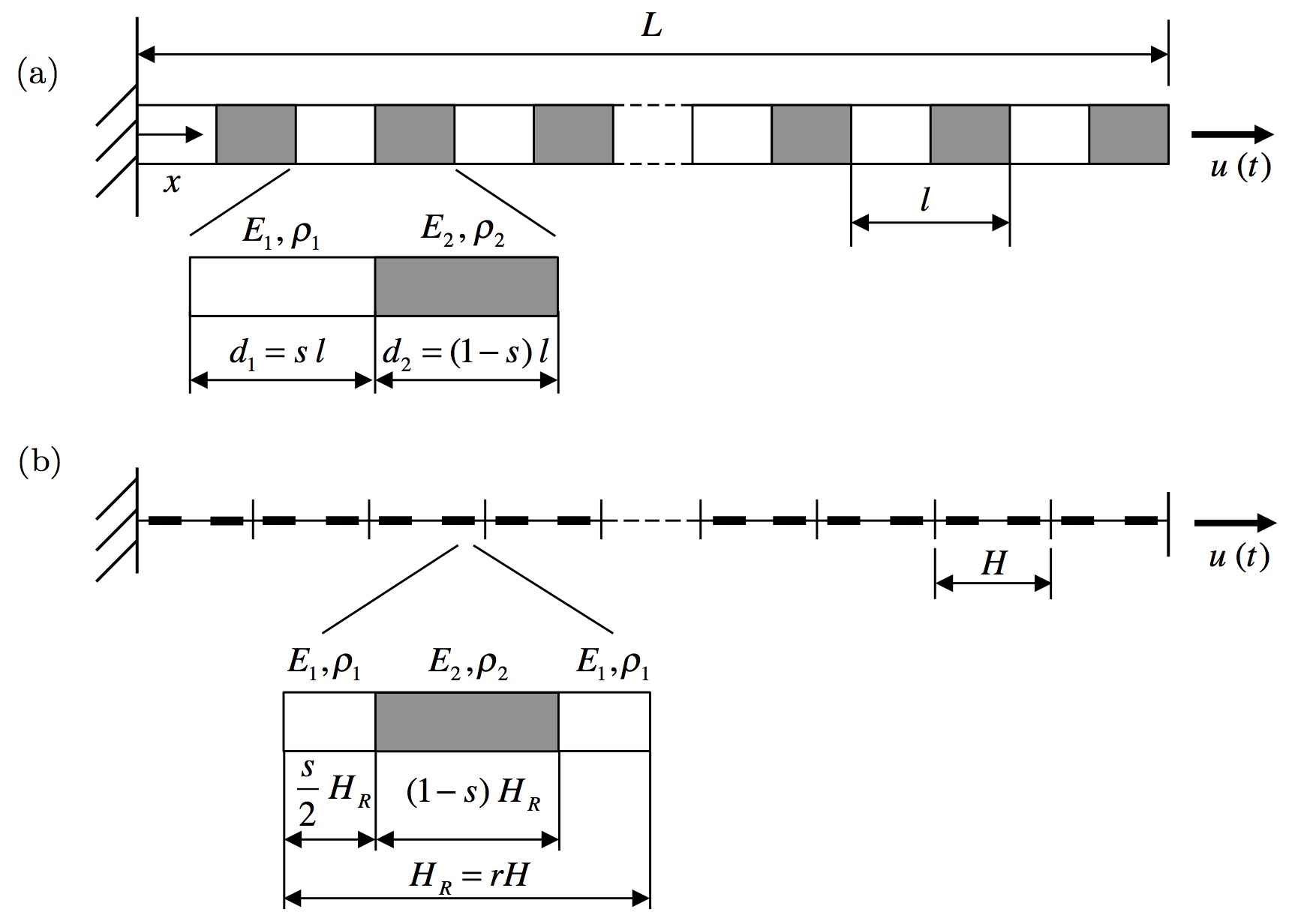}}
    \caption[]{(a) 1-D model for standard finite element analyses. (b) Macroscopic 1-D model and RVE for multiscale simulations. } \label{sf}
\end{center}
\end{figure}

We study the dynamic behavior of a one-dimensional heterogeneous elastic bar with periodic microstructure, as shown in Figure \ref{sf}. The periodic unit cell, of size $l$, is composed of two sequential layers of distinct materials, with Young's modulus $E_1$ and $E_2$, densities per unit length $\rho_1$ and $\rho_2$, and volume fractions $s$ and 1$-s$, respectively. The total length of the bar is $L$ and the period of the microstructure $l$ is considered to satisfy $l \ll L$. The bar is initially unloaded with zero initial displacement and velocity. The left boundary of the bar is fixed, while an external time-dependent displacement boundary condition $u(t)$=$A$[1$-$cos(2$\pi ft$)] is applied on the right end, where $A$ is the displacement amplitude and $f$ denotes the loading frequency. 

\begin{figure}[htbp]
\begin{center}
    {\includegraphics[width=\textwidth]{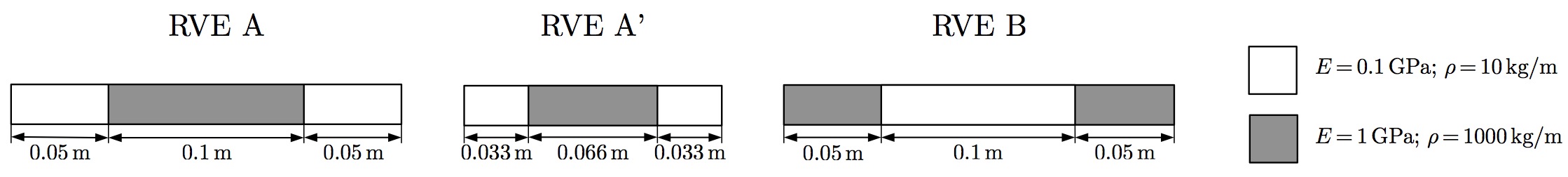}}
    \caption[]{Representative volume elements used in the numerical analyses. They consist of alternating layers of two materials, whose properties are depicted on the right of the figure. RVE A and A' are self-similar and only differ on a length scale. RVE A and B represent identical microstructures and exclusively differ on the order of the sublayers. All RVEs have identical static effective properties, with $\rho_{\text{eff}}$\,=\,505\,kg/m, $E_\text{eff}$\,=\,0.182\,GPa and $c_{\text{eff}}$\,=\,600\,m/s. } \label{RVE}
\end{center}
\end{figure}

This problem will be analyzed with both, a standard finite element solver and with the FE$^2$ procedure (QEMS), cf.~Figures~\ref{sf}(a) and (b) respectively. For the standard solver, the bar is discretized into $n_s-$1 elements of identical size $H_s=L/(n_s-$1) and linear shape functions are used to approximate the displacement field. Similarly, for the FE$^2$ method, the domain is divided at the macroscopic scale into $N-$1 elements of size $H=L/(N-$1); and linear shape functions are used as a basis for the discretization of the macro-fields. Two quadrature points are then considered in each element to obtain an exact integration of all the matrices participating in the numerical strategy \citep{stroud1974numerical}; and a representative volume element (RVE) is assigned to each of them, cf.~Figure~\ref{sf}(b). It should be noted that the RVE should be large enough to be statistically representative of the microstructure and at the same time remain small enough so that the assumption of scale separation is not violated. For periodic structures, the RVE can be appropriately selected as a unit cell, i.e.,~$H_R=rH=l$, where $r=H_R/H$ is the ratio between the size of the RVE and the size of the macro element. The long wavelength assumption requires that the thickness of each constituent, and thus $H_r$, is much smaller than the corresponding wavelength, i.e., $H_R=l\ll\lambda_{\text{eff}}=\sqrt{E_{\text{eff}}/\rho_{\text{eff}}}/f$. Here, the subscript `eff' indicates the effective (static) material properties\footnote{The effective dynamic properties are in general different. See for instance \citet{milton2007modifications} for the effective dynamic density.}, and they can be computed in this example as $E_{\text{eff}}^{-1}=s E_1^{-1}+(1-s)E_2^{-1}$, and $\rho_{\text{eff}}=s\rho_1 +(1-s) \rho_2$. In addition, each RVE is divided into $n-$1 elements, and the origin of the local coordinate system is set in the center of mass of the RVE. Linear shape functions are as well used for approximating the microscopic displacement field. The numerical scheme for the QEMS then proceeds as described in Section \ref{Sec:MultiscaleSolver} and summarized in Figure~\ref{flow}. Convergence of the Newton-Raphson procedure in one iteration is verified numerically. In particular, Eq.~\eqref{Eq::Dynamic} reveals that such convergence is achieved if the numerical error in $\mathcal{F}_{Bi}=0$ satisfies error$\,=\,$machine precision$/(\Delta t)^2$. Next, we consider symmetric RVEs, as shown in Figure~\ref{sf}(b). RVE A of Figure~\ref{RVE} will be the primary microstructure used, whereas RVE A' and B will be used in Sections \ref{Sec:Microstructure effect} and \ref{Sec:ChoiceRVE} to analyze the effect of the microstructure and the RVE choice, respectively, on the dynamic response of the layered composite.

\subsection{Convergence analyses and self-consistency} \label{Sec:ConvergenceAnalyses}

\begin{figure} 
\begin{center}  
\subfigure[ ] { \label{Fig:Convergence_H}    
\includegraphics[width=0.481\columnwidth]{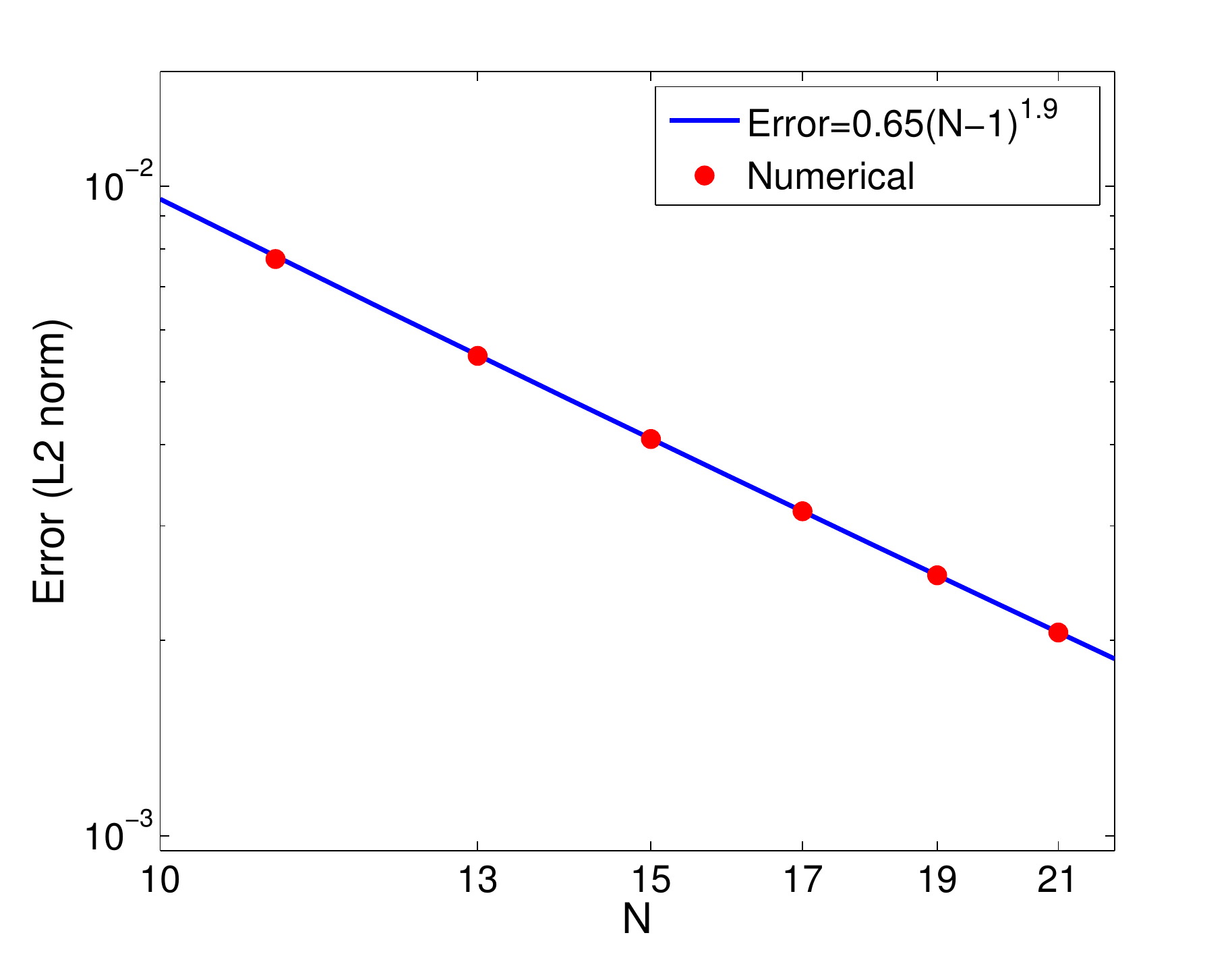} 
}    
\subfigure[ ] { \label{ET}    
\includegraphics[width=0.481\columnwidth]{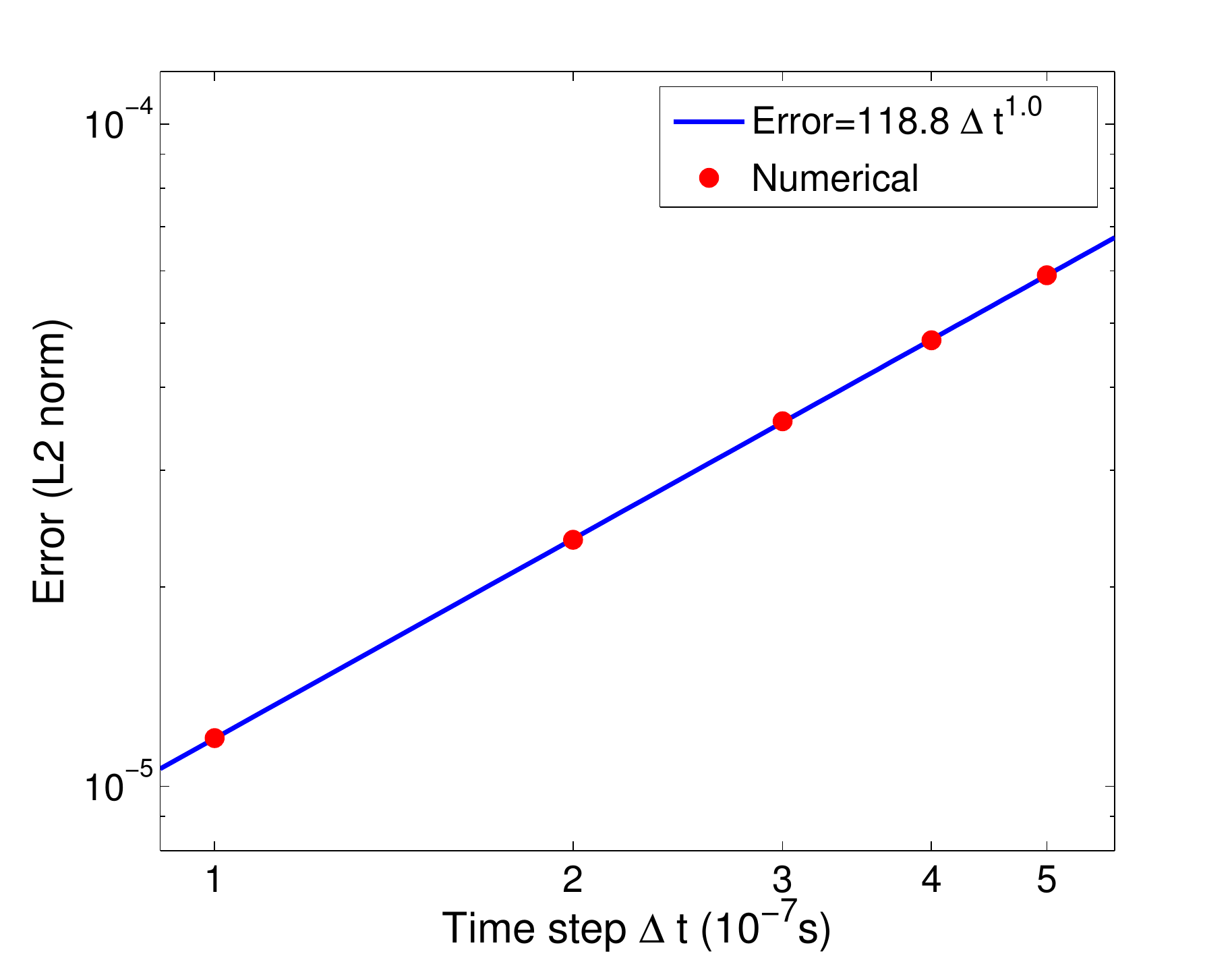}    
}    
\caption{(a) Spatial and (b) temporal convergence analysis of the $L^2$ norm of the displacement error for the QEMS solver. The parameters for the simulation are $L$\,=\,6\,m, RVE A, $n$\,=\,5, $f$\,=\,140\,Hz and $A$\,=\,0.2\,m. For (a) the time step is fixed at  $\Delta t$\,=\,$1\times 10^{-7}$\,s, while for (b) the number of macroscopic nodes is fixed at $N$\,=\,19, $r$\,=\,0.6. } \label{RVEs}    
\end{center}
\end{figure}

The numerical convergence of the multiscale scheme is here studied for both the spatial and temporal discretization. For the analyses, the microstructure and RVE discretization are held fixed, with $L$\,=\,6\,m, material properties and microstructure as in RVE A, $n$\,=\,5 and the external excitation considered is characterized by $A$\,=\,0.2\,m and $f$\,=\,140\,Hz. We first examine the convergence of the QEMS solution at time $t=1/f$ as a function of the number of macroscopic nodes $N$. The time step is considered fixed and set to a small value ($\Delta t=1\times 10^{-7}$\,s) to exclude (reduce) the error induced by the temporal discretization. The $L^2$ norm\footnote{The $L^2$ norm of the displacement error is defined as $\|\varphi^h-\varphi_{\text{exact}}\|_{L^2}=\left(\int_{\Omega_0}|\varphi^h-\varphi_{\text{exact}}|^2dX\right)^{1/2}$, where $\varphi^h$ represents the numerical solution and $\varphi_{\text{exact}}$ the exact one.} is used to assess the difference between the approximate and exact solution (here estimated from the convergence analysis). The error, represented in log-log scale in Figure~\ref{Fig:Convergence_H} reveals a convergence rate of 1.9, close to the quadratic convergence expected for single scale finite element schemes \citep{strang1973analysis}.

For the temporal convergence analysis, the number of macro nodes is fixed at $N$\,=\,16 and the $L^2$ norm of the displacement error is investigated with respect to the time step $\Delta t$ in the range 1$\times10^{-7}-5\times10^{-7}$\,s. The analysis shows, see Figure~\ref{ET}, that the multiscale solver inherits the linear convergence in time from the underlying explicit dynamic integration scheme that has been spatially homogenized \citep{zienkiewicz1967finite,wood1990practical}. 
 
\begin{figure} [H]
\begin{center}  
\subfigure[ ] { \label{25HzDispH} 
\includegraphics[width=0.481\columnwidth]{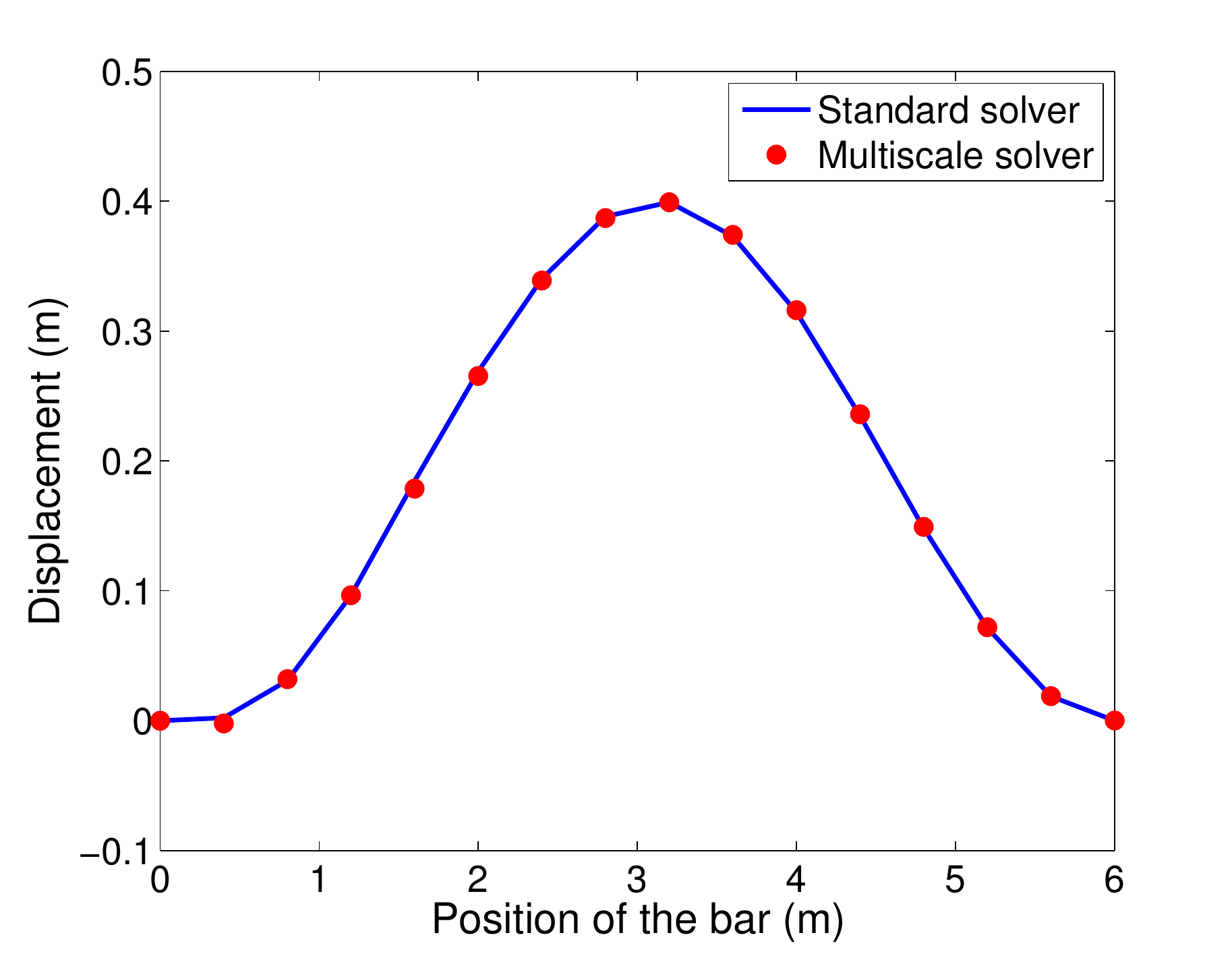} 
}    
\subfigure[ ] {\label{MvsSvsHz}   
\includegraphics[width=0.481\columnwidth]{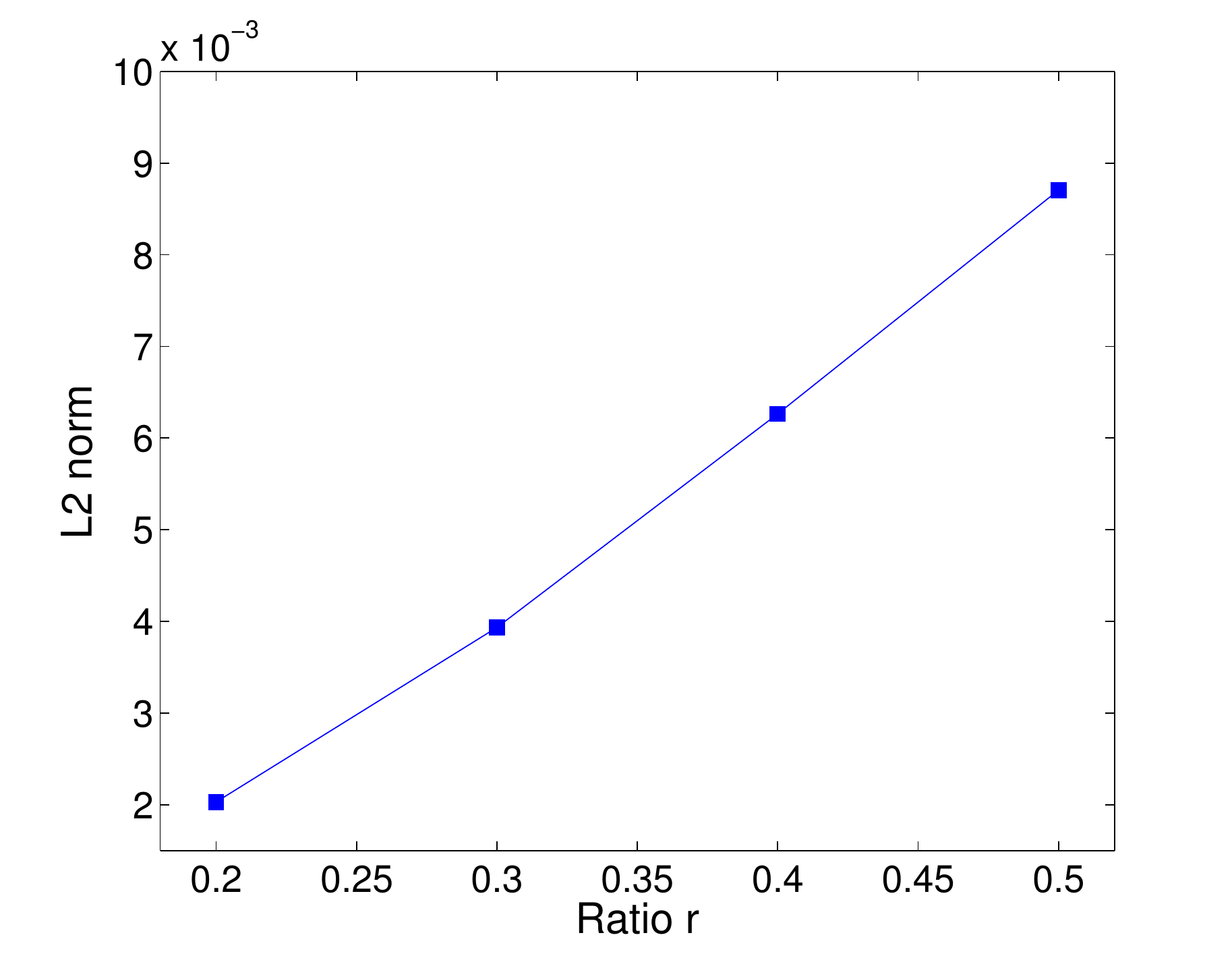}    
}    
\caption{Consistency between the QEMS and the standard solver for a homogeneous bar: (a) comparison of the macroscopic displacement along the bar, with $r$\,=\,0.5; (b) $L^2$ norm of the displacement difference as a function of $r$. The parameters for the simulation are $N$\,=\,$n_s$\,=\,16, $n$\,=\,5, $E$\,=\,0.182\,GPa, $\rho$\,=\,505\,kg/m, $f$\,=\,105\,Hz, $A$\,=\,0.2\,m, $\Delta t$\,=\,1$\times10^{-7}$\,s and $t=1/f$. }     
\end{center}
\end{figure}

Additionally, as a sanity check, the results of the quasi-explicit multiscale solver (QEMS) and the explicit standard solver are compared 
%
for a homogeneous material, where any sensible spatial coarse-graining procedure shall leave the results invariant up to numerical errors. The effective material properties of the microstructure previously analyzed are chosen for the homogeneous bar, $E_{\text{eff}}$\,=\,0.182\,GPa, $\rho_{\text{eff}}$\,=\,505\,kg/m, and the system is excited at a loading frequency of $f$\,=\,105\,Hz. For consistency, both solvers use an identical macroscopic discretization, with the number of nodes chosen as $N$\,=\,$n_s$\,=\,16. The displacement along the homogeneous bar at $t=1/f$ is plotted in Figure~\ref{25HzDispH} for the the standard and the multiscale solver (with $r$\,=\,0.5, and $n$\,=\,5), showing an excellent agreement between them. The difference in the solution, although not visible to the eye, can be quantified, and is shown in Figure~\ref{MvsSvsHz} as a function of the ratio $r$ in the multiscale solver. As expected, the (numerical) error converges to zero as the ratio $r$ is decreased.

\subsection{Dynamic response and consistency with standard solver solution} \label{Sec:DispersionProperty}

\begin{figure} [H]
\begin{center}  
\subfigure[Low frequency ] {\label{200Max}   
\includegraphics[width=0.481\columnwidth]{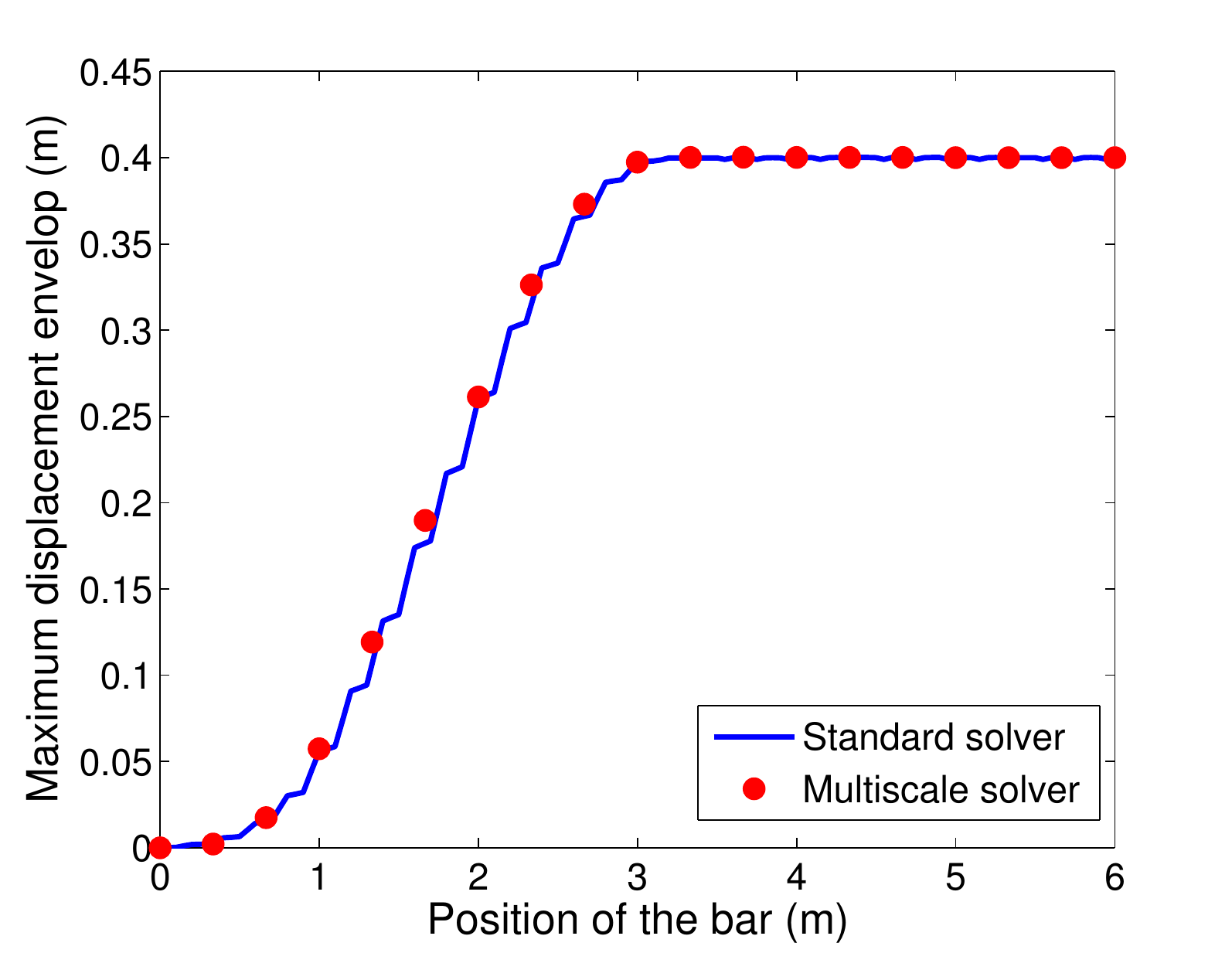} 
}    
\subfigure[High frequency ] { \label{MD480Hz}   
\includegraphics[width=0.481\columnwidth]{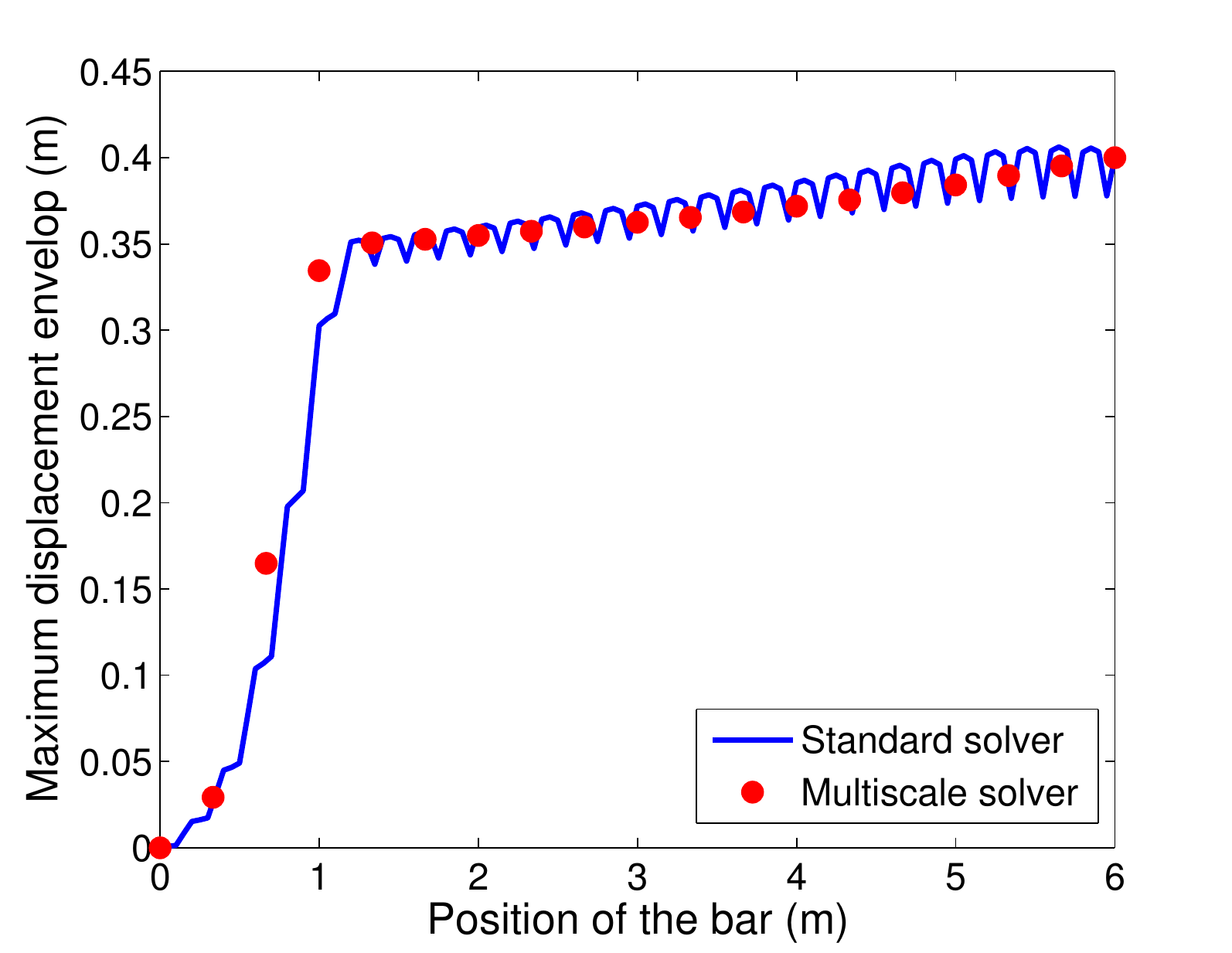}    
}    
\caption{Envelope of maximum displacement along the heterogeneous bar by the standard and the multiscale solver (a) at low frequency, $f$\,=\,105\,Hz, over a time period of $t$\,=\,$[0,\,1/f]$; and (b) at high frequency, $f$\,=\,480\,Hz, over $t$\,=\,$[0,\,4.5/f]$. The parameters for the simulations are RVE A, $N$\,=\,19, $n$\,=\,5, $n_s$\,=\,121, $r$\,=\,0.6, 
$\Delta t$\,=\,1$\times 10^{-7}$\,s.}
\end{center}
\end{figure}

Heterogeneous periodic materials are known to have a frequency-dependent behavior to wave propagation. In some cases, there may even be distinct frequency ranges in which no wave can propagate \citep{kushwaha1993acoustic,liu2000locally,aravantinos2014elastodynamic}. These are known as band gaps or stop bands, in contrast to the remaining frequency ranges, which are denoted as pass bands. For one dimensional systems, these stop bands occur at frequencies that are out of the range of applicability of FE$^2$ methods, which are constructed on the premises of separation of length scales. Yet, the dispersive nature of periodic materials is noticeable at lower frequencies, and is here examined by plotting the maximum displacement envelope along the bar over a given time interval. For these analyses, the external loading is set to last for one period and the decay of the amplitude of the impulse is examined. We consider the microstructure and the multiscale scheme associated to RVE A, for which
 the lower edge of the first stop band of the corresponding infinite layered structure can be obtained with the transfer matrix method \citep{camley1983transverse} and results in $f$\,=\,988.5\,Hz. We note that the finite periodic structure studied here numerically does not destroy the frequency-banded nature of the dynamic response and the calculation of infinite layered structure remains valid for a finite structure as long as the number of layers is large enough \citep{hussein2006dispersive}. Figure \ref{200Max} shows the maximum displacement response in the entire heterogeneous structure during a time span $t$\,=\,1/$f$ for a relatively low frequency of $f$\,=\,105\,Hz. The final time is set to guarantee that the wave propagation does not reach the fixed end. It can be seen that the amplitude of the displacement remains the same as that of the input excitation, $A$\,=\,0.4\,m, indicating that no dispersion occurs during the wave propagation. Only at the left end of the bar, the amplitude of the bar sharply decreases to zero to comply with the displacement boundary conditions at that end.  At a higher loading frequency, wave scattering takes place. Figure~\ref{MD480Hz} shows that there exists approximately 12.5\% dispersion in the maximum displacement in the structure at time $4.5/f$. As indicated by the figures, the results at both, low and high frequency, are in excellent agreement with the standard solver solution. The zigzag shape obtained for the single scale FE solution is due to the discontinuous distribution of Young's modulus along the bar, while for the multiscale solver, the homogenized response of the bar is presented, which is, as expected, smooth. The numerical results where obtained with a fixed time step of $\Delta t = 1\times 10^{-7}$\,s, and the number of nodes in the finite element schemes were chosen such that the single and the multiscale solver have an identical discretization for each sublayer. In particular, the number of macro nodes is $N$\,=\,16 and $n$\,=\,5 for each RVE, while for the standard solver, the number of nodes is set as $n_s$\,=\,121.

\subsection{Effect of the microstructure on the dynamic response of the material} \label{Sec:Microstructure effect}

\begin{figure} [H]
\begin{center}
\subfigure[Low frequency ] {\label{Fig:LengthScale_Low_f}   
\includegraphics[width=0.481\columnwidth]{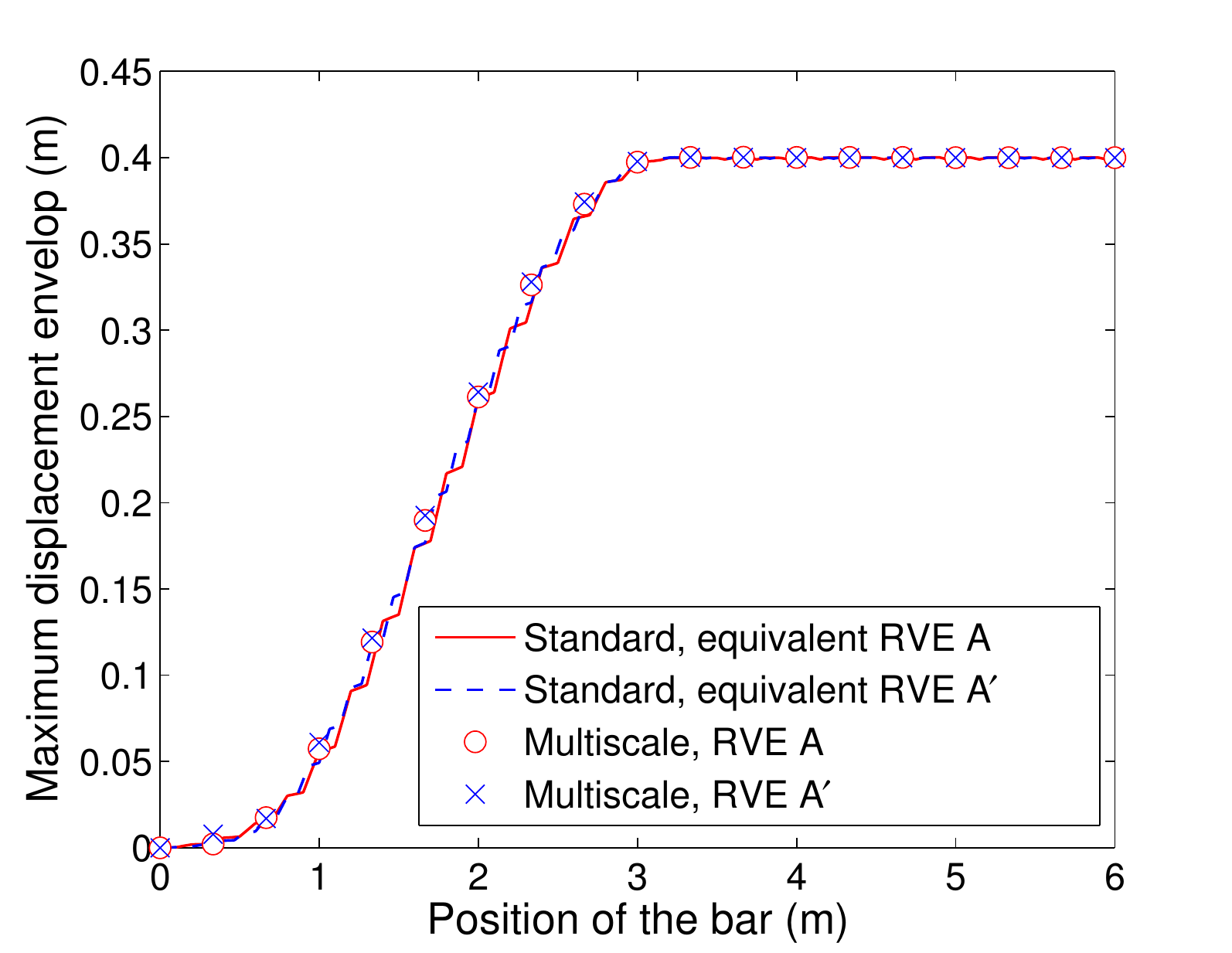} 
}  
\subfigure[High frequency ] { \label{MaxRatio} 
\includegraphics[width=0.481\columnwidth]{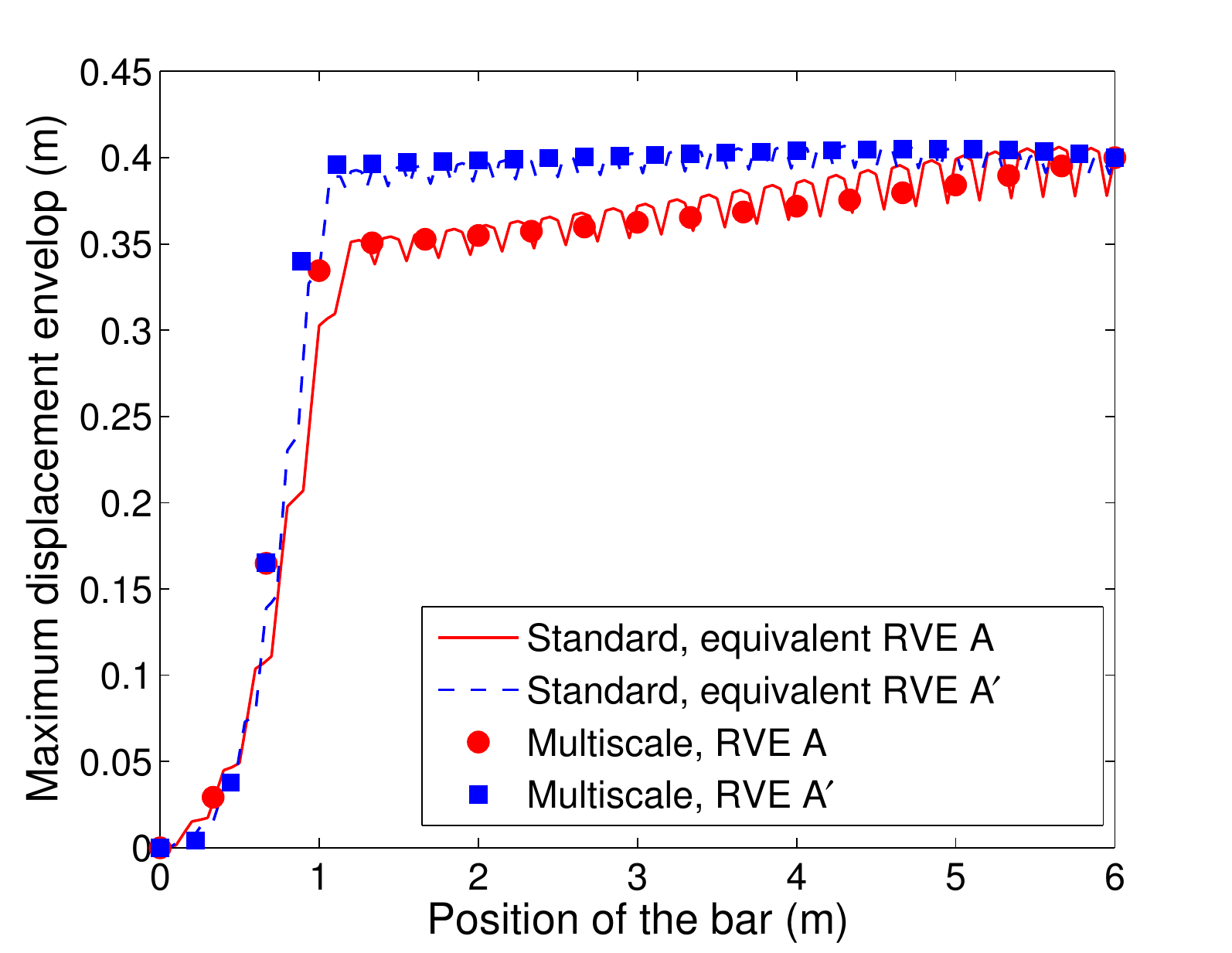}    
}    
\caption[]{Size effect on the dynamic response of heterogeneous materials. The envelope of maximum displacement along the bar for RVE A and A' is compared at (a) low frequency ($f$\,=\,105\,Hz) over a time period of $t$\,=\,$[0,\,1/f]$; and (b) high frequency ($f$\,=\,480\,Hz) within time period of $t$\,=\,$[0,\,4.5/f]$, These RVEs correspond to self-similar microstructures, with a different length scale. The parameters for the simulation associated to RVE A are $N$\,=\,19, $r$\,=\,0.6, $n$\,=\,5, $n_s$\,=\,121, $\Delta t$\,=\,1$\times 10^{-7}$\,s, and the parameters associated to RVE A' are $N$\,=\,28, $r$\,=\,0.6, $n$\,=\,5, $n_s$\,=\,181, $\Delta t$\,=\,1$\times 10^{-7}$\,s.} \label{Fig:LengthScale}
\end{center} 
\end{figure}

It is well known that materials with an equivalent static behavior, but distinct microstructures, do not necessarily have an identical response under dynamic loading. In this section we analyze such differences with both the standard and the QEMS solver. In particular, we consider RVE A and A', depicted in Figure~\ref{RVE}, which are self-similar and therefore have the same effective (static) properties, but they differ on a length scale. Figure \ref{Fig:LengthScale_Low_f} depicts the envelope of maximum displacement along the bar, when it is excited at a low frequency ($f$\,=\,105\,Hz). It can be seen that for both, QEMS and standard solver, the results with the two RVEs are indistinguishable, as expected. However, when the loading frequency is increased to 480\,Hz, the response of both microstructures significantly differ, cf.~Figure~\ref{MaxRatio}. In particular, as the microstructure becomes finer (RVE A') the heterogeneous material gradually looses its exotic properties and resembles more closely a homogeneous material where no dispersion occurs.  

\begin{figure} [H]
\begin{center}  
{   \label{DispersionAandA'}
\includegraphics[width=0.482\columnwidth]{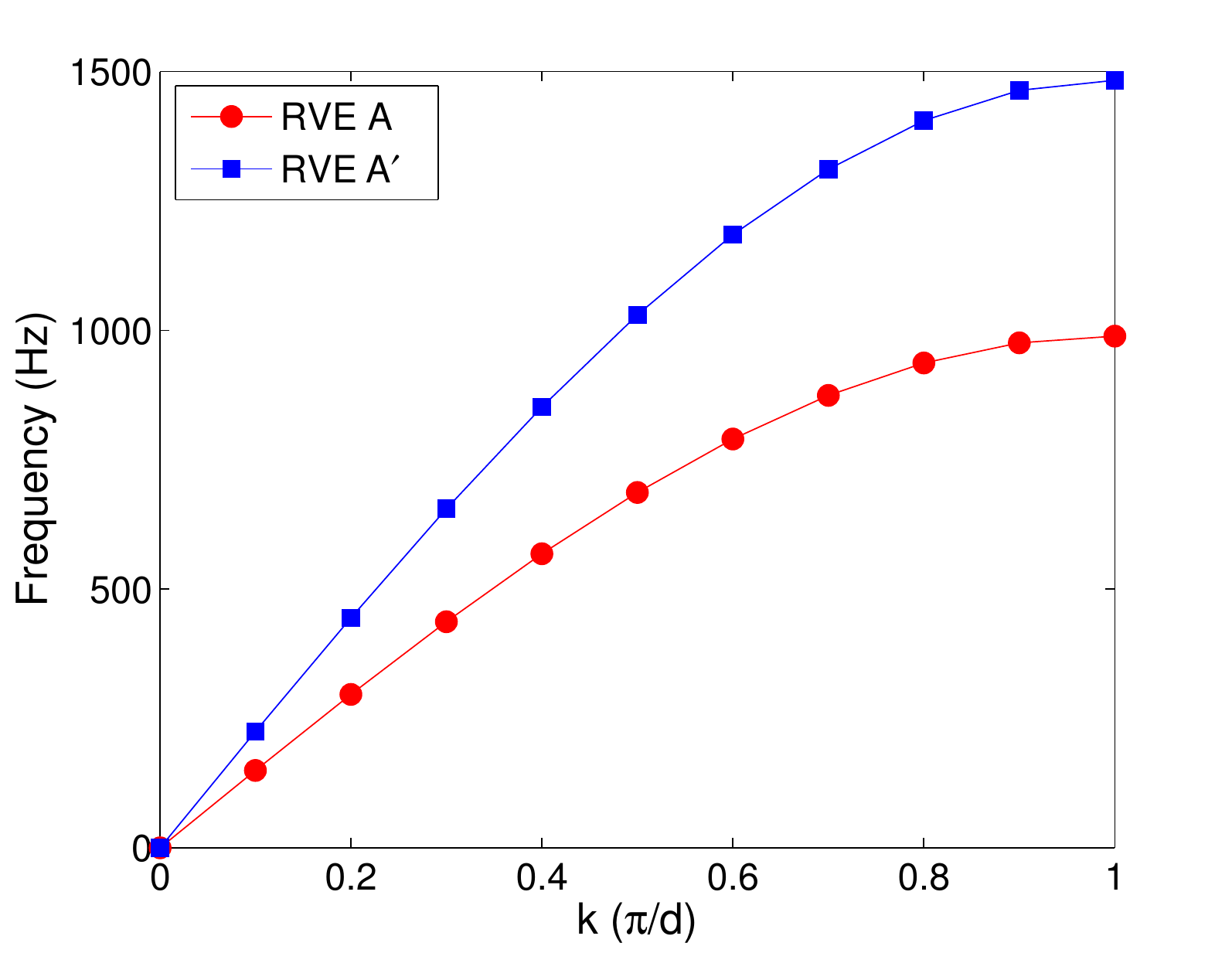} 
}    
\caption{Dispersion curves of RVE A and RVE A' obtained with transfer matrix method.} \label{181vs818at125Hz}    
\end{center}
\end{figure}

This result is consistent with the dispersion analysis of these two structures, computed as if they were 1D infinite periodic mediums. As shown in Figure \ref{DispersionAandA'}, the first band gap for RVE A starts from 988.5\,Hz, while the lower edge of the first band gap for RVE A' is 1482.8\,Hz. Therefore, at any given frequency, the dispersion expected for RVE A' is lower than that of RVE A, and the differences between both RVEs vanish at lower frequencies, for which the effective homogeneous non-dispersive response is recovered. 
We note that this comparison is meaningful, since only a very small number of unit cells in a finite bar are typically sufficient for frequency bandedness to carry over from the infinite periodic case \citep{hussein2006dispersive,aaberg1997usage}.

\subsection{Choice of the representative volume element} \label{Sec:ChoiceRVE}

\begin{figure} [H] 
\begin{center} 
\subfigure[Low frequency ] { 
\includegraphics[width=0.481\columnwidth]{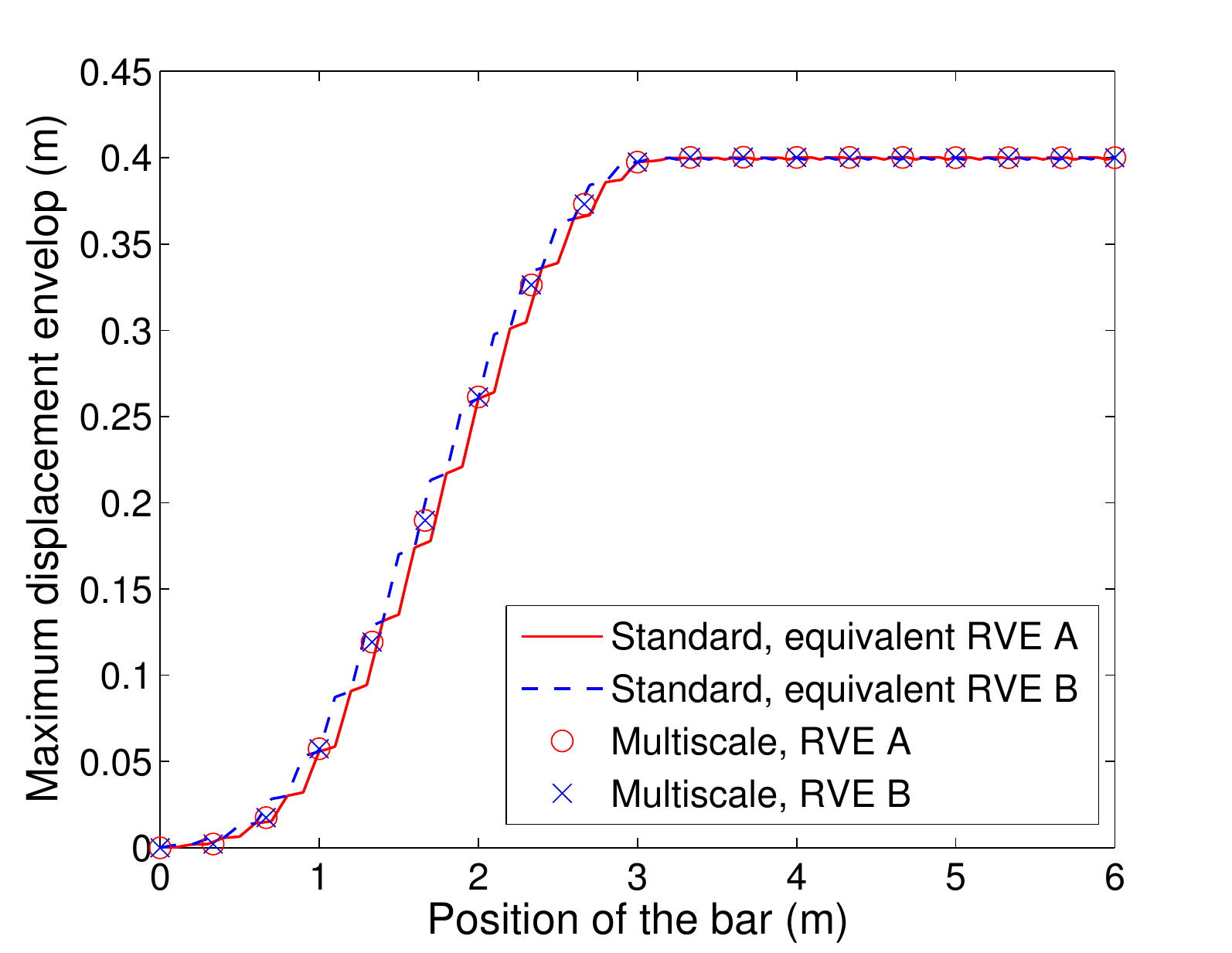} 
}  
\subfigure[High frequency ] { 
\includegraphics[width=0.481\columnwidth]{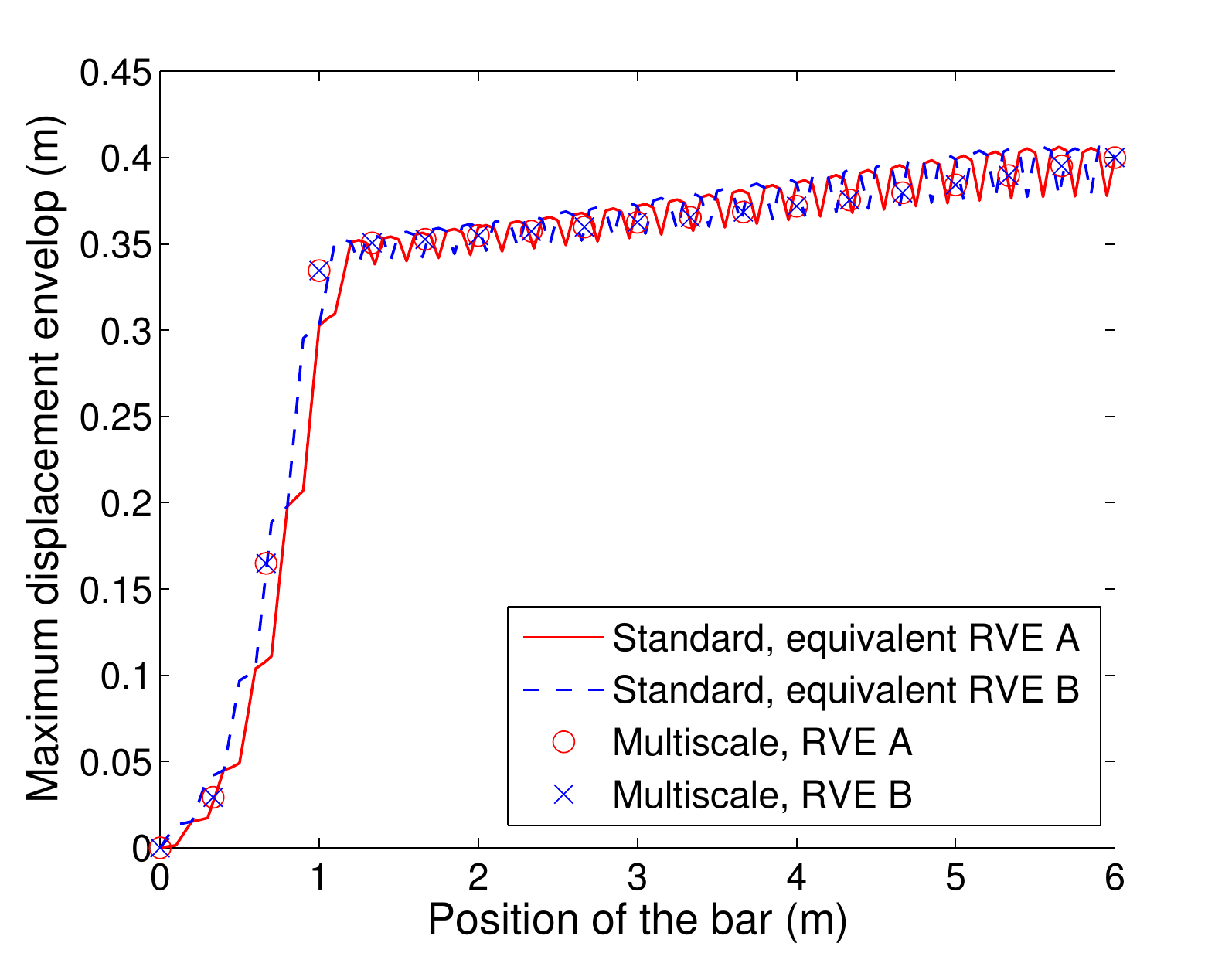}    
}    
\caption[]{Effect of the RVE choice on the envelope of maximum displacement for a given microstructure. The parameters for the simulation are $f$\,=\,480\,Hz, $N$\,=\,19, $n$\,=\,5, $n_s$\,=\,121, $t$\,=\,4.5/$f$, $\Delta t$\,=\,1$\times 10^{-7}$\,s.  } \label{MD480AB}
\end{center} 
\end{figure}

The selection of a suitable representative volume element for complex microstructures is generally a difficult problem by itself that, even in the static setting, and it may involve a detailed optimization procedure, see for example \cite{Balzani2014}. The periodicity of the layered material studied here naturally simplifies the choice of the RVE to a unit cell from which the periodic structure may be recovered. Yet, there still exists multiple choices of such cells, actually, an infinite number of them. All of these choices are equivalent in the classical setting (static with no body forces), but they are a priori distinct for general loading conditions. From all the possible RVE choices, two are favored because of their symmetry and they correspond to RVE A and B in Figure~\ref{RVE}. We note that an artificial asymmetry in the RVE construction would lead, for example, in the static case with constant body forces, to an artificial internal torque that translates into a non-symmetric macroscopic Cauchy stress tensor. The two chosen RVEs only differ in the order of the sublayers: RVE A has the stiffer and denser sublayer (represented in grey) in the middle section of the RVE, while RVE B has the softer and lighter sublayer arranged in the central region. The dynamic response of the multiscale solver with both RVEs is shown in Figure~\ref{MD480AB} for two frequencies, 105Hz and 480Hz, and compared with the standard solver solution. The small differences between the RVE A and B with the single finite element scheme is due to the order of sublayers, and they will coincide with each other after every unit cell. As for the multiscale solver, the order of sublayers does not influence the dispersion results, validating the robustness of the framework with respect to, a priori, indistinguishable RVE choices for the same microstructure.

\subsection{Static and dynamic part of the stress tensor} \label{Sec:StaticDynamicStress}

\begin{figure} [H]
\begin{center}  
\subfigure[Static part of stress] { \label{fig:a}    
\includegraphics[width=0.481\columnwidth]{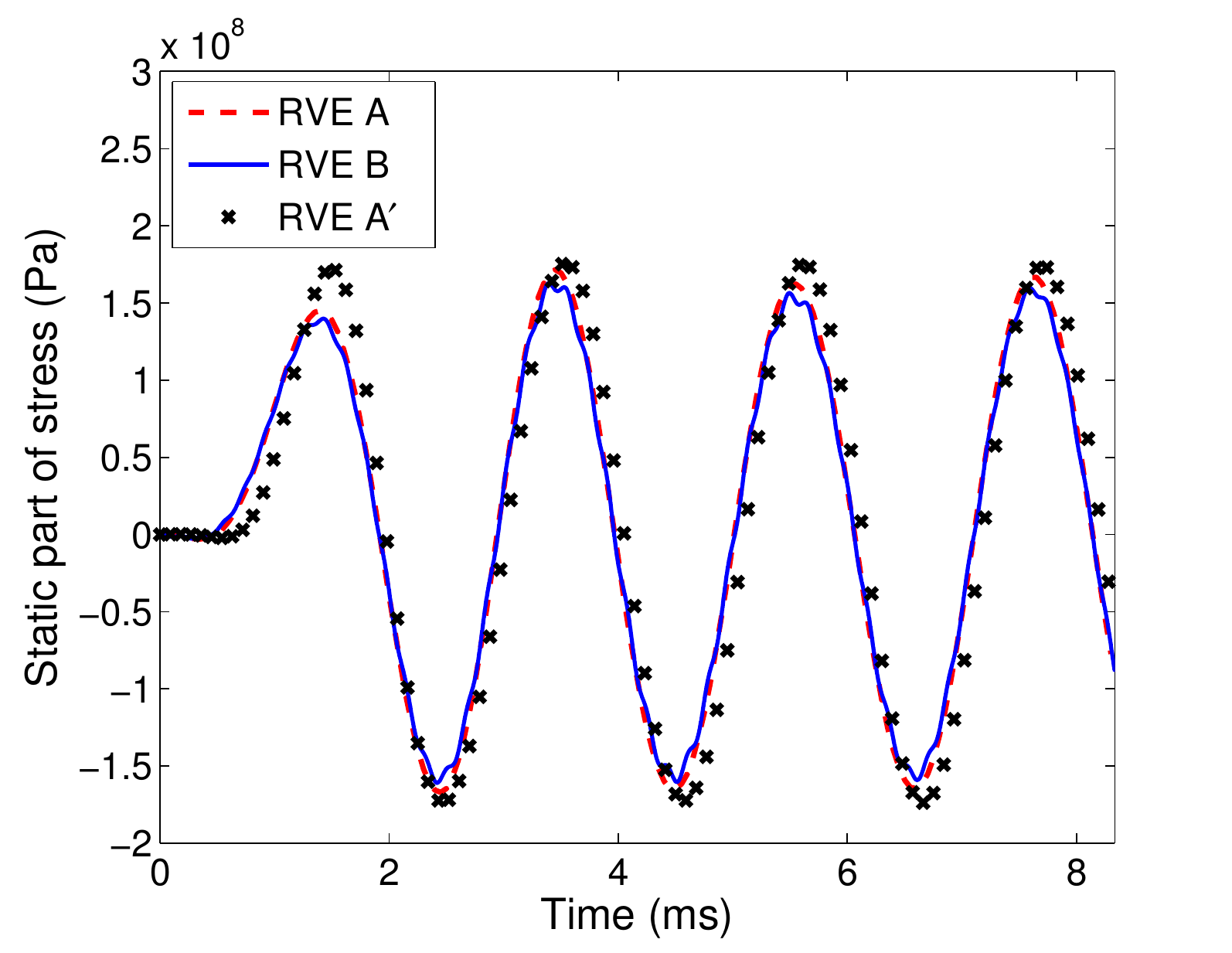} 
}    
\subfigure[Dynamic part of stress] { \label{fig:b}    
\includegraphics[width=0.481\columnwidth]{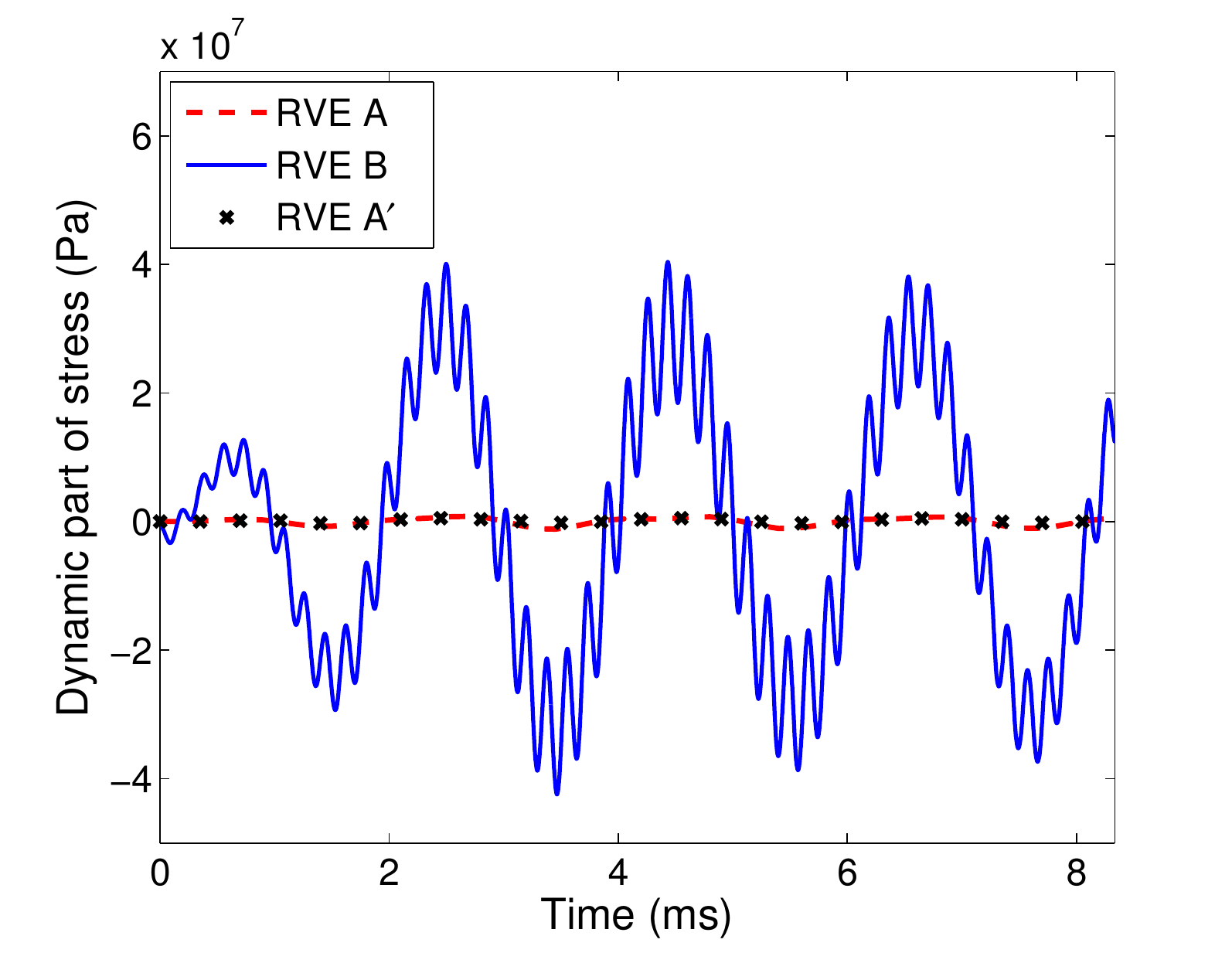}    
}   
\caption{Time evolution of the static and dynamic stress for the microstructures associated to RVE A, A' and B for a periodic excitation with frequency $f$\,=\,480\,Hz. The results of RVE A and B correspond to the first RVE in the 17th macro element in the multiscale solver, while the results of RVE A' correspond to the first RVE in the 25th macro element. The parameters for the simulation associated to RVE A and B are $N$\,=\,19, $r$\,=\,0.6, $n$\,=\,5, $\Delta t$\,=\,1$\times 10^{-7}$\,s, and the parameters associated to RVE A' are $N$\,=\,28, $r$\,=\,0.6, $n$\,=\,5, $\Delta t$\,=\,1$\times 10^{-7}$\,s} \label{StressRVEsAB}    
\end{center}
\end{figure}

\begin{figure} [H]
\begin{center}  
\subfigure[] {\label{fig:Ratio}   
\includegraphics[width=0.481\columnwidth]{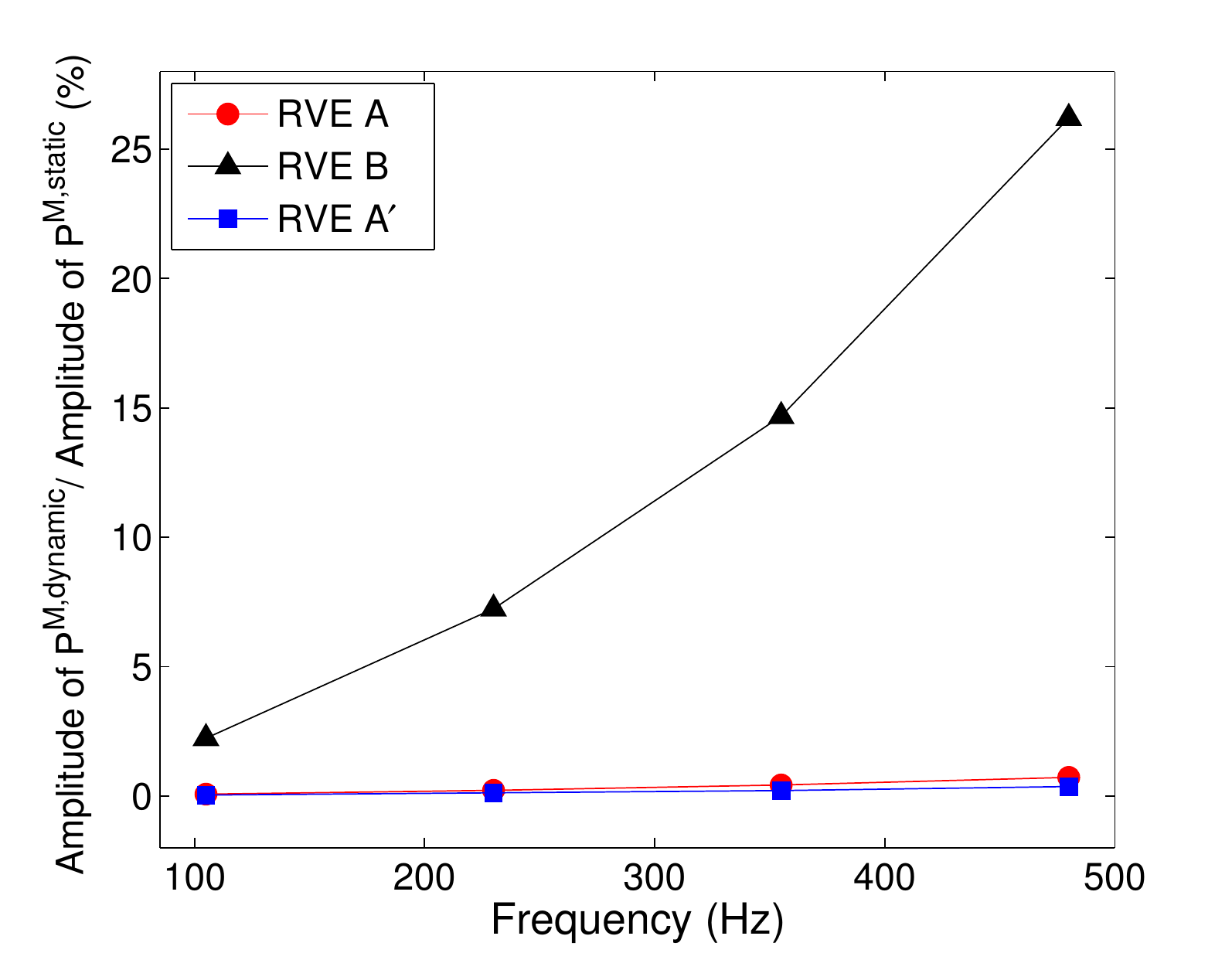} 
}    
\subfigure[] { \label{DSandSSandDisp}
\includegraphics[width=0.481\columnwidth]{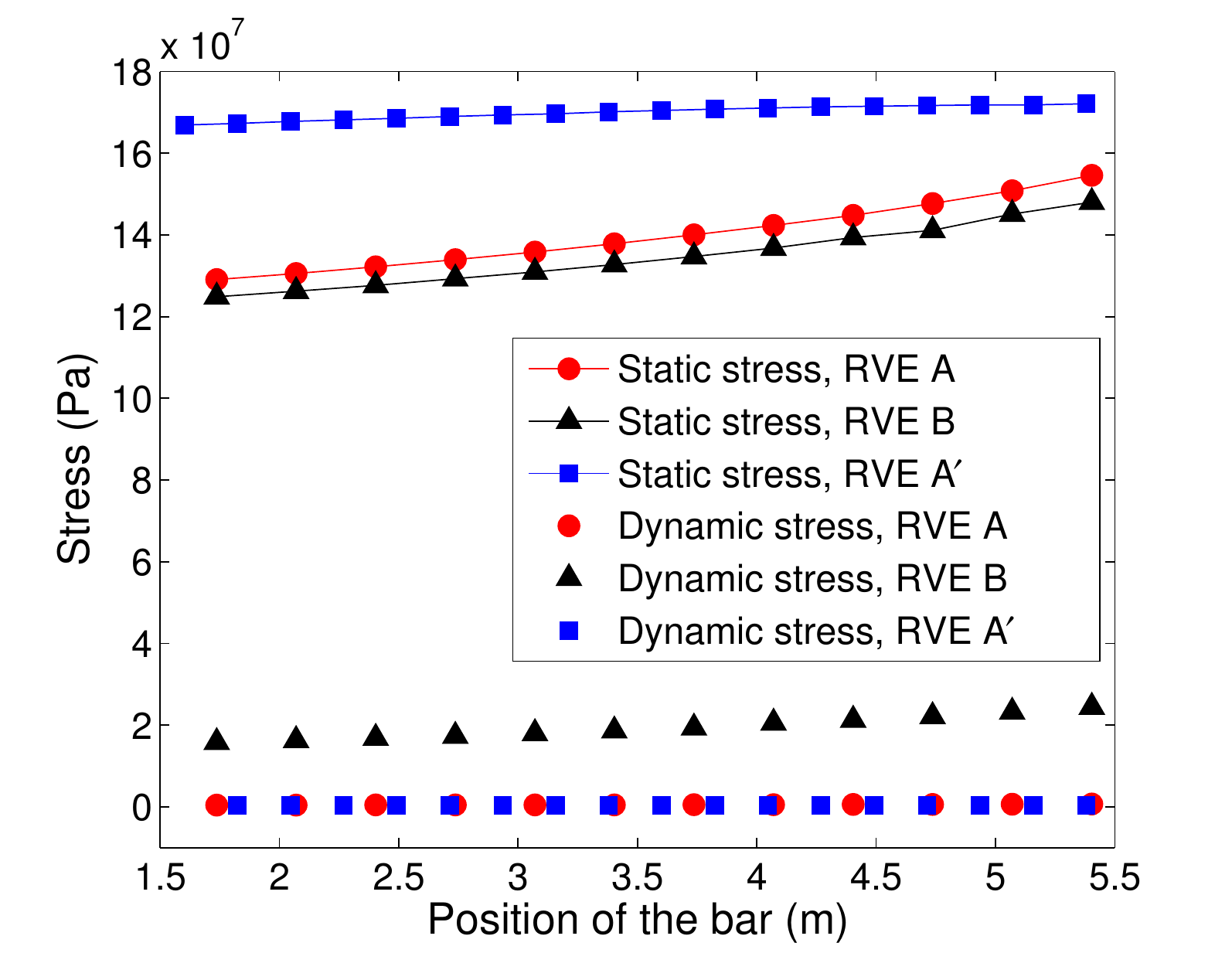}    
}   
\caption{(a) Ratio between the amplitude of dynamic and static part of stress with varying external frequencies for the microstructures associated to RVE A, A' and B. (b) Envelope of amplitude of static and dynamic stress along the bar within the time period $t$\,=\,4.5/$f$ for RVEs A, A' and B. The results of RVE A and B correspond to the first RVE in the 17th macro element in the multiscale solver, while the results of RVE A' correspond to the first RVE in the 25th macro element. The parameters for the simulation associated to RVE A and B are $N$\,=\,19, $r$\,=\,0.6, $n$\,=\,5, $\Delta t$\,=\,1$\times 10^{-7}$\,s, and the parameters associated to RVE A' are $N$\,=\,28, $r$\,=\,0.6, $n$\,=\,5, $\Delta t$\,=\,1$\times 10^{-7}$\,s.}
\end{center}
\end{figure}

As shown in the general framework of the multiscale numerical scheme, the macroscopic stress associated to an RVE can be computed as
\begin{equation} \label{Eq:StressDecomposition}
\mathbf{P}^\M = \langle \mathbf{P} \rangle + \langle \rho_0\left(\ddot{\p}-\mathbf{B} \right) \otimes \mathbf{X} \rangle = \mathbf{P}^{\M,\text{static}} + \mathbf{P}^{\M,\text{dynamic}}.
\end{equation}
It is comprised of the classical microscopic stress average, often called the static part of the stress, $\mathbf{P}^{\M,\text{static}}$, and an inertia-dependent term denoted the dynamic part, $\mathbf{P}^{\M,\text{dynamic}}$ \citep{molinari2001micromechanical,jacques2012effects}. At low frequencies, the stress tensor of the material can be rigorously shown to be equal to the static stress tensor, cf.~Proposition 12.9 \cite{cioranescu1999introduction}, and thus the contribution of the dynamic part is negligible. For a higher loading frequency, for instance, $f$\,=\,480\,Hz, the dynamic part is expected to become more important. Figure \ref{StressRVEsAB}(a) and (b) show a measurable contribution of the static and dynamic stress, respectively. For the results of RVE A and B, the RVE depicted is located in the first integration point in the 17th macro element, i.e., at the position of $X$\,=\,5.40\,m, and the closest RVE in the case of RVE A' is located in the first integration point in the 25th macro element with $X$\,=\,5.38\,m. Interestingly, the maximums of the static and dynamic stress occur approximately at the same instance of time and they are in anti-phase. 
In order to quantify the contribution of both stress terms, Figure \ref{fig:Ratio} shows the ratio between the amplitude of dynamic stress and the static stress as a function of the loading frequencies. As expected, there is a monotonic increase of such ratio with frequency. 


Although the dynamic stress becomes increasingly more important at larger frequencies, it is important to emphasize that the static part of the stress is also intrinsically dependent on whether the micro displacement field follows the static or dynamic evolution equation. In other words, the expression of the static part of the stress is identical to the total stress in a static setting, but its numerical value will differ in both scenarios. The amplitude of the static and dynamic stresses for RVE A, A' and B are shown in Figure \ref{fig:Ratio}, in which the amplitude is defined as the average of the maximum and the minimum stress within the time evolution. It can be seen that the results are consistent with Eq.~\eqref{Eq:StressDecomposition}, which indicates that the layers with larger densities have a predominant role in the amplitude of the dynamic stresses. As a result, RVE B has a larger amplitude of $\mathbf{P}^{\M,\text{dynamic}}$ than RVE A and A'. It is however difficult to infer from the information of the figure the overall response of the material and its dispersive nature; recall that RVE A and B correspond to an identical microstructure, and have the same temporal response and dispersive behavior.

For a perfect periodic excitation of a homogeneous bar with linear elastic materials, 
the amplitude of the stress is proportional to the amplitude of the displacement. It is thus of interest to represent the maximum stress envelope for both contributions to the stresses along the bar to understand whether the dispersive nature of the heterogeneous material is mainly induced by the static or dynamic component of the stress tensor. Figure \ref{DSandSSandDisp} precisely shows these results for RVE A, A' and B with an impulse excitation, and, interestingly, the static stresses have a more significant contribution to the decay of the total stresses along the bar. These results thus indicate that the labeling of the two contributions to the stresses as `static' and `dynamic' could be misleading, as the so-called `static' part plays a crucial role in the dispersive nature of heterogeneous media at high frequencies.

%
\subsection{Numerical remarks}\label{Sec:NumericalRemarks}\label{Sec:Error_CoarseGraining}

\begin{figure} [H]
\begin{center}  
\subfigure[Low frequency ] {
\includegraphics[width=0.481\columnwidth]{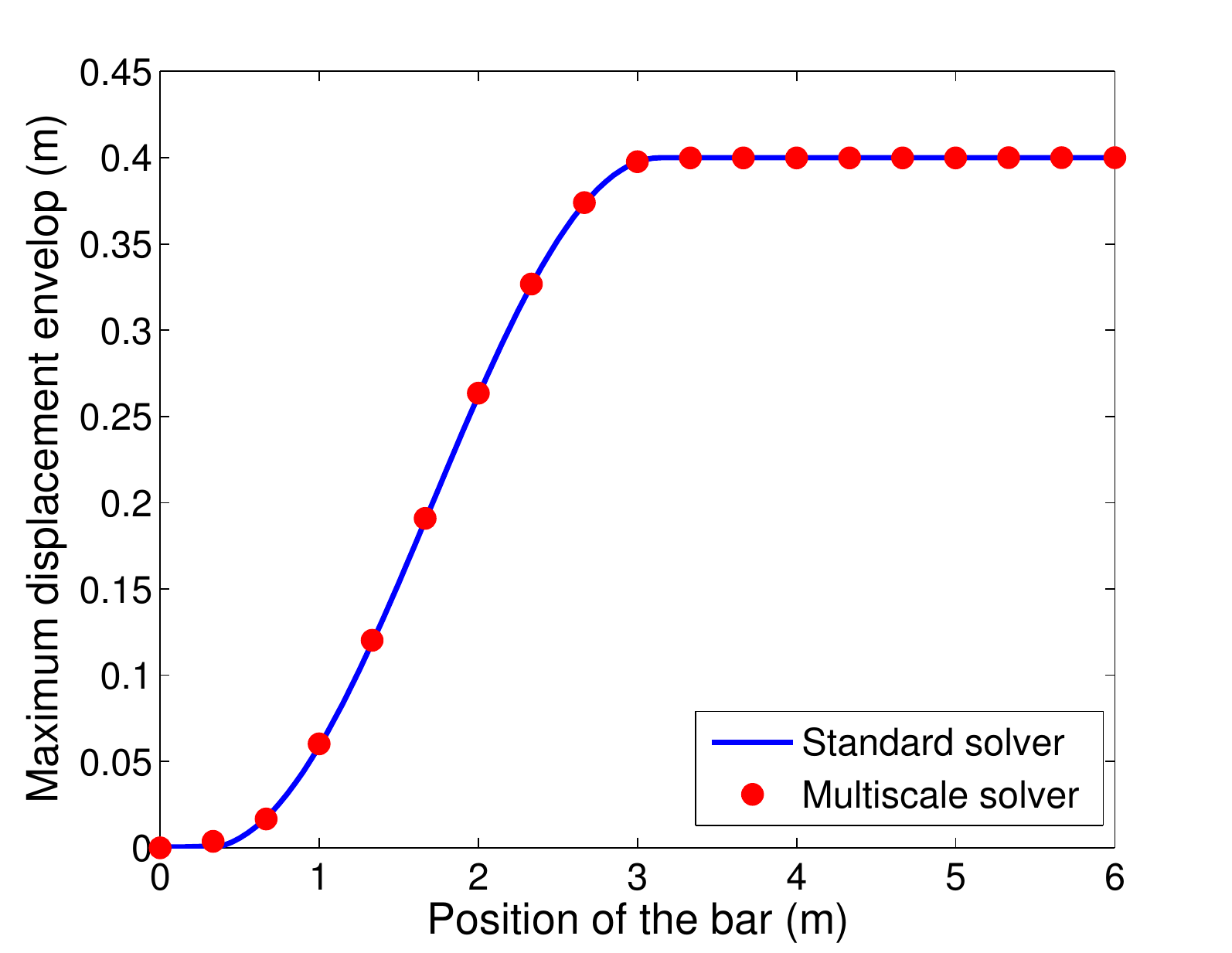} 
}    
\subfigure[High frequency ] { 
\includegraphics[width=0.481\columnwidth]{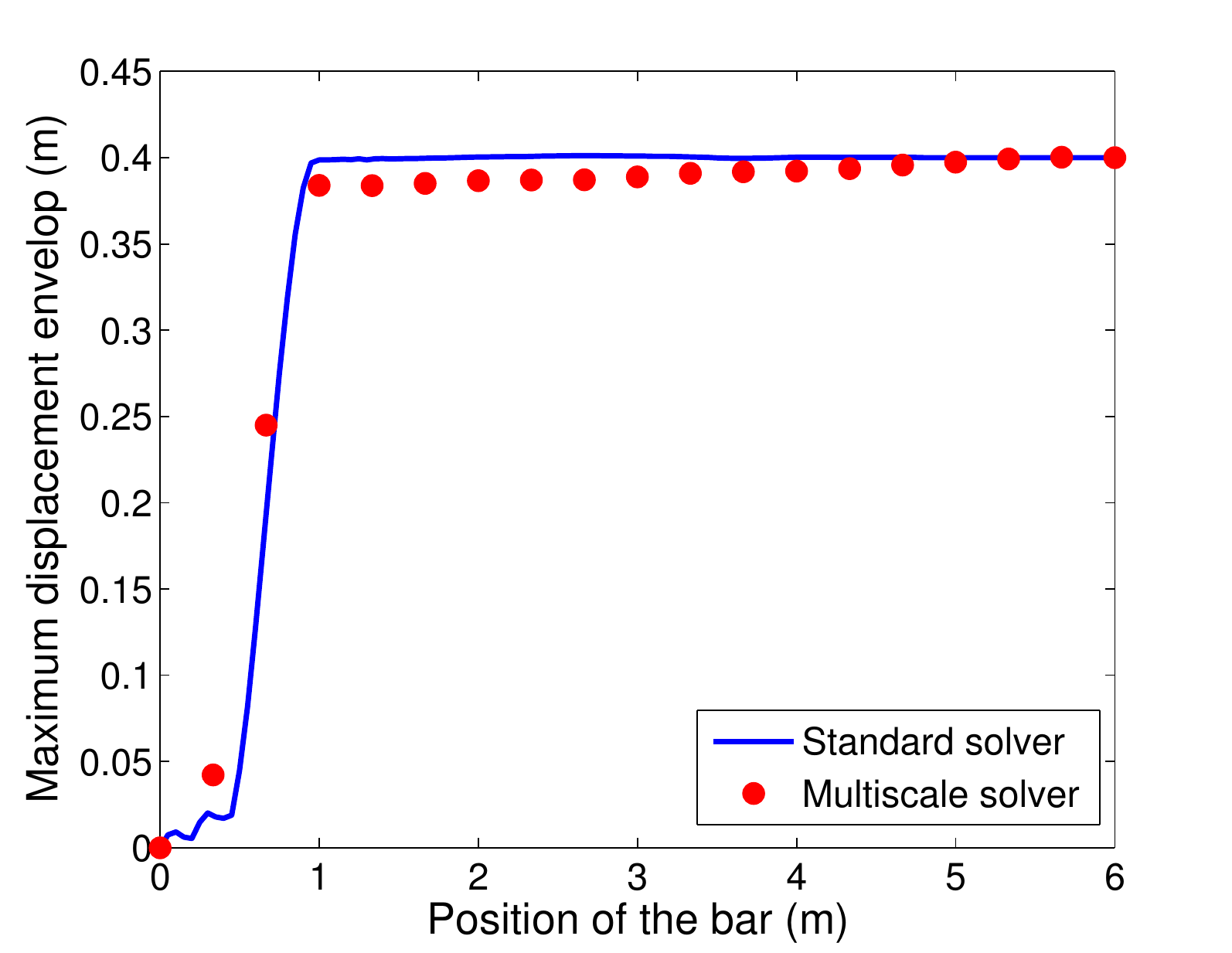}    
}     
\caption{Envelope of maximum displacement along a homogeneous bar with the multiscale solver (a) at low frequency, $f$\,=\,105\,Hz, over a time period of $t$\,=\,$[0,\,1/f]$; and (b) at high frequency $f$\,=\,480\,Hz, over $t$\,=\,$[0,\,4.5/f]$. The parameters for the simulation are $E_{\text{eff}}$\,=\,0.182\,GPa, $\rho_{\text{eff}}$\,=\,505\,kg/m, $r=0.6$, $s$\,=\,1/2, $N$\,=\,19, $n$\,=\,5, $n_s$\,=\,121,
$\Delta t$\,=\,1$\times 10^{-7}$\,s.}  \label{fig:NumericalDispersion}
\end{center}
\end{figure}

It was demonstrated in Section \ref{Sec:ConvergenceAnalyses} that wave propagation solutions by QEMS converge as the spatial and temporal discretization are refined, and that the converged response is identical to that of a single scale finite element solution, cf.~Section \ref{Sec:ConvergenceAnalyses} and \ref{Sec:DispersionProperty}. At finite resolution though, numerical errors will exist and those are inherent to any approximation scheme. For a fixed discretization, these errors will be more significant at high frequencies and they can be attributed to two (related) factors: the decrease of separation of length scales as the frequency increase, and the reduction on the number of degrees of freedom per wavelength at higher frequencies. The first factor is inherent to the multiscale nature of any FE$^2$ solver and will induce errors in the scale transition. In particular, it will lead to spurious numerical dispersion effects at high frequencies. To quantigy those, we exercise the multiscale solver with a homogeneous material at various frequencies and analyze the maximum displacement envelope. As shown in Figure \ref{fig:NumericalDispersion}, the numerical dispersion is vanishingly small at low frequencies, as expected, and only reaches a value of 2.5\% at a high frequency. We note that at such frequency, the separation of scales is $H_R/\lambda_\text{eff}$\,=\,6.25 ($\lambda_\text{eff}=v_\text{eff}\,f$\,=\,1.25\,m, $H_R$\,=\,0.2\,m), and we are therefore at the limit of applicability of the FE$^2$ procedure.

\begin{figure} [htbp]
\begin{center}  
\subfigure[Temporal evolution] {\label{MSvsSSat125Hz}     
\includegraphics[width=0.481\columnwidth]{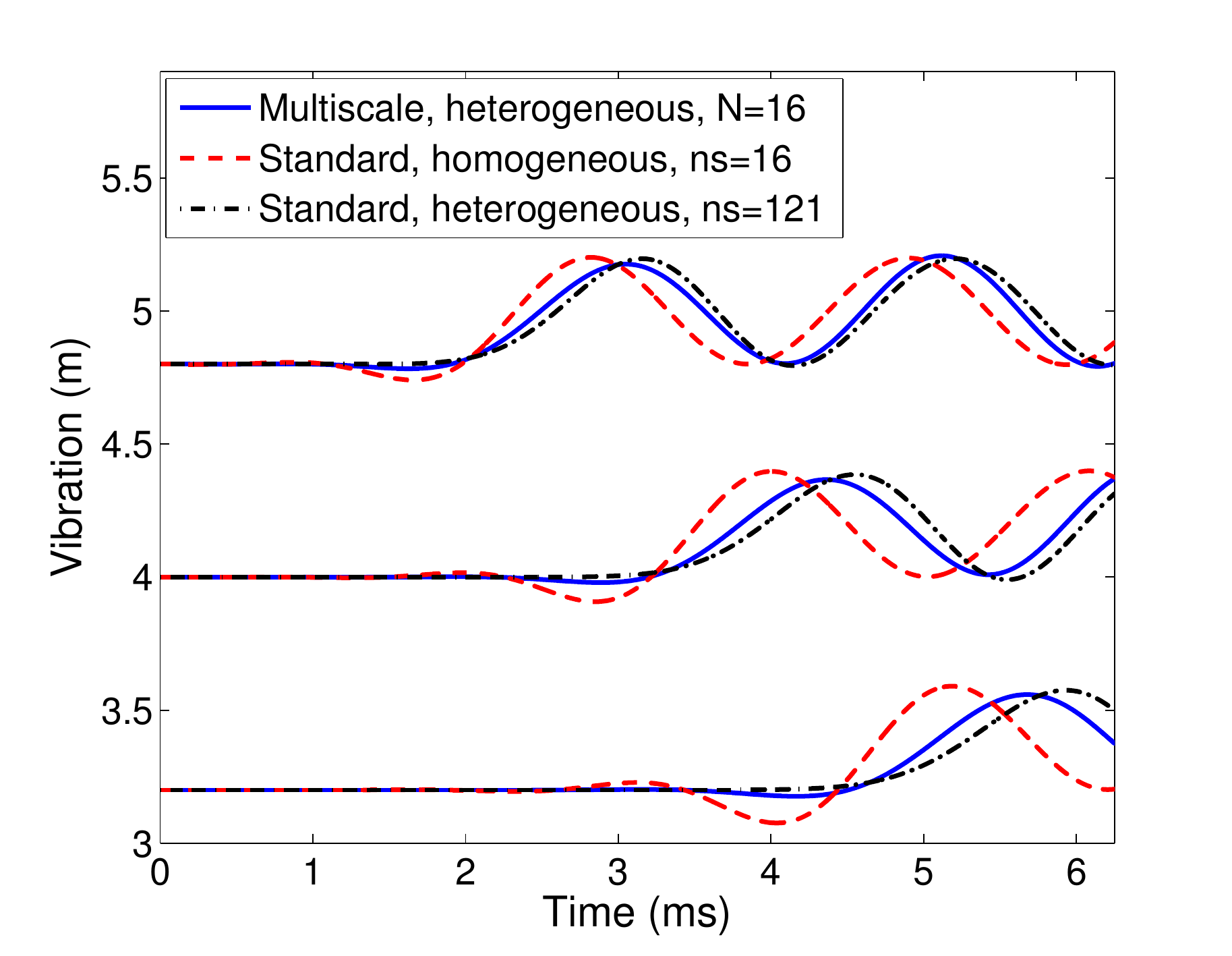} 
}    
\subfigure[Frequency dependence] { \label{Fig:NumericalErrors_Wave} 
\includegraphics[width=0.481\columnwidth]{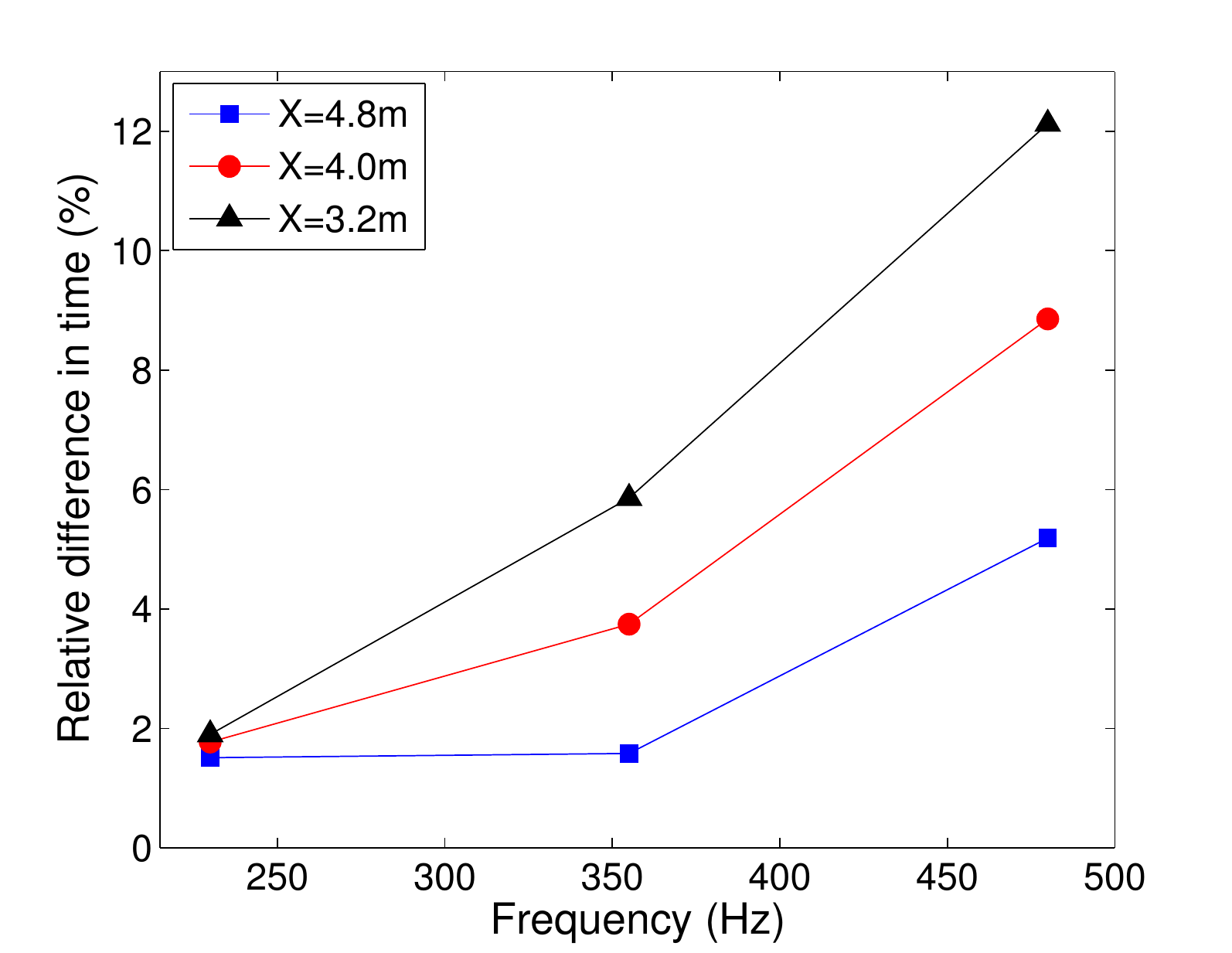}    
}    
\caption{(a) Vibration at $X$\,=\,3.2\,m, 4.0\,m, and 4.8\,m of a heterogeneous bar by different solvers at $t$\,=\,3/$f$, for a periodic excitation with $f$\,=\,480\,Hz. (b) Relative difference in time between the maxima of the wave at $t=3/f$ by the multiscale and standard solver, normalized by the period of the external excitation. The simulations were performed for RVE A, and the parameters of the simulation are $N$\,=\,16, $r$\,=0.5, $n$\,=\,5, $\Delta t$\,=\,1$\times 10^{-7}$\,s. } \label{MSvsSSat125Hz}    
\end{center}
\end{figure}

The second source of numerical error (reduction in the number of degrees of freedom) applies, in principle, to any discretization scheme regardless of their single or multi scale nature. The number of degrees of freedom used to approximate the displacement field will affect the effective stiffness of the material and therefore the speed of the waves propagating through the composite \citep{bathe2006finite}. To examine this effect, we compare the temporal evolution of the displacement field at three points of the bar, cf.~Figure \ref{MSvsSSat125Hz}, with the multiscale solver and standard solver with the same number of elements, i.e., $n_s$\,=\,$N$, and with higher resolution. To obtain the results for $n_s$\,=\,$N$ with the standard solver, the bar had to be considered homogeneous (with the effective static properties), since the mesh size exceeds, by construction (separation of length scales in the multiscale solver), the sublayers' thickness. The results, shown in Figure \ref{MSvsSSat125Hz} for RVE A, indicate that the QEMS solution lies in between the fully resolved standard FE solution for the heterogeneous bar, and the coarse standard finite element response, both in terms of wave speed, and displacement profile. This is to be expected, as the number of degrees of freedom of the multiscale discretization (counting the macro and micro nodes) lies in between the other two discretizations. The error between the high resolution standard FE solution and that given by the coarser (multiscale) discretization is measured quantitatively with the difference in time between the two maxima at a given instant of time (note that the velocity is not constant and can thus not be represented as a unique value per frequency), and is shown in Figure~\ref{Fig:NumericalErrors_Wave} as a function of the frequency.  As expected form the above reasoning in terms of degrees of freedom per wavelength, the error decreases with the frequency. 

In general FE$^2$ methods, one has to additionally account for the stiffening or softening effects that different boundary conditions on the RVE have on the multiscale response: stress boundary conditions, affine or periodic displacement boundary conditions \citep{peric2011micro,terada2000simulation,coenen2012multi}. Periodic boundary conditions are known to deliver most accurate results. However, in the one dimensional example studied here, both type of displacement boundary conditions (affine and periodic) are equivalent, and consequently, the only stiffnening factor in the results above is the reduction of degree of freedoms intrinsic to any multiscale or coarse-graining procedure. 

We finally note that numerical errors may be quantified with appropriate convergence analyses as the ones done in Section \ref{Sec:ConvergenceAnalyses} and with dispersion analyses as those of Figure~\ref{fig:NumericalDispersion}. These errors may be reduced, if the separation of length scales permit, by increasing the spatio-temporal resolution. We note that better separation of length scales can be achieved for higher-dimensional structures than for one dimensional systems, and those will be analyzed in a separate publication.


\section{Conclusion} \label{Sec:Conclusions}
We have developed a general scheme for the micro-macro transition for general heterogeneous materials under arbitrary loading conditions, including dynamics and the presence of body forces. The homogenization procedure is based on a discrete version of Hill's averaging relations \citep{chenchen2015}. This discrete setting naturally allows to recast any static or dynamic problem, as an incremental minimum principle, which is fully analogous to the classical static setting with no body-forces. This observation enables a general variational treatment of the coarse-graining procedure, that directly provides the macroscopic equilibrium equations and the micro-macro relations without the need of any additional kinematic assumption. This general scheme provides a unified treatment of the various micro-macro relations obtained for different loading conditions; and, interestingly, it may also be applied to discrete systems, such as atomistic simulations, delivering the classical Virial stress formula as the continuum Cauchy stress for an equilibrium sample.

The spatial homogenization strategy has been applied to an explicit dynamic time integration scheme, delivering a concurrent quasi-explicit {$\textup{FE}^2$} procedure. The coupling between the two scales forbids a purely explicit method. However, the resulting formulation has the noteworthy property that the Hessian underlying the numerical scheme is constant and thus enables an exact implicit solution with a single Newton-Raphson iteration. This holds true and exactly, regardless of the non-linear or history dependent nature of the microconstituents. The Hessian only depends on the micro and macro discretization and can be precomputed at the beginning of the simulation. This leads to an efficient quasi-explicit multiscale solver (QEMS). We further note that QEMS concurrently solves for the micro and macro degrees of freedom, in contrast to the sequential minimization traditionally used to solve FE$^2$ problems. This leads not only to a potentially faster algorithm, but may also deliver a more reliable solution, as demonstrated for quasi-continuum methods \citep{sorkin2014local}. The precise computational cost of the QEMS solver with respect to standard FE$^2$ procedures and its scalability with the number of degrees of freedom will be analyzed in a separate publication where a complex three dimensional dynamic problem will be examined with inelastic constituents. As a first step, the present study primarily focused on the applicability of the general framework and its physical reliability. Towards that endeavor, we implemented this multiscale solver and applied it to a one-dimensional boundary problem of an elastic layered composite under periodic dynamic excitation. 
Conventional spatial and temporal convergence analyses demonstrated the accuracy and stability of the multiscale solver. As expected, the computational calculation showed that the coupled micro-macro boundary problem can be solved in a single Newton-Raphson iteration, where the numerical error is certified to be pure system error induced by machine precision. The consistency between QEMS and a standard single scale finite element implementation is demonstrated for both homogeneous and heterogeneous systems over the range of frequencies permitted by the separation of length scales. As expected, the homogenization procedure over a material with spatially uniform properties, leaves its dynamic behavior invariant. For the heterogeneous case, an excellent agreement of the dispersive behavior of the material is obtained for different microstructures, even for relatively coarse discretizations, validating the multiscale scheme. Detailed analyses of the so-called `static' and `dynamic' part of the stresses show that both terms are significantly affected by the dynamic nature of the microscopic equilibrium equation used at the RVE level; and surprisingly, the `static' contribution appears to have a crucial role in the dispersion of the material.  Several numerical aspects of the multiscale scheme are also analyzed in detail for this one-dimensional example. In particular, the equivalence of different RVE choices for the same microstructure is demonstrated for a large range of frequencies. Additionally,  
awareness is drawn on the numerical errors pertaining to any coarse-graining computational scheme to capture wave propagation, as the separation of length scales draws a lower bound on the potential mesh refinement to reduce numerical errors. 

This one-dimensional example provides a proof of concept for the proposed QEMS scheme and delivers
detailed understanding of the impact that the parameters of the multiscale framework have on the effective dynamic response of the material. Its application towards more realistic examples in higher dimensions will be treated elsewhere.

\section*{Appendix A: Discrete averaging results with periodic boundary conditions }
 
 We derive in this appendix the relations $\frac{\partial \langle W_{\text{eff}} \rangle}{\partial \p^\M}$ and $\frac{\partial \langle W_{\text{eff}} \rangle}{\partial \F^\M}$ for general static and dynamic finite element problems with periodic boundary conditions, i.e., the boundary nodes $\{c\}$ satisfy
 \begin{equation} \label{Eq_AA_bc}
\delta\varphi^{n+1}_{ci} = \delta\varphi^{\M,n+1}_{i} +\delta F^{\M,n+1}_{iQ} X_{cQ}+ \delta\tilde{\varphi}^{n+1}_{ci},\quad \text{with $\tilde{\p}_c$ periodic.}
\end{equation}

\subsection*{ Static problem with body forces.}

Minimization of the potential energy given in Eq.~(\ref{Eq:Pi_body_discrete}), gives 
\begin{equation} \label{Eq_AA_S1}
\begin{split}
&\frac{\partial \Pi}{\partial \varphi_{bi}} = 0 = \int_{\Omega_0} P_{iJ} N_{b,J} \, dV - \int_{\Omega_0} \rho_0 B_i  N_b \, dV,\\
& \frac{\partial \Pi}{\partial \varphi_{ci}} = 0 = \int_{\Omega_0} P_{iJ} N_{c,J} \delta \tilde{\varphi}_{ci}\, dV - \int_{\Omega_0} \rho_0 B_i  N_c \delta \tilde{\varphi}_{ci}\, dV
\end{split}
\end{equation}
for the interior nodes $\{b\}$ and outer nodes $\{c\}$, respectively.

Variation of the macroscopic effective energy density with respect to the macroscopic degrees of freedom reads

\begin{equation}
\begin{split}
|\Omega|&\delta \langle W_{\text{eff}}\rangle= \int_{\Omega_0} P_{iJ} \sum_a \delta \varphi_{ai} N_{a,J} \, dV -\int_{\Omega_0} \rho_0 B_i \sum_a \delta \varphi_{ai} N_a \, dV =\sum_b \delta \varphi_{bi} \left[\int_{\Omega_0} \left(P_{iJ} N_{b,J}-\rho_0 B_i N_b \right) \, dV \right]  \\
&\phantom{=}+ \sum_c \delta \varphi_{ci} \left[\int_{\Omega_0} \left(P_{iJ} N_{c,J}-\rho_0 B_i N_c \right) \, dV \right]   =  \sum_c \delta \varphi_{ci} \left[\int_{\Omega_0} \left(P_{iJ} N_{c,J}-\rho_0 B_i N_c \right) \, dV \right]  \\
& =  \sum_c \delta \varphi^\M_{i} \left[\int_{\Omega_0} \left(P_{iJ} N_{c,J}-\rho_0 B_i N_c \right) \, dV \right]+ \sum_c \delta F^\M_{iQ} X_{cQ} \left[\int_{\Omega_0} \left(P_{iJ} N_{c,J}-\rho_0 B_i N_c \right) \, dV \right]  \\ 
& \phantom{=}+ \sum_c  \left[\int_{\Omega_0} \left(P_{iJ} N_{c,J}-\rho_0 B_i N_c \right)\delta \tilde{\varphi}_{ci} \, dV \right],
\end{split}
\end{equation}
where we have used the microscopic equilibrium equations, cf.~Eqs.~\eqref{Eq_AA_S1} and the boundary conditions, cf.~Eq.~\eqref{Eq_AA_bc}. Then, by further application of the microscopic equilibrium equations and the properties of the shape functions, cf.~Eqs.~\eqref{Eq:Properties_Shape_Functions} and \eqref{Eq:ShapeFunctions_properties},
\begin{equation}
\begin{split}
|\Omega|&\delta \langle W_{\text{eff}}\rangle=  \sum_a \delta \varphi^\M_{i} \left[\int_{\Omega_0} \left(P_{iJ} N_{a,J}-\rho_0 B_i N_a \right) \, dV \right]  + \sum_a \delta F^\M_{iQ} X_{aQ} \left[\int_{\Omega_0} \left(P_{iJ} N_{a,J}-\rho_0 B_i N_a \right) \, dV \right]  \\
& =  \delta \varphi^\M_{i} \left[\int_{\Omega_0} \Big(P_{iJ} \sum_a N_{a,J}-\rho_0 B_i \sum_a N_a \Big) \, dV \right] +  \delta F^\M_{iQ}  \left[\int_{\Omega_0} \Big(P_{iJ} \sum_a N_{a,J} X_{aQ}-\rho_0 B_i \sum_a N_a X_{aQ} \Big) \, dV \right]  \\
& =  \delta \varphi^\M_{i} \left[\int_{\Omega_0} -\rho_0 B_i  \, dV \right]  +  \delta F^\M_{iJ}  \left[\int_{\Omega_0} \left(P_{iJ} -\rho_0 B_i X_J \right) \, dV \right].  \\
\end{split}
\end{equation}
Therefore,
\begin{equation}
\frac{\partial \langle W_{\text{eff}}\rangle}{\partial \boldsymbol \varphi^\M} = \langle - \rho_0 \mathbf{B} \rangle \quad \quad \text{and} \quad \quad \frac{\partial \langle W_{\text{eff}}\rangle}{\partial \mathbf{F}^\M} = \langle \mathbf{P} - \rho_0 \mathbf{B} \otimes \mathbf{X} \rangle
\end{equation}
which is identical to the results obtained in Section \ref{Sec:BodyForces} for affine boundary conditions on the RVE.

\subsection*{Dynamic problem with body forces}
The equilibrium equations for the interior and exterior nodes are, respectively,
\begin{align}
&0=\frac{\partial \Pi}{\partial \varphi_{bi}} = \int_{\Omega_0} \left[ 2\rho_0 \frac{ \sum_a (\varphi^{n+1}_{ai}-\varphi^{n+1,pre}_{ai})N_a}{\Delta t^2}  N_b + P^n_{iJ} N_{b,J} - \rho_0 B_i^{n} N_b \right] \, dV, \\ \label{BCc1c2}
&0=\frac{\partial \Pi}{\partial \varphi_{ci}} \quad \rightarrow  \quad \int_{\Omega_0} \left[ 2\rho_0 \frac{ \sum_a (\varphi^{n+1}_{ai}-\varphi^{n+1,\text{pre}}_{ai})N_a}{\Delta t^2}  N_c + P^n_{iJ} N_{c,J} - \rho_0 B_i^{n} N_c \right] \delta \tilde{\varphi}_{ci}\, dV=0.
\end{align}
Variations of the potential energy evaluated at the equilibrium solution lead
\begin{equation}
\begin{split}
|\Omega_0|\delta \langle W_{\text{eff}}\rangle = & \int_{\Omega_0} \sum_b \delta \varphi^{n+1}_{bi} \left[ 2\rho_0 \frac{ \sum_a (\varphi^{n+1}_{ai}-\varphi^{n+1,\text{pre}}_{ai})N_a}{\Delta t^2}  N_b + P^n_{iJ} N_{b,J} - \rho_0 B_i^{n} N_b \right] \, dV \\
&+ \int_{\Omega_0} \sum_c \delta \varphi^{n+1}_{ci} \left[ 2\rho_0 \frac{ \sum_a (\varphi^{n+1}_{ai}-\varphi^{n+1,\text{pre}}_{ai})N_a}{\Delta t^2}  N_c + P^n_{iJ} N_{c,J} - \rho_0 B_i^{n} N_c \right] \, dV \\
=& \int_{\Omega_0} \sum_c \delta \varphi^{n+1}_{ci} \left[ 2\rho_0 \frac{ \sum_a (\varphi^{n+1}_{ai}-\varphi^{n+1,\text{pre}}_{ai})N_a}{\Delta t^2}  N_c + P^n_{iJ} N_{c,J} - \rho_0 B_i^{n} N_c \right] \, dV, 
\end{split}
\end{equation}
where we have made use of the equilibrium equations. Then, applying the boundary conditions for nodes $c$, cf.~\eqref{Eq_AA_bc}, and the equilibrium equations for the boundary nodes, leads
\begin{equation}
\begin{split}
|\Omega_0|&\delta \langle W_{\text{eff}}\rangle =  \int_{\Omega_0} \sum_c \delta \varphi^{\M,n+1}_{i} \left[ 2\rho_0 \frac{ \sum_a (\varphi^{n+1}_{ai}-\varphi^{n+1,\text{pre}}_{ai})N_a}{\Delta t^2}  N_c + P^n_{iJ} N_{c,J} - \rho_0 B_i^{n} N_c \right] \, dV \\
&+ \int_{\Omega_0} \sum_c\delta F^{\M,n+1}_{iQ} X_{cQ} \left[ 2\rho_0 \frac{ \sum_a (\varphi^{n+1}_{ai}-\varphi^{n+1,\text{pre}}_{ai})N_a}{\Delta t^2}  N_c + P^n_{iJ} N_{c,J} - \rho_0 B_i^{n} N_c \right] \, dV \\
&+ \int_{\Omega_0} \sum_c \delta \tilde{\varphi}^{n+1}_{ci} \left[ 2\rho_0 \frac{ \sum_a (\varphi^{n+1}_{ai}-\varphi^{n+1,\text{pre}}_{ai})N_a}{\Delta t^2}  N_c + P^n_{iJ} N_{c,J} - \rho_0 B_i^{n} N_c \right] \, dV \\
= & \int_{\Omega_0} \sum_a \delta \varphi^{\M,n+1}_{i} \left[ 2\rho_0 \frac{ \sum_{a'} (\varphi^{n+1}_{a'i}-\varphi^{n+1,\text{pre}}_{a'i})N_{a'}}{\Delta t^2}  N_a + P^n_{iJ} N_{a,J} - \rho_0 B_i^{n} N_a \right] \, dV \\
&+ \int_{\Omega_0} \sum_a\delta F^{\M,n+1}_{iQ} X_{aQ} \left[ 2\rho_0 \frac{ \sum_{a'} (\varphi^{n+1}_{a'i}-\varphi^{n+1,\text{pre}}_{a'i})N_{a'}}{\Delta t^2}  N_a + P^n_{iJ} N_{a,J} - \rho_0 B_i^{n} N_a \right] \, dV. \\
\end{split}
\end{equation}
Next, we simplify the equations above by making use of the properties of the shape functions, cf.~Eqs.~\eqref{Eq:Properties_Shape_Functions} and \eqref{Eq:ShapeFunctions_properties}, and obtain
\begin{equation}
\begin{split}
|\Omega_0|&\delta \langle W_{\text{eff}}\rangle =\delta \varphi^{\M,n+1}_{i} \int_{\Omega_0} \ \left[ 2\rho_0 \frac{ \sum_{a'} (\varphi^{n+1}_{a'i}-\varphi^{n+1,\text{pre}}_{a'i})N_{a'}}{\Delta t^2}   \sum_a N_a + P^n_{iJ}   \sum_a N_{a,J} - \rho_0 B_i^{n}   \sum_a N_a \right] \, dV \\
&+ \delta F^{\M,n+1}_{iQ} \int_{\Omega_0}  \left[ 2\rho_0 \frac{ \sum_{a'} (\varphi^{n+1}_{a'i}-\varphi^{n+1,\text{pre}}_{a'i})N_{a'}}{\Delta t^2}  \sum_a X_{Qa} N_a + P^n_{iJ} \sum_a X_{aQ} N_{a,J} - \rho_0 B_i^{n}  \sum_a X_{aQ} N_a \right] \, dV \\
 =&\delta \varphi^{\M,n+1}_{i} \int_{\Omega_0} \ \left[ 2\rho_0 \frac{ \sum_{a'} (\varphi^{n+1}_{a'i}-\varphi^{n+1,\text{pre}}_{a'i})N_{a'}}{\Delta t^2}  - \rho_0 B_i^{n}  \right] \, dV \\
&+ \delta F^{\M,n+1}_{iQ} \int_{\Omega_0}  \left[ 2\rho_0 \frac{ \sum_{a'} (\varphi^{n+1}_{a'i}-\varphi^{n+1,\text{pre}}_{a'i})N_{a'}}{\Delta t^2}  X_{Q}  + P^n_{iQ}  - \rho_0 B_i^{n} X_{Q} \right] \, dV. 
\end{split}
\end{equation}
Therefore,
\begin{align}
&\frac{\partial \langle W_{\text{eff}}\rangle}{\partial \p^{\M,n+1}} = \bigg \langle 2\rho_0 \frac{ \sum_{a} (\p^{n+1}_{a}-\p^{n+1,pre}_{a})N_{a}}{\Delta t^2}  - \rho_0 \B^{n}  \bigg \rangle\\
&\frac{\partial \langle W_{\text{eff}}\rangle}{\partial \F^{\M,n+1}} = \bigg \langle \Pb^n + \left( 2\rho_0 \frac{ \sum_{a} (\p^{n+1}_{a}-\p^{n+1,pre}_{a})N_{a}}{\Delta t^2}   - \rho_0 \B^{n} \right) \otimes \X \bigg \rangle.
\end{align}
These results are identical to those of Section \ref{Sec:DiscreteDynamicProblem} for affine displacement boundary conditions.

\section*{Appendix B: Continuum averaging results with periodic boundary conditions }
Analogous results to those of Appendix A, in the continuum setting (in space), are derived below.

\subsection*{Static case with body forces}
Minimization of the potential energy given by Eq.~(\ref{Eq: PiBodyforceCon}) under periodic boundary conditions gives
\begin{equation} 
\delta \Pi = \int_{\Omega_0} P_{iJ} \delta \varphi_{i,J} \, dV -  \int_{\Omega_0} \rho_0 B_i \delta \varphi_i \, dV= \int_{\partial \Omega_0} P_{iJ}  \delta \tilde{\varphi}_i N_J \, dS  -\int_{\Omega_0} \big( P_{iJ,J} +  \rho_0 B_i \big) \delta \varphi_i \, dV =0, 
\end{equation}
where we have used the relation $\delta \varphi_i=\delta \tilde{\varphi}_i$ on the boundary. Then, according to the periodicity of $\tilde{\varphi}$, the equilibrium equations give
\begin{equation}\label{EQ}
\nabla \cdot \mathbf{P} + \rho_0 \mathbf{B} = 0, \quad \text{for } \mathbf{ X} \in \Omega_0.
\end{equation}
\begin{equation}\label{PN}
P_{iJ}N_J|_{L}=-P_{iJ}N_J|_{L^*} \quad \text{for } \mathbf{ X} \in \partial \Omega_0,
\end{equation}
where $L$ and $L^*$ are two opposite sides of the RVE, for instance, left and right, or top and bottom, for a square domain in two dimensions.

Next, we consider the average of the microscopic potential energy density, $W_{\text{eff}}=W(\nabla \boldsymbol \varphi,\mathbf{X})- \rho_0 \mathbf{B} \cdot \boldsymbol \varphi$, evaluated at the equilibrium solution just derived, and take variations with respect to the macroscopic fields $\boldsymbol \varphi^\M$ and $\mathbf{F}^\M$, 
\begin{equation}
\begin{split}
|\Omega_0|\delta \langle W_{\text{eff}} \rangle&= \int_{\Omega_0} P_{iJ} \delta \varphi_{i,J} \, dV -\int_{\Omega_0} \rho_0 B_i \delta \varphi_i\, dV =\int_{\partial \Omega_0} P_{iJ} N_J  \delta \varphi_i \, dS - \int_{\Omega_0} \left(P_{iJ,J} + \rho B_i\right)\delta \varphi_i \, dV  \\
&=  \left[ \int_{\partial \Omega_0} P_{iJ} N_J \, dS \right] \delta \varphi^\M_i+ \left[ \int_{\partial \Omega_0} P_{iJ} N_J X_P \, dS \right] \delta \varphi^\M_{i,P}+ \int_{\partial \Omega_0} P_{iJ} N_J \delta \tilde{\varphi_i}\, dS.  \\
\end{split}
\end{equation}
The last term vanish by the periodicity of the field $\delta \tilde{\varphi_i}$ and Eq.~\eqref{PN}. Thus,
\begin{equation}
\frac{\partial \langle W_{\text{eff}} \rangle}{\partial \p^\M} = \langle \nabla \cdot \Pb\rangle = \langle -\rho_0 \B \rangle\quad  \quad \text{and} \quad \quad \frac{\partial \langle W_{\text{eff}} \rangle}{\partial \F^\M} = \langle \Pb + \nabla \cdot \Pb \otimes \X \rangle= \langle \Pb -\rho_0 \B\otimes \X \rangle.
\end{equation}

\subsection*{Dynamic case with body forces}

The minimum of the incremental potential energy given by Eq.~\eqref{Eq:Pi_dynamic}, gives for the case of periodic boundary conditions
\begin{equation} 
 \begin{split}
\delta \Pi &= \int_{\Omega_0}\bigg[2\rho_0\frac{ \varphi^{n+1}_i-\boldsymbol \varphi^{n+1,\text{pre}}_i}{\Delta t^2}\delta \varphi_i^{n+1}+ P_{iJ}^n \delta \varphi_{i,J}^{n+1} - \rho_0 B_i^n \delta \varphi_i^{n+1} \bigg]\, dV\\
&=  \int_{\partial \Omega_0} P_{iJ}^{n}N_J  \delta \tilde{\varphi}_i^{n+1} \, dS+ \int_{\Omega_0} \bigg[2\rho_0\frac{ \varphi_i^{n+1}-\boldsymbol \varphi^{n+1,\text{pre}}_i}{\Delta t^2} - P_{iJ,J}^{n} -  \rho_0 B_i^n \bigg] \delta \varphi_i^{n+1} \, dV.
 \end{split}
 \end{equation}
The equilibrium equations then read
\begin{equation}
2 \rho_0 \frac{\boldsymbol \varphi^{n+1}-\boldsymbol \varphi^{n+1,\text{pre}}}{\Delta t^2} = \nabla \cdot \mathbf{P}^n + \rho_0 \mathbf{B}^n, \quad \text{for } \mathbf{X} \in \Omega_0,
\end{equation}
and
\begin{equation}
 \int_{\partial \Omega_0} P_{iJ}^{n}N_J  \delta \tilde{\varphi}_i^{n+1} \, dS=0 \quad \text{for } \mathbf{X} \in \partial \Omega_0,
\end{equation}

Consider now the average of incremental effective strain energy density, $W^{n+1}_{\text{eff}}$, defined in Section \ref{SubSec:DynamicContinuum}. Its variation with respect to the macroscopic fields $\boldsymbol \varphi^\M$ and $\mathbf{F}^\M$ gives
\begin{equation}
\begin{split}
 |\Omega_0|\delta \langle W^{n+1}_{\text{eff}} \rangle&=  \int_{\Omega_0}\bigg[2\rho_0\frac{ \varphi^{n+1}_i-\boldsymbol \varphi^{n+1,\text{pre}}_i}{\Delta t^2}\delta \varphi_i^{n+1}+ P^n_{iJ} \delta \varphi_{i,J}^{n+1} - \rho_0 B_i^n \delta \varphi_i^{n+1} \bigg]\, dV \\
 & = \int_{\Omega_0} \bigg[2\rho_0\frac{ \varphi_i^{n+1}-\boldsymbol \varphi^{n+1,\text{pre}}_i}{\Delta t^2} - P_{iJ,J}^{n} -  \rho_0 B_i^n \bigg] \delta \varphi_i^{n+1} \, dV + \int_{\partial \Omega_0}P^n_{iJ} N_J \delta \varphi_i^{n+1} \, dV \\
 & = \delta \varphi_i^{\M,n+1} \bigg[\int_{\Omega_0}P^n_{iJ,J} \, dV \bigg] + \delta F^{\M,n+1}_{iQ}  \bigg[\int_{\Omega_0} \left(P^n_{iJ}X_Q \right)_{,J} \, dV \bigg] + \int_{\Omega_0}P^n_{iJ} N_J \delta \tilde{\varphi}_i^{n+1}dV.
 \end{split}
\end{equation}
The last term vanishes by the periodic boundary conditions and associated equilibrium equations and
\begin{align}
&\frac{\partial \langle W^{n+1}_{\text{eff}} \rangle}{\partial \p^{\M,n+1}} =\langle \nabla \cdot \Pb^n \rangle = \bigg \langle 2 \rho_0 \frac{ \p^{n+1}-\p^{n+1,\text{pre}}}{\Delta t^2} -  \rho_0 \B^n \bigg \rangle\\
&\frac{\partial \langle W^{n+1}_{\text{eff}} \rangle}{\F^{\M,n+1}} =\langle \Pb^n +  \nabla \Pb^n \otimes \X \rangle  = \Big \langle \Pb^n + \Big(2 \rho_0 \frac{ \p^{n+1}-\p^{n+1,\text{pre}}}{\Delta t^2} -  \rho_0 \B^n \Big)\otimes \X \Big \rangle.
\end{align}

\section*{Acknowledgements}
C. Reina thanks Prof. Michael Ortiz for interesting discussions.

\end{document}